\documentclass[11pt]{article}
\usepackage[T1]{fontenc}
\usepackage[utf8]{inputenc}
\usepackage{lmodern}
\usepackage[margin=1in]{geometry}
\usepackage{amsmath,amssymb,amsthm}
\usepackage{microtype}
\usepackage{parskip}
\usepackage{booktabs,tabularx}
\usepackage{flafter}
\usepackage{graphicx}
\usepackage[numbers,sort&compress]{natbib}
\usepackage[hidelinks]{hyperref}
\usepackage{algorithm}
\usepackage{algpseudocode}
\usepackage{xcolor}
\usepackage{array}
\allowdisplaybreaks[1]
\makeatletter
\newcommand{\printcompactcontents}{
  \clearpage
  \begingroup
  \small
  \setlength{\parskip}{0pt}
  \renewcommand*{\l@section}[2]{
    \addvspace{3pt}
    \@dottedtocline{1}{0em}{1.6em}{\bfseries ##1}{\bfseries ##2}
  }
  \tableofcontents
  \endgroup
  \clearpage
}
\makeatother

\newtheorem{theorem}{Theorem}[section]
\newtheorem{lemma}[theorem]{Lemma}

\newtheorem{corollary}[theorem]{Corollary}
\newtheorem{fact}[theorem]{Fact}
\theoremstyle{definition}
\newtheorem{definition}{Definition}[section]

\theoremstyle{plain}
\newcommand{\R}{\mathbb R}
\newcommand{\E}{\mathbb E}
\newcommand{\Prob}{\mathbb P}
\newcommand{\Law}{\mathcal L}
\newcommand{\Normal}{\mathcal N}
\newcommand{\TV}{\operatorname{TV}}
\newcommand{\tr}{\operatorname{tr}}
\newcommand{\one}{\mathbf 1}
\newcommand{\op}{\mathrm{op}}
\hypersetup{
  pdftitle={Optimal Private Gaussian Multi-sampling},
  pdfsubject={Complete appendix proofs with exact independent outputs}
}
\title{Optimal Private Gaussian Multi-sampling}
\author{
Raoof Zare Moayedi \quad
Mohammad Hossein Yassaee \quad
Javad B. Ebrahimi\\[0.5em]
\normalsize Sharif University of Technology\\[0.3em]
\texttt{raoofmoayedi2000@gmail.com},\quad
\texttt{yassaee@sharif.edu},\quad
\texttt{javad.ebrahimi@sharif.ir}
}
\date{}
\makeatletter
\renewenvironment{proof}[1][\proofname]{\par
  \pushQED{\qed}
  \normalfont \topsep6\p@\@plus6\p@\relax
  \trivlist
  \item[\hskip\labelsep\itshape #1\@addpunct{.}]\leavevmode\ignorespaces
}{\popQED\endtrivlist\@endpefalse}
\makeatother
\begin{document}
\maketitle

\begin{abstract}
We give polynomial-time algorithms for Gaussian multi-sampling under
approximate $(\varepsilon,\delta)$-differential privacy, with optimal
sample complexity up to logarithmic factors.
Our main construction uses
$\widetilde O(m+d\sqrt m/\varepsilon)$ independent input samples
from a $d$-dimensional Gaussian $P$ with unknown mean and covariance
to generate $m$ exactly independent outputs.
Their joint distribution is $\alpha$-close in total variation to
$P^{\otimes m}$.
Our lower bounds hold even for dependent outputs under the same
joint accuracy guarantee, showing that exact independence incurs
no additional sample cost beyond logarithmic factors.

The single-sampler of Iverson, Kamath, and Mouzakis [COLT 2025]
uses $\widetilde O(d/\varepsilon)$ input samples.
Running it on a separate input dataset for each of $m$ outputs
therefore costs $\widetilde O(md/\varepsilon)$ in total.
Our construction reduces this dimension-dependent sample cost
from linear to square-root dependence on the number of outputs.

Our approach builds on the stable weighted statistics of Brown,
Hopkins, and Smith [COLT 2023], also used for Gaussian single-sampling
by Iverson, Kamath, and Mouzakis.
To control privacy loss when the entire batch shares these statistics,
we compare the joint output densities and obtain a privacy bound
that scales with the square root of the batch size.
The stability step downweights influential records and can reject
the entire batch. Both actions can make the outputs dependent.
Our correction restores exact independence while preserving
differential privacy and joint accuracy.
\end{abstract}

\printcompactcontents

\section{Introduction}\label{main:introduction}
Suppose we are given independent samples from an unknown distribution.
Without privacy, releasing the first $m$ input samples gives $m$
independent outputs with the exact product distribution.
When the input contains sensitive data, however, direct release
exposes the selected records.
Differential privacy requires the output distribution to change
little when any one input sample is replaced, on every input
dataset~\cite{DMNS06,DworkRoth2014}.
Generating such a batch while protecting the contributing
individuals is a basic goal of private synthetic data
generation~\cite{BLR08,HLM12}.
We ask how much input data is needed to achieve this goal.

A natural approach is to privately learn a distribution and then
draw from the learned model~\cite{BKSW21,AAK21,AL22}.
Once such an estimate is released, postprocessing protects every
subsequent draw without further privacy cost. Its accuracy determines
the quality of the generated batch. Even with independent sampling
randomness, all outputs depend on the same fitted model and therefore
need not be independent.
Private single-sampling, introduced in~\cite{RSSS21}, establishes
a separation from learning: for several discrete distribution
families and certain parameter regimes, generating one output
requires asymptotically fewer input samples than privately learning
the distribution.

We study this question for Gaussian distributions with unknown mean
and covariance. For input drawn from a Gaussian $P$, we require
exactly independent outputs whose joint distribution is
$\alpha$-close in total variation to
$P^{\otimes m}$~\cite{CheuNayak2025}.
For a $d$-dimensional Gaussian, the single-sampler of Iverson,
Kamath, and Mouzakis~\cite{IKM25} uses
$\widetilde O(d/\varepsilon)$ input samples.
A direct approach to obtain $m$ independent outputs is to run
a single-sampler on a separate input dataset for each output,
using independent randomness and error $\alpha/m$ for every call.
Applied to~\cite{IKM25}, this uses
$\widetilde O(md/\varepsilon)$ input samples in total.
Sharing the input could reduce this cost, but the same randomly
drawn records then influence every call.

\begin{quote}
\emph{Can the input be shared to reduce the sample cost of private
Gaussian multi-sampling while preserving joint accuracy and exact
independence?}
\end{quote}

\paragraph{This work.}
Our main result gives polynomial-time algorithms for private
Gaussian multi-sampling with near-optimal sample complexity.
The outputs are exactly independent on Gaussian input.
For $0<\varepsilon\le1$, our main $(\varepsilon,\delta)$-private
construction generates $m$ outputs using
$\widetilde O(m+d\sqrt m/\varepsilon)$ input samples, without
bounds on the unknown mean or covariance.
For independent input samples from any Gaussian $P$ with positive
definite covariance, the joint output distribution is $\alpha$-close
in total variation to $P^{\otimes m}$.

For unrestricted Gaussians under approximate privacy, this
answers Cheu and Nayak's \cite{CheuNayak2025} question of whether Gaussian multi-sampling
can avoid the linear factor in $m$ from repeating a
single-sampler~\cite{CheuNayak2025}. Our
dimension-dependent sample cost grows as $\sqrt m$ up to
logarithmic factors.

Cheu and Nayak \cite{CheuNayak2025} also ask how much allowing dependence can help under
joint total variation accuracy. Our main lower bound shows that
allowing dependent outputs can improve the optimal sample
complexity by at most logarithmic factors.
Together with our upper bounds, it gives the characterization
$
 \widetilde\Theta_\alpha\!\left(
 m+\min\left\{\frac{d\sqrt m}{\varepsilon+\delta},
                    \frac m\delta\right\}\right)
$
for every fixed $0<\alpha<1$, throughout
$0\le\varepsilon\le1$ and $0<\delta<1$.

Ordinary empirical means and covariances can change arbitrarily
when one input record is replaced.
Our main algorithmic contribution is a method for restoring exact
product structure after stabilizing the resampling rule for privacy.
We build on the stable weighted statistics of Brown, Hopkins,
and Smith~\cite{BHS23}, also used for Gaussian single-sampling
in~\cite{IKM25}.
Their methods downweight influential records and privately test
whether the resulting statistics can be used.
Our starting point is to resample the input dataset conditional
on its empirical mean and covariance, retaining $m$ resampled
records as the output.
For Gaussian input, this conditional distribution does not depend on the
unknown parameters, and averaging over the input recovers
$P^{\otimes m}$.
Downweighting records and rejecting batches when the private test
fails change this sampling distribution.
We prove privacy for the resulting joint distribution and construct
an efficient exact correction that restores the required product
structure.

Our \textbf{first }contribution is a joint privacy comparison with
square-root dependence on the batch size.
Density comparisons already underlie the Gaussian single-sampler
of~\cite{IKM25}.
In our resampling procedure, the outputs within a candidate batch
are generally dependent.
We therefore analyze how the density of the entire batch changes
when one input record is replaced.
The resulting bound allows the entire batch to share the same
stable statistics and gives a dimension-dependent sample cost of
$\widetilde O(d\sqrt m/\varepsilon)$.

Our \textbf{second} contribution is an efficient correction that
restores exact independence after weighting and rejection. Weighting
changes the distribution of accepted batches. A single rejection
affects all $m$ positions, so even a rare rejection can introduce
dependence across the batch.

To include rejection in the mixture, we represent it by an all-blank
batch and allow independent omissions in exact resampling. At fixed
empirical statistics $\tau=(\bar x,S)$, let $F_\tau$ be the distribution
of exact resampling with omissions. Let $G_\tau$ be the distribution
of the private sampler conditional on those statistics, with the same
omissions on acceptance and all positions blank on rejection. Both
distributions describe the batch before filling the blanks.

Write $w$ for the overall probability of taking the private branch.
We construct a correction distribution $H_\tau$ such that
\[
 F_\tau=wG_\tau+(1-w)H_\tau.
\]
The correction branch adds at most $1-w$ to the additive privacy
parameter, so $w$ must be close to one. For that same $w$, the private
contribution $wG_\tau$ must fit within $F_\tau$ on every event, including
the all-blank outcome. We construct and sample the correction using
public simulations at fixed empirical statistics, for which the
conditional distribution of the Gaussian dataset is independent of
the unknown parameters. Averaging the corrected batches over the
input statistics and filling the blanks independently gives the
desired product distribution.

Our \textbf{third} contribution is a matching lower bound for jointly accurate
Gaussian multi-sampling that allows dependent outputs. In our Gaussian
family, even a small number of jointly accurate outputs reveals a
direction correlated with the unknown cross-covariance. We relate this
correlation to agreement with individual input records, which privacy
limits. The additive privacy parameter $\delta$ allows unusually large
agreement on a small amount of probability mass in each record
comparison. By bounding these contributions together across records,
we obtain the $m/\delta$ branch of the lower bound up to logarithmic
factors.

\subsection{Our results}\label{main:intro:results}

We now describe our results in more detail, beginning with the
privacy and sampling guarantees.

\begin{definition}[Differential privacy~\cite{DMNS06,DworkRoth2014}]
\label{main:def:dp}
Let $\varepsilon\ge0$ and $\delta\in[0,1]$.
A randomized mechanism $M$ is
$(\varepsilon,\delta)$-differentially private if, for every pair
of datasets $x,x'$ of the same size differing in one record
and every measurable output event $E$,
$\Pr\{M(x)\in E\}\le
e^\varepsilon\Pr\{M(x')\in E\}+\delta$.
The probabilities are over the randomness of $M$.
\end{definition}

Throughout, we consider Gaussian distributions
$P=\mathcal N(\mu,\Sigma)$ on $\R^d$, with arbitrary unknown
$\mu$ and positive definite $\Sigma$, without bounds on either
parameter. We write $\Law(Z)$ for the distribution of $Z$.
Two distributions are $\alpha$-close if their total variation
distance is at most $\alpha$.

\begin{definition}[Joint accuracy and independent multi-sampling]
\label{main:def:sampling}
Let $d,m,n\ge1$ be integers and $0\le\alpha<1$.
A randomized mechanism $M:(\R^d)^n\to(\R^d)^m$ is jointly
$\alpha$-accurate for Gaussians if, for every such $P$ and
$X\sim P^{\otimes n}$,
$\TV(\Law(M(X)),P^{\otimes m})\le\alpha$.
It is an independent $\alpha$-multi-sampler if, additionally,
$\Law(M(X))=Q_P^{\otimes m}$ for some distribution $Q_P$.
These distributions include both input and mechanism randomness.
\end{definition}

The independent version is strong multi-sampling in the
terminology of~\cite{CheuNayak2025}.
Privacy is required on every fixed dataset; joint accuracy and
independence concern Gaussian input.
Our upper bound achieves exactly independent outputs, whereas
our lower bound assumes only joint accuracy.
Table~\ref{main:table:comparison} compares our sample complexity
with prior bounds for Gaussian learning and sampling.

\paragraph{Independent Gaussian multi-sampling.}
Our main algorithmic result is the following.

\begin{theorem}[Independent Gaussian multi-sampling, informal]
\label{main:intro:upper}
\label{main:intro:theorem}
Let $d,m\ge1$ be integers, $0<\alpha,\delta<1$, and
$0<\varepsilon\le1$.
There is an $(\varepsilon,\delta)$-differentially private mechanism
using $n=\widetilde O(m+d\sqrt m/\varepsilon)$ input samples
such that, for every Gaussian $P$ under consideration and
$X\sim P^{\otimes n}$,
$\Law(M(X))=Q_P^{\otimes m}$, where
$Q_P=(1-\eta)P+\eta\mathcal N(0,I_d)$ and
$\eta=\alpha/(2m)$.
Consequently,
$\TV(Q_P^{\otimes m},P^{\otimes m})\le\alpha/2$.
The mechanism uses $O(n^3)$ operations.
\end{theorem}

Throughout, tildes suppress logarithmic factors in $d$, $m$, and
$1/\delta$. For upper bounds they also suppress logarithmic factors in
$1/\alpha$, and for the positive-$\varepsilon$ construction in
$1/\varepsilon$. A subscript $\alpha$ permits constants depending on
$\alpha$.

At $m=1$, Theorem~\ref{main:intro:upper} recovers the
$\widetilde O(d/\varepsilon)$ single-sampling bound.
For $m\ge d^2/\varepsilon^2$, it uses only
$\widetilde O(m)$ input samples, matching the linear cost required
even without privacy, as our lower bound shows.

\paragraph{Lower bounds and optimality.}
We complement our algorithmic result with a lower bound that
permits dependent outputs.

\begin{theorem}[Gaussian multi-sampling lower bound, informal]
\label{main:intro:lower}
Fix $0\le\alpha<1$. Let $d,m,n\ge1$ be integers,
$0\le\varepsilon\le1$, and $0<\delta<1$.
Every $(\varepsilon,\delta)$-differentially private mechanism
that is jointly $\alpha$-accurate in the sense of
Definition~\ref{main:def:sampling} requires
$n=\widetilde\Omega_\alpha\!\left(
m+\min\left\{
\tfrac{d\sqrt m}{\varepsilon+\delta},
\tfrac m\delta
\right\}\right)$.
The outputs need not be independent or identically distributed.
\end{theorem}

Cheu and Nayak~\cite{CheuNayak2025} study lower bounds for weak
multi-sampling of Gaussians with bounded mean and covariance under
zero-concentrated differential privacy. Their reduction uses $d+1$
generated samples as public observations for a Gaussian learner that
also receives private data. The learner's guarantee assumes Gaussian
public observations.
Our theorem addresses joint accuracy under approximate differential
privacy and permits arbitrary dependence among the outputs.
It applies to every batch size, with no restriction relating
$\delta$ to the input sample size.

\begin{table*}[t]
\centering
\caption{Sample complexity of Gaussian learning and sampling with
unknown mean and fixed small total variation error $\alpha$.
For multi-sampling, outputs are exactly independent and accuracy
is measured jointly over the entire batch.
Learning succeeds with probability at least $0.9$.
Here $0<\varepsilon\le1$ and $0<\delta<1$; our characterization
also covers $\varepsilon=0$.}
\label{main:table:comparison}
\small
\setlength{\tabcolsep}{4pt}
\renewcommand{\arraystretch}{1.15}
\begin{tabular}{@{}lllll@{}}
\toprule
Task & Reference & Covariance & Privacy & Sample Complexity\\
\midrule
Learn
 & \cite{AshtianiEtAl2020,DMR20}
 & Unbounded & None
 & $\widetilde\Theta(d^2)$ \\
Learn
 & \cite{AL22,HKMN23}
 & Unbounded & $(\varepsilon,\delta)$
 & $\widetilde O(d^2/\varepsilon)$ \\
\midrule
Single
 & \cite{GHKM23}
 & Known $\Sigma$ & $(\varepsilon,\delta)$
 & $\widetilde O(\sqrt d/\varepsilon)$ \\
Single
 & \cite{GHKM23}
 & Bounded $I_d\preceq\Sigma\preceq\kappa I_d$
 & $(\varepsilon,\delta)$
 & $\widetilde O(\kappa^2d/\varepsilon)$ \\
Single
 & \cite{GHKM23}
 & Unbounded & $(\varepsilon,\delta)$
 & $\widetilde O(d^{3/2}/\varepsilon)$ \\
Single
 & \cite{IKM25}
 & Unbounded & $(\varepsilon,\delta)$
 & $\widetilde O(d/\varepsilon)$ \\
\midrule
Multi
 & \cite{CheuNayak2025}
 & Known $\Sigma$ & $(\varepsilon,\delta)$
 & $\widetilde O(m\sqrt d/\varepsilon)$ \\
Multi
 & \cite{CheuNayak2025}
 & Bounded $I_d\preceq\Sigma\preceq\kappa I_d$
 & $(\varepsilon,\delta)$
 & $\widetilde O_\kappa(md/\varepsilon)$ \\
Multi
 & \cite{IKM25}, repeated
 & Unbounded & $(\varepsilon,\delta)$
 & $\widetilde O(md/\varepsilon)$ \\
\addlinespace[2pt]
Multi
 & \textbf{This work}
 & Unbounded & $(\varepsilon,\delta)$
 & $\widetilde\Theta_\alpha\!\left(
      m+\min\left\{\tfrac{d\sqrt m}{\varepsilon+\delta},
                       \tfrac m\delta\right\}\right)$ \\
\bottomrule
\end{tabular}
\par\smallskip
\begin{minipage}{\linewidth}
\footnotesize
Bounds from~\cite{GHKM23} are as stated in its published theorems.
The approximate-DP bounds shown for~\cite{CheuNayak2025} follow
by running those single-samplers on separate input blocks;
repetition uses error $\alpha/m$ per output.
The subscript $\kappa$ permits dependence on $\kappa$.
\end{minipage}
\end{table*}

\subsection{Additional related work}\label{main:related}

\paragraph{Private learning and multi-sampling.}
Private Gaussian estimation has been studied through confidence
intervals~\cite{KV18}, recursive
preconditioning~\cite{KLSU19,BDKU20}, and hypothesis
selection~\cite{BKSW21}.
Sample-efficient learners without parameter bounds were developed
in~\cite{AAK21}, followed by polynomial-time
algorithms~\cite{KMSSU22,AL22}.
Private sampling relaxes learning, with separations for discrete
distributions~\cite{RSSS21} and subsequent work on Gaussian
single-sampling~\cite{GHKM23,IKM25} and
multi-sampling~\cite{CheuNayak2025}.
For Gaussians, constant-accuracy learning requires
$\widetilde\Theta(d^2)$ inputs even without
privacy~\cite{AshtianiEtAl2020,DMR20}, whereas we generate constant-size
batches from $\widetilde O(d)$ inputs for fixed small joint error,
constant $\varepsilon>0$, and inverse-polynomial $\delta$ in $d$.
Conversely, our joint guarantee lets any nonprivate learner use the
batch with failure probability increased by at most its total
variation error. Taking $m=\Theta(d^2)$ and sufficiently small constant
joint error recovers the $\widetilde O(d^2/\varepsilon)$ private
learning benchmark~\cite{AL22,HKMN23}.

\paragraph{Stability and private estimation.}
Propose-test-release~\cite{DL09} privately tests stability before
release. Covariance-aware mean estimation can avoid private covariance
release~\cite{BGSUZ21}; stable-statistic methods give efficient
covariance-adaptive estimators~\cite{BHS23,DHK23}.
We use the stable weighted mean and covariance statistics
of~\cite{BHS23}, also used in~\cite{IKM25}, with a
propose-test-release-style test.
Gaussian sampling also privatizes stabilized covariance
estimates~\cite{AlabiEtAl2023}; our joint comparison and exact
correction ensure an exact product distribution on Gaussian input, including
input randomness.
The broader connection between robustness and privacy draws on
contamination models~\cite{Huber1964}, efficient robust
estimators~\cite{DKKLMS16,LRV16}, and Gaussian and sum-of-squares
methods~\cite{DKKLMS18,KMZ22}; see~\cite{DK23}.
Private mean estimation addresses bounded moments~\cite{KSU20}
and contamination~\cite{LKKO21}.
General approaches use resilience~\cite{SCV18,LKO22}, stabilized
relaxations~\cite{KMV22}, sum-of-squares exponential
mechanisms~\cite{HKM22}, and robustness-to-privacy
reductions~\cite{HKMN23,AUZ23}; strong privacy and success guarantees
can also imply robustness~\cite{GH22}.
Other methods use local sensitivity~\cite{NRS07}, inverse
sensitivity~\cite{AD20}, or stable-core aggregation~\cite{TCKMS22}.

\paragraph{Gaussian sufficiency and sample amplification.}
Sample amplification enlarges datasets while approximating the joint
distribution of independent draws~\cite{AGSV20,AGHSV24}, potentially
using fewer inputs than learning.
Our Gaussian resampling uses the sufficiency-based approach
of~\cite{AGHSV24}, conditioning on the empirical mean and covariance.
This uses the classical separation of these statistics from a uniform
orientation~\cite{James1954,Muirhead1982}.
We additionally require privacy on every fixed dataset and exact
output independence on Gaussian input, beyond the amplification
objective.

\paragraph{Lower bounds for differential privacy.}
Fingerprinting codes~\cite{BS98} yield approximate-DP
lower bounds~\cite{BUV2018,SU17Between}, with tracing arguments for
estimation~\cite{DSSUV15}.
Analytical fingerprinting covers selection~\cite{SU17Selection}
and exponential families~\cite{KMS22}; score attacks give a general
statistical formulation~\cite{CWZ2023}.
Our lower bound extracts correlation with the Gaussian parameter
directly from jointly accurate, possibly dependent batches, without
requiring enough outputs for covariance learning.
Score-based tracing then bounds record contributions using privacy.
We control adaptively selected exceptional contributions collectively
through selection entropy, connecting our estimate to
information-theoretic bounds on adaptive selection~\cite{RZ16}.
\paragraph{Organization.}
Section~\ref{sec:proof-overview} explains the main proof ideas.
Section~\ref{sec:preliminaries} collects the facts used in the proofs.
Section~\ref{sec:upper-bound} constructs the sampler and proves its
guarantees. Section~\ref{sec:lower-bound} proves the lower bound.
The appendices give the supporting estimates.

\paragraph{Private synthetic data.}
A synthetic dataset lets analysts run multiple analyses on a single private release. One objective is to reproduce answers to a specified class of queries on the observed dataset, such as low-order marginals~\cite{BLR08,HLM12,VietriEtAl22}. This empirical benchmark also appears in methods that use private density estimation to generate synthetic data~\cite{BojkovicLoh24}. Another objective is to control the \(1\)-Wasserstein distance between the input and synthetic empirical distributions, which bounds the error in the empirical average of every \(1\)-Lipschitz function~\cite{BSV24,HVZ23}. The resulting error bounds can reflect proximity to a low-dimensional affine subspace or the growth of packing numbers of the data support across scales~\cite{HSVZ25,ZMAYA26}.

\section{Techniques}\label{main:technical}\label{sec:proof-overview}

We explain how to share input records across the output batch while
preserving privacy and exact independence. We then show how joint
accuracy alone gives a matching lower bound.

\subsection{Independent Gaussian multi-sampling}

To illustrate our main idea, consider the Gaussian sampling rule
of~\cite{GHKM23,IKM25} before its privacy modifications.
Let $\bar X$ be the mean of the first $r$ records, set
$D_i=(X_{r+2i-1}-X_{r+2i})/\sqrt2$ for $i\in[s]$, and return
$
 Y=\bar X+\sqrt{1-\frac1r}\sum_{i=1}^s a_iD_i,
$
where $a\in\R^s$ is a uniformly random unit vector.
For Gaussian input from $P=\mathcal N(\mu,\Sigma)$, this gives
$Y\sim P$. However, repetitions with independent randomness on the
same input have cross-covariance $\Sigma/r$ through their common
empirical mean.
We instead coordinate the entire batch, then stabilize its
statistics and correct the resulting distribution.

\paragraph{Step 1: Resample an independent batch.}
Assume $n>\max\{m,d\}$, as in the construction below. Let $x$ be the
$n\times d$ input matrix, with row $x_i^\top$ for record $x_i$,
using the superscript $\top$ for transpose.
Independently of the input, draw $Q$ uniformly from the $n\times n$
orthogonal matrices that fix the all-ones vector $\mathbf1_n$.
Keep the first $m$ rows of $Qx$.
When the input matrix $X$ has independent rows drawn from $P$, the
rows of $QX$ are jointly Gaussian conditional on $Q$. Because $Q$ fixes
$\mathbf1_n$, each row has mean $\mu$. Orthogonality gives each row
covariance $\Sigma$ and gives distinct rows zero cross-covariance.
Thus the retained rows are independent draws from $P$.

The rotation preserves the empirical mean \(\bar x\) and the empirical scatter
$
S=\sum_i(x_i-\bar x)(x_i-\bar x)^\top.
$
To see how to resample using only these two statistics, first subtract the mean. Each centered column sums to zero, so the rotation acts on the \((n-1)\)-dimensional subspace \(\mathbf1_n^\perp\). In an orthonormal basis of this subspace, the rotated centered matrix has the distribution of \(HS^{1/2}\), where \(H\) is a uniform \((n-1)\times d\) matrix with orthonormal columns.

To recover the first \(m\) rows in the original coordinates, we project this frame onto those coordinates. The principal \(m\times m\) block of the projector onto \(\mathbf1_n^\perp\) is \(I_m-\mathbf1_m\mathbf1_m^\top/n\). By rotational invariance, the projected frame has the distribution of \((I_m-\mathbf1_m\mathbf1_m^\top/n)^{1/2}Z\), where \(Z\) consists of the first \(m\) rows of \(H\). Adding the mean back gives
\[
Y=\mathbf1_m\bar x^\top+
(I_m-\mathbf1_m\mathbf1_m^\top/n)^{1/2}ZS^{1/2}.
\]
\begin{figure}[!t]
  \centering
  \includegraphics[width=\linewidth]{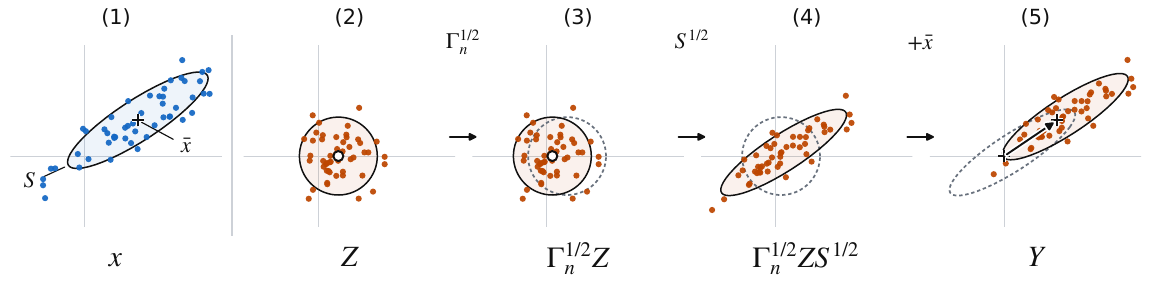}
  \caption{Conditional Gaussian resampling in two dimensions.
  (1) The observed data (blue) determine the empirical mean $\bar{x}$
  and the empirical scatter $S$.
  (2) The frame block $Z$ (orange) consists of the first $m$ rows of a
  uniform $(n-1)\times d$ matrix with orthonormal columns.
  (3) Left multiplication by $\Gamma_n^{1/2}$ moves the batch center
  toward zero while preserving all pairwise displacements, where
  $\Gamma_n=I_m-\mathbf{1}_m\mathbf{1}_m^\top/n$.
  (4) Right multiplication by $S^{1/2}$ stretches and compresses the
  cloud along the principal axes of $S$.
  (5) Adding $\bar{x}$ to each row translates the cloud to obtain $Y$.
  Dashed outlines show the cloud before each transformation.}
  \label{fig:conditional-resampling}
\end{figure}
This is \emph{conditional resampling} at fixed \((\bar x,S)\), as in sample amplification~\cite{AGHSV24}. The rows are generally dependent at fixed empirical statistics, so Step~2 compares their joint densities. Averaging over the empirical statistics of the Gaussian input gives the product distribution \(P^{\otimes m}\). Figure~\ref{fig:conditional-resampling} illustrates the three
transformations in the resampling formula.

\paragraph{Step 2: Prove privacy of the stabilized batch.}
Changing one record can alter the empirical mean and scatter
without bound. We adapt the \emph{stable weighted statistics} of
Brown, Hopkins, and Smith~\cite{BHS23}, also used in~\cite{IKM25},
to limit the influence of each record. The test adds
Laplace noise clipped to a bounded interval and accepts if the noisy
score is below a public threshold. Clipping bounds the noise, so
acceptance certifies that the stability bounds apply to the input
and its neighbors.

Write $\rho_i\in[0,1]$ for the weight of record $i$ and
$b=\sum_i\rho_i$ for the total weight. On acceptance, $b>m$. Form
the weighted mean $\theta=b^{-1}\sum_i\rho_i x_i$ and the weighted
scatter $C=\sum_i\rho_i(x_i-\theta)(x_i-\theta)^\top$, which is
positive definite on acceptance. Draw an independent copy of $Z$
from Step~1 and form the candidate batch
\[
 Y=\mathbf1_m\theta^\top+
 (I_m-\mathbf1_m\mathbf1_m^\top/b)^{1/2}ZC^{1/2}.
\]
Setting every weight to one recovers the
resampling formula from Step~1. If the test fails, the private sampler returns the rejection symbol $\bot$. In Step~3, we
represent $\bot$ by leaving all $m$ output positions blank.
We write $\mathcal A$ for this private sampler.

Iverson, Kamath, and Mouzakis~\cite{IKM25} compare
densities for a single output, bounding the normalizing and
output-dependent terms separately in their scalar analysis.
The conditional dependence of the candidate rows calls for a
comparison of the full batch densities.

\textbf{We prove a joint density comparison with square-root
dependence on $m$ in the privacy-loss bound.}
In the joint log-density ratio, the first-order contributions from
normalization and from the output-dependent determinant cancel in
expectation even when mean, scatter, and total weight change together.

To see the cancellation for scalar data, fix the weighted mean
$\theta$ and total weight $b$ and vary only the scatter $C>0$.
Set $\Gamma_b=I_m-\mathbf1_m\mathbf1_m^\top/b$.
By the candidate formula~\eqref{eq:full-candidate}, subtracting
the mean from each entry of $Y$ and multiplying by
$\Gamma_b^{-1/2}$ gives
\[
 y=\Gamma_b^{-1/2}(Y-\mathbf1_m\theta)=\sqrt C\,Z.
\]
Thus $y$ consists of the first $m$ coordinates of a uniform unit
vector in $\mathbb R^N$, where $N=n-1$, scaled by $\sqrt C$.
Its density is
$f_C(y)\propto C^{-m/2}(1-\|y\|^2/C)^\gamma$ on
$\|y\|^2<C$, where $\gamma=(N-m-2)/2$.

For this illustration, take $N\ge2m+8$, so
$\gamma>0$. Hold $y$ fixed with $\|y\|^2<C$ and increase the
scatter from $C$ to $C(1+h)$ for a small $h>0$. The larger
scatter spreads probability over a wider region, decreasing the
normalizing factor $C^{-m/2}$. The same batch has smaller squared
length relative to the larger scatter, so the factor
$(1-\|y\|^2/C)^\gamma$ increases for $y\ne0$. Thus the normalization
contributes positively to $\log(f_C(y)/f_{C(1+h)}(y))$, and the
output-dependent factor contributes nonpositively.

Since $\theta$ and $b$ are fixed, both batches undergo the same
transformation, which preserves their density ratio. The coefficient
of $h$ in the first-order expansion of the log-density ratio is
\[
 D(y):=\left.\frac{d}{dh}\log\frac{f_C(y)}{f_{C(1+h)}(y)}
 \right|_{h=0}
 =\frac m2-\gamma\frac{\|y\|^2}{C-\|y\|^2}.
\]
Under $f_C$, the term being subtracted has mean
$m/2$, equal to the first term. Bounding the two terms separately
gives an $O(m)$ bound on $D(y)$ at constant confidence.
Their difference has mean zero and variance
$m(N-2)/(2(N-m-4))=\Theta(m)$, so the linear privacy loss $hD(y)$
has standard deviation $\Theta(|h|\sqrt m)$.

\smallskip
\noindent\textbf{From scalar cancellation to joint privacy.}
We encode the weighted mean $\theta$, scatter $C$, and total weight
$b$ in a single matrix. Append a $1$ to each record to form
$\widetilde x_i=(x_i^\top,1)^\top$ and take the weighted sum of
their outer products,
\[
 B=\sum_{i=1}^n \rho_i\widetilde x_i\widetilde x_i^\top
 =\begin{pmatrix}
 C+b\theta\theta^\top & b\theta\\
 b\theta^\top & b
 \end{pmatrix}.
\]
Let $B'$ be the corresponding matrix on a neighboring dataset.
To measure the change on the scale of $B$, multiply both matrices
on the left and right by $B^{-1/2}$. This sends $B$ to the identity
and gives $B^{-1/2}(B'-B)B^{-1/2}$ as the difference on that scale.
We write $\Delta$ for the nuclear norm of this difference, the sum
of the absolute values of its eigenvalues. The
scalar cancellation extends to this joint change. The first-order
terms again cancel in expectation, including the contribution
from the normalizing factor that depends on $b$.

A change in the mean introduces a signed projection in the
first-order term. Bounding this term by vector lengths can give
a bound linear in $m$. Using the random direction of the projected
vector gives the square-root scale for this term as well.

To combine the directional bounds, let $h_j$ be
the eigenvalues of the relative change between $B$ and $B'$.
The corresponding eigenvectors are fixed before drawing $Y$.
In each fixed eigenvector direction,
the centered fluctuation is $O(\sqrt{mt}+t)$ for a tail parameter
$t\ge1$. Summing with weights $|h_j|$ gives
$O(\Delta(\sqrt{mt}+t))$ because $\sum_j|h_j|=\Delta$.

Under the conditions of Lemma~\ref{lem:full-candidate-privacy},
including that $\Delta(\sqrt{mt}+t)$ is sufficiently small, the
directional estimates ensure that the neighboring density is
positive at the sampled batch with high probability. Factoring
the ratio of the output-dependent determinants bounds the Taylor
remainder by $O(\Delta^2(m+t))$. Write $f_B^{(m)}$ and $f_{B'}^{(m)}$
for the joint candidate densities under $B$ and $B'$, respectively.
The resulting bound is
\[
 \left|\log\frac{f_B^{(m)}(Y)}{f_{B'}^{(m)}(Y)}\right|
 =O\bigl(\Delta(\sqrt{mt}+t)\bigr)
\]
except with probability $O((d+1)e^{-t})$ under
either candidate distribution. This joint comparison handles
dependence within the batch and, with $\Delta=\widetilde O(d/n)$
from stability, gives the $\widetilde O(d\sqrt m/\varepsilon)$
sample term. Section~\ref{sec:joint-privacy} combines the candidate
comparison with the stability test to prove that $\mathcal A$ is
$(\varepsilon,\delta_0)$-private, where $\delta_0=\delta/4$.

\paragraph{Step 3: Restore the exact product distribution.}
Weighting changes the distribution of accepted batches, and a
single rejection affects every position at once. To recover exact
independence, we use a mixture that runs the private sampler most
of the time and corrects its distribution on the remaining runs.
We draw a temporary batch, called the \emph{proposal},
by exact resampling at fixed public reference statistics, with the
omissions described below. The selector uses the proposal and
additional public randomness to choose a branch.
The \emph{private branch} discards the proposal and runs the private sampler on the observed dataset with independent randomness. The
\emph{correction branch} keeps the proposal and shifts and rescales
its vectors using the observed empirical statistics. Table~\ref{tab:correction-notation} summarizes the notation for the
public proposal, the separate reference simulation, and the branch decision.

\smallskip
\noindent\textbf{Giving rejection a place in the mixture.}
Exact resampling always returns a batch of vectors. The target for
the mixture must also assign positive probability to an outcome
that represents rejection. Filling every position from a fallback
on rejection can introduce dependence through the decision shared
by all positions. For example, start with two
independent draws from $P$. If one shared decision replaces both
positions with independent draws from a fixed distribution $R$
with probability $r$, the batch has distribution
$(1-r)P^{\otimes 2}+rR^{\otimes 2}$. This generally differs from the
product of its marginals, $((1-r)P+rR)^{\otimes 2}$.
Iverson, Kamath, and Mouzakis~\cite[Appendix~E]{IKM25} account for
rare weighting and test effects through the total variation error
of a single output. Here we account for both effects in the mixture. 

Before filling the batch, each position holds either
a vector or a blank $\star$. The exact resampling
rule independently leaves each position blank with probability
\(\eta=\alpha/(2m)\). The private branch applies the same
omission rule on acceptance and leaves every position blank on
rejection. The target therefore assigns probability \(\eta^m\)
to the all-blank batch. After either branch, we fill each blank
independently from the fixed distribution \(R=\mathcal N(0,I_d)\),
where \(I_d\) is the identity in dimension \(d\).

At fixed empirical statistics \(\tau=(\bar x,S)\)
with \(S\succ0\), let \(F_\tau\)
be the exact resampling distribution with omissions. Let \(G_\tau\)
be the private branch's distribution, averaged over Gaussian
datasets conditional on \(\tau\) and over the sampler's randomness.
Let \(H_\tau\) be the distribution conditional on taking the
correction branch at these statistics. All three distributions
describe the batch before filling the blanks. If \(w\in(0,1)\) is
the overall probability of taking the private branch, we require
\[
 F_\tau=wG_\tau+(1-w)H_\tau.
\]
The private branch contributes \(wG_\tau\). For the remaining mass
to be nonnegative, we need \(wG_\tau(E)\le F_\tau(E)\) for every
batch event \(E\). In particular, the private branch can contribute
at most \(\eta^m\) to the all-blank outcome. The correction branch
must sample the remaining mass after normalizing it by \(1-w\).
We need the same input-independent \(w\) close to one for all
statistics, because the correction branch can add up to \(1-w\)
to the additive privacy parameter.

\smallskip
\noindent\textbf{Which proposals should we replace?}
To make the branch choice independent of the observed input, we
prepare the proposal and the selection rule at public reference
statistics \(\tau_0=(0,I_d)\). Gaussian conditioning lets us simulate
a dataset with empirical mean zero and scatter \(I_d\) without knowing
\(P\). We generate it by centering a standard Gaussian
matrix and rescaling it to have scatter \(I_d\). Its columns are
orthonormal, and the entries in each column sum to zero. We call
this matrix a \emph{centered frame}.
Applying \(v\mapsto\bar x+S^{1/2}v\) to its records
gives the distribution of a Gaussian dataset conditional on
\(\tau=(\bar x,S)\). Both resampling rules are affine equivariant,
so the same map carries their output distributions from \(\tau_0\)
to \(\tau\) by transforming the vectors and preserving blanks.

Draw the proposal \(\zeta\) from \(F_{\tau_0}\) by taking the first
\(m\) rows of a uniformly sampled centered frame and applying the
independent omissions. Let \(\mathsf{Priv}=1\) mean
that we replace the proposal, and let \(\mathsf{Priv}=0\) mean that
we keep it. The selected proposals must carry exactly the mass
that the private branch will supply. Thus, for every measurable
batch event \(E\), we require
\[
 \Pr\{\zeta\in E,\ \mathsf{Priv}=1\}=wG_{\tau_0}(E).
\]
Taking \(E\) to be the whole batch space gives the
overall replacement probability \(w\). Conditional on replacement,
the proposal has distribution \(G_{\tau_0}\).

Let \(f_0\) and \(f_{\mathrm{priv}}\) be the densities of
\(F_{\tau_0}\) and \(G_{\tau_0}\), using counting measure for the
blank pattern and Lebesgue measure for the retained vectors.
Accepted outputs lie in the proposal support by
the density comparison in Section~\ref{sec:exact-correction},
and the proposal gives positive mass to rejection's all-blank
outcome. Dividing the selected density \(w f_{\mathrm{priv}}(\zeta)\)
by the proposal density gives the required replacement probability
\(w f_{\mathrm{priv}}(\zeta)/f_0(\zeta)\) at each proposal with
\(f_0(\zeta)>0\). This probability depends on the proposal, while
its average \(w\) is independent of the observed input.
The affine map from the reference statistics to the observed
statistics carries the retained proposals to the mass
\(F_\tau-wG_\tau\). A private call on the observed input supplies
the complementary mass \(wG_\tau\), giving the required mixture.

The private density is an average over reference datasets and
stability-test noise. To see the difficulty with this average,
consider an ordinary reference dataset drawn independently of
the proposal, together with a draw of the test noise. Once these
are fixed, we can evaluate the conditional private density at
\(\zeta\). Multiplying it by \(w\) and dividing by
\(f_0(\zeta)\) gives a score. With the proposal fixed, averaging
over the reference dataset and the stability-test randomness gives

\[
 \mathbb E[\text{score}\mid\zeta]
 =\frac{w f_{\mathrm{priv}}(\zeta)}{f_0(\zeta)}.
\]

The score can exceed one. For an all-blank proposal, a rejected
reference run assigns probability one to that outcome, giving a
score of \(w\eta^{-m}\). Whenever scores exceed one with positive
probability, capping them at one lowers their expectation and breaks
the mixture identity.

\smallskip
\noindent{\textbf{Importance sampling for exact correction.}
We use \emph{importance sampling} to reduce large
scores while preserving their exact expectation. Increasing the
sampling density of a reference dataset and dividing its score by
the same factor leaves its contribution to the expectation
unchanged. We choose the reweighted density using a determinant
ratio that controls how large the score can be.}
This ratio divides the determinant of the sum of
$\widetilde x_i\widetilde x_i^\top$ over all records by the
determinant of the same sum over the records left after deletion.
Both determinants account for scatter and record count.
When the simulated test accepts, a power of the ratio bounds how
much weighting can increase the density of the retained vectors.
When it rejects, the clipped noise lets the test certify a minimum
number of influential deletions. Each such deletion increases the
determinant ratio, giving the growth needed to offset the
all-blank ratio \(\eta^{-m}\).

The reweighted distribution needs the deletion set
before drawing the dataset, and the simulated test records that
set after the draw.
We therefore guess the set before drawing and use a simulation for
replacement only if the guess matches. On a match, we divide the
score by the probability of the guess and by the density of the
reweighted draw relative to the ordinary draw.
A mismatch gives replacement probability zero.
Each reference run contributes through its one matching guess,
so averaging over the guesses and reweighted draws cancels both
sampling factors and recovers the required expectation.

For a fixed deletion set, the determinant reweighting changes one
exponent in the density of a frame block in coordinates determined
by that set. We can draw this block from a uniform frame with fewer
rows and complete the rest of the reference dataset.
Section~\ref{sec:exact-correction}
shows that the determinant bounds give valid probabilities on
both simulated acceptance and rejection. The number of deletions
needed on rejection accounts for the term linear in \(m\) in the
sample bound, up to logarithmic factors.

The parameters give \(\delta_0+1-w\le\delta\), so the mechanism is
\((\varepsilon,\delta)\)-private. On Gaussian input, the mixture
recovers \(F_\tau\) at the observed statistics. Averaging over
those statistics gives independent draws from \(P\) with independent
omissions. Filling the blanks independently from \(R\) gives
\(((1-\eta)P+\eta R)^{\otimes m}\), with joint total variation
error at most \(m\eta=\alpha/2\).

\subsection{Lower-bound ideas}

Earlier reductions for Gaussian sampling use generated observations
for private estimation. Ghazi et al.~\cite{GHKM23} construct private
estimators from repeated single-sampler calls. Cheu and
Nayak~\cite{CheuNayak2025} study weak multi-sampling of Gaussians
with bounded mean and covariance under zero-concentrated differential
privacy. Their reduction uses \(d+1\) generated samples as public
observations for a learner that also receives private data and whose
guarantee assumes Gaussian public observations.
Our argument uses joint total variation accuracy to
extract a direction correlated with the unknown cross-covariance,
even from a short batch whose outputs are dependent.

\paragraph{Extracting a covariance direction from a short batch.}
\textbf{We construct a Gaussian family whose noise variance grows
with the batch size while the batch still reveals a direction aligned
with the unknown cross-covariance.}
For the high-dimensional argument, set \(k=\lfloor d/2\rfloor\)
and split the first \(2k\) coordinates of each record into two blocks.
Let \(z,g\in\R^k\) be independent standard Gaussian vectors. The
blocks are \(z\) and \(w=Bz+\sigma g\), where \(B\in\R^{k\times k}\)
is unknown and \(\sigma>0\) is the noise standard deviation.
The cross-covariance is \(\E[wz^\top\mid B]=B\).
If \(d\) is odd, append an independent standard
Gaussian coordinate. Write \(P_B\) for the resulting distribution.

Write \(\widehat z_j,\widehat w_j\) for the blocks of released vector
\(j\). Write \(\langle\cdot,\cdot\rangle\) for the matrix inner
product, which sums products of corresponding entries, and
\(\|\cdot\|_F\) for the resulting Frobenius norm.
Since \(wz^\top\) has conditional mean \(B\), we use
the corresponding sum over the released batch to extract a direction.
\[
 A=\sum_{j=1}^m\widehat w_j\widehat z_j^\top,
 \qquad T=\frac{A}{\|A\|_F}\quad\text{if }A\ne0.
\]
Set \(T=0\) if \(A=0\). We measure alignment by
\(\langle B,T\rangle\). Let \(T^\star\) be the
direction computed by the same rule from an ideal batch of \(m\)
independent draws from \(P_B\).

In the analysis, we draw the entries of \(B\) independently from
\(\mathcal N(0,1)\).
We increase the noise with \(m\) to make each record
less informative about \(B\). The argument below turns a given
amount of batch alignment into a lower bound on record agreement
proportional to \(\sigma\). We choose \(\sigma^2=km\), the expected
total squared length of the first blocks in an ideal batch. At this
noise level, the ideal batch still reveals a direction aligned
with \(B\).
By Lemma~\ref{lem:lb:direction}, the ideal direction satisfies
\(\langle B,T^\star\rangle\ge\sqrt k/8\) with probability at least
\(1-26/k\) over the prior and the ideal draws. For each fixed \(B\),
joint total variation accuracy transfers this event to \(T\) with
a loss of at most \(\alpha\) in probability.
Averaging the fixed-\(B\) comparison over the prior
gives
\[
 \Pr\{\langle B,T\rangle\ge\sqrt k/8\}\ge 1-\alpha-26/k.
\]
For every fixed \(\alpha<1\), the bound is positive in sufficiently
large dimension and applies to dependent outputs.

\paragraph{From alignment to agreement with input records.}
We multiply \(T\) by a weight so that negative alignment cannot
cancel positive alignment in expectation. Write \(\widetilde T\)
for the weighted direction. The weight is zero when
\(\langle B,T\rangle\le0\), one when
\(\langle B,T\rangle\ge\sqrt{k}/8\), and linear in between.
It depends on \(B\) and is used only in the analysis.

Thus \(\|\widetilde T\|_F\le1\) and
\(\langle B,\widetilde T\rangle\ge0\) everywhere. On the event
\(\{\langle B,T\rangle\ge\sqrt{k}/8\}\), we have
\(\widetilde T=T\), so \(\langle B,\widetilde T\rangle\ge\sqrt{k}/8\).
The preceding bound on the probability of this event gives
\(\mathbb E[\langle B,\widetilde T\rangle]=\Omega_\alpha(\sqrt{k})\)
in sufficiently large dimension. The expectation is over \(B\),
the input records, and the mechanism's randomness.

For input record \(X_i\), write \(z_i,w_i\) for its two blocks and
\(g_i=(w_i-Bz_i)/\sigma\) for its standard Gaussian residual.
Let \(p_B\) be the density of one record under \(P_B\). The score
of \(X_i\) is
\[
 \nabla_B\log p_B(X_i)=\frac{g_i z_i^\top}{\sigma}.
\]
We measure agreement with record \(i\) by
\(\langle\widetilde T,g_i z_i^\top\rangle\), using the scaled
score matrix obtained by multiplying the score by \(\sigma\).
Gaussian integration by parts~\cite{KMS22,CWZ2023,Narayanan25}
relates the total expected agreement to the alignment with \(B\).
\[
 \frac{1}{\sigma}\sum_{i=1}^n\E\langle\widetilde T,g_i z_i^\top\rangle
 \ge \E\langle B,\widetilde T\rangle-\frac{8}{\sqrt k}.
\]
The last term bounds the contribution from differentiating the
weight. For fixed \(\alpha<1\) and sufficiently large \(d\), the
alignment dominates this term.
Multiplying by \(\sigma=\sqrt{km}\) turns alignment
of order \(\sqrt k\) into expected total agreement of order
\(\Omega_\alpha(k\sqrt m)\).

\paragraph{Bounding the contribution allowed by approximate privacy.}
Privacy bounds the same sum. Let \(\widetilde T^{(i)}\) be the
weighted direction obtained after replacing record \(i\) by an
independent draw conditional on \(B\). The original matrix
\(g_i z_i^\top\) is then independent of \(\widetilde T^{(i)}\)
conditional on \(B\), so
\(\E\langle\widetilde T^{(i)},g_i z_i^\top\rangle=0\) and
\(\E|\langle\widetilde T^{(i)},g_i z_i^\top\rangle|\le1\).
The bound by one on expected absolute agreement
makes the multiplicative contribution at most \(e^\varepsilon-1\)
per record, hence \(O(n\varepsilon)\) in total for \(\varepsilon\le1\).
The additive allowance \(\delta\) can concentrate on outcomes with
large agreements. Narayanan~\cite{Narayanan25} bounds this
contribution separately for each record in a covariance-estimation
lower bound. In our setting, Gaussian tails applied record by
record give \(\widetilde O(nk\delta)\) for the additive contribution.
This gives a \(\sqrt m/\delta\) alternative when the additive
contribution accounts for a constant fraction of the required
agreement, up to logarithmic factors.

\textbf{We bound the additive contribution through a sum of
scaled score matrices with data-dependent coefficients.}
Swapping the original and replacement records expresses the excess
mass from both directions of the privacy comparison against the
same direction \(\widetilde T\). For each record, this mass gives
a random weight \(b_i\in[0,1]\) with \(\E b_i\le\delta\). To see
how the sum arises, take the original-record contribution and set
\(c_i=b_i\operatorname{sgn}\langle\widetilde T,g_i z_i^\top\rangle\),
where \(\operatorname{sgn}\) is the sign function with
\(\operatorname{sgn}(0)=0\). Since
\(\|\widetilde T\|_F\le1\),
\[
 \sum_{i=1}^n b_i|\langle\widetilde T,g_i z_i^\top\rangle|
 =\left\langle\widetilde T,\sum_{i=1}^n c_i g_i z_i^\top\right\rangle
 \le\left\|\sum_{i=1}^n c_i g_i z_i^\top\right\|_F.
\]
The replacement-record contribution has the same form. The
coefficients satisfy \(|c_i|\le1\) and \(\E|c_i|\le\delta\) and may
depend on all the matrices. Each signed term has nonnegative
inner product with \(\widetilde T\), while components orthogonal
to \(\widetilde T\) can still cancel in the sum.

The selected coefficients can favor unusually large sums. We
round each coefficient to its sign with probability \(|c_i|\)
and to zero otherwise. Given the records and coefficients, each
rounded value has mean \(c_i\). By convexity, the original
expected norm is at most the expected norm of the rounded sum.
Let \(V\) be the resulting sign pattern and
\(p_v=\Pr(V=v)\) the probability of pattern \(v\).

For a fixed pattern \(v\) selecting \(j\ge1\)
records, Lemma~\ref{lem:lb:fixed-selection} bounds the norm of
its sum by \(O(k\sqrt{j+u}+j+u)\) except with probability
\(3e^{-u}\). Since the pattern can depend on the records,
conditioning on \(V=v\) can increase this failure probability by a
factor of \(1/p_v\). Taking the tail parameter to be
\(u+\log(1/p_v)\) compensates for this factor. Integrating the
conditional tail gives a bound in terms of \(j+\log(1/p_v)\).
Concavity of the square root lets us average these two quantities
inside the square root. The expected number of selected records is
at most \(n\delta\), and the average of \(\log(1/p_v)\) is the
entropy of \(V\)~\cite{RZ16}, which is at most
\(n\delta\log(2e/\delta)\). The zero pattern contributes zero.
This yields
\[
 \E\left\|\sum_{i=1}^n c_i g_i z_i^\top\right\|_F
 \le \widetilde O\!\left(k\sqrt{n\delta}+n\delta\right)
\]
for the original matrices, with the same bound for the replacement
matrices.

Privacy thus bounds the sum of expected agreements by
\(\widetilde O(n(\varepsilon+\delta)+k\sqrt{n\delta})\).
At least one of these two terms must supply a
constant fraction of the \(\Omega_\alpha(k\sqrt m)\) agreement
forced by accuracy. The linear term gives the
\(d\sqrt m/(\varepsilon+\delta)\) alternative, since
\(k=\lfloor d/2\rfloor\). The square-root term gives the
\(m/\delta\) alternative. Thus their minimum is a lower bound,
up to logarithmic factors.

A Gaussian family with one unknown mean coordinate gives the linear
lower bound in \(m\). A jointly accurate batch localizes that mean
to order \(m^{-1/2}\) with constant probability at fixed accuracy.
With \(n\) input observations, uniform localization over means at
constant success probability requires a radius of order
\(n^{-1/2}\) or larger. Matching the two scales
therefore requires \(n=\Omega_\alpha(m)\) even without privacy.
Section~\ref{sec:lower-bound} combines these bounds and covers
the remaining dimensions.

\section{Preliminaries}\label{sec:preliminaries}\label{ub-bg:background}

\subsection{Notation and Gaussian setup}\label{sec:notation-gaussian}

\paragraph{Datasets and empirical statistics.}
We write $d$ for the dimension, $n$ for the number of input records,
and $m$ for the number of output vectors. A fixed dataset
$x=(x_1,\ldots,x_n)$ is identified with the matrix whose $i$th row is
$x_i^\top$. We use $X$ for a random dataset. For $[n]=\{1,\ldots,n\}$
and $I\subseteq[n]$, the submatrix $x_I$ retains the rows indexed by $I$
in increasing order. The empirical mean and unnormalized scatter are
\[
  \bar x=\frac{1}{n}\sum_{i=1}^n x_i,
  \qquad
  S(x)=\sum_{i=1}^n (x_i-\bar x)(x_i-\bar x)^\top,
  \qquad
  \tau(x)=(\bar x,S(x)).
\]
For a nonempty subset of records \(x_I\), the scatter \(S(x_I)\)
uses the empirical mean of that subset.

\paragraph{Matrix notation.}
We write $\one_r$ for the all-ones vector in $\R^r$, $I_r$ for the
$r\times r$ identity, and $e_i$ for the $i$th standard basis vector.
For symmetric matrices, $A\preceq B$ means that $B-A$ is positive
semidefinite, and $A\succ0$ means that $A$ is positive definite.
If $A\succeq0$, then $A^{1/2}$ denotes its symmetric positive
semidefinite square root. We use $\|v\|$ for the Euclidean norm
and $\|A\|_{\op}$, $\|A\|_F$, and $\|A\|_*$ for the operator,
Frobenius, and nuclear norms. The Frobenius inner product is
$\langle A,B\rangle=\tr(A^\top B)$. For symmetric $H$, the
nuclear norm is the sum of the absolute eigenvalues of $H$.
For $0\prec A\preceq B$ we have
$B^{-1}\preceq A^{-1}$ and $\det A\le\det B$. For positive definite $A,B$ and any invertible $L$ of the same size,
the matrices $A^{-1/2}BA^{-1/2}$ and
$(LAL^\top)^{-1/2}(LBL^\top)(LAL^\top)^{-1/2}$
have the same eigenvalues. For a matrix $A$ and a set $J$ of row indices, $A_J$ denotes
the submatrix formed by taking the rows of $A$ indexed by $J$,
in increasing index order. In particular, $A_{[m]}$ consists
of the first $m$ rows of $A$.

\paragraph{Probability notation and Gaussian facts.}
We write \(P^{\otimes r}\) for the distribution of \(r\)
independent draws from \(P\). We use
$\stackrel{\mathrm d}{=}$ for equality in distribution and
$\one_E$ for the indicator of an event $E$. All logarithms are
natural. Let the rows of $G$ be independent $\Normal(0,\Sigma)$
vectors. If $L$ is independent of $G$ and has orthonormal rows
almost surely, then $LG$ has independent $\Normal(0,\Sigma)$ rows
and is independent of $L$.

An $N\times p$ frame $Q$ satisfies $Q^\top Q=I_p$. A uniform
frame consists of the first $p$ columns of a uniformly random
$N\times N$ orthogonal matrix. Its distribution is invariant under
left multiplication by any $N\times N$ orthogonal matrix.
An $n\times d$ frame $U$ is centered if $\one_n^\top U=0$.
For $n>d$, the uniform distribution on centered frames is invariant
under orthogonal transformations that fix $\one_n$.
Lemma~\ref{lem:prelim-frame-projection} gives the projection rule
for uniform frames, and Lemma~\ref{ub-bg:haar-block-density}
gives the density of a block.

For a standard Gaussian vector $g$ and a symmetric matrix $A$ of the
same dimension,
$\E(g^\top A g)=\tr A$ and
$\operatorname{Var}(g^\top A g)=2\tr(A^2)$.
For independent standard Gaussian vectors $g,h$ of compatible
dimensions and a fixed matrix $C$, $g^\top C h$ has mean zero
and second moment $\|C\|_F^2$. For a scalar standard Gaussian $Z$
and $t\ge0$, $\Prob\{|Z|>t\}\le2e^{-t^2/2}$.

\subsection{Comparing distributions}\label{sec:standard-facts}

\paragraph{Total variation.}
For distributions $P,Q$ on the same space, let
$\TV(P,Q)=\sup_E|P(E)-Q(E)|$, where the supremum is over measurable
events. If $(U,V)$ is a coupling of $P,Q$, then
$\TV(P,Q)\le\Prob\{U\ne V\}$. A maximal coupling attains equality
on the spaces used here by Lemma~\ref{lem:prelim-coupling}.
For a fixed randomized map $K$,
$\TV(\Law(K(U)),\Law(K(V)))\le\TV(P,Q)$. For two mixtures with
the same mixing distribution, their TV distance is at most the average
of the paired TV distances~\cite[Appendix~B]{IKM25}.
For two Gaussians with common covariance $\Sigma\succ0$, we use
\[
  \TV\bigl(\Normal(\mu,\Sigma),\Normal(\nu,\Sigma)\bigr)
  \le \frac12\bigl\|\Sigma^{-1/2}(\mu-\nu)\bigr\|.
\]
See Lemma~\ref{lem:prelim-gaussian-tv}.

\begin{definition}[Hockey-stick divergence]
\label{def:prelim-hockey-stick}
Let \(P,Q\) have densities \(p,q\) with respect to a common
dominating measure \(\rho\), for example \(\rho=P+Q\).
For \(\varepsilon\ge0\), define
\begin{equation}\label{eq:prelim-hockey-stick}
  D_{e^\varepsilon}(P\|Q)
  =\sup_E\{P(E)-e^\varepsilon Q(E)\}
  =\int(p-e^\varepsilon q)_+\,d\rho,
\end{equation}
where $a_+=\max\{a,0\}$ and $E$ ranges over measurable events.
\end{definition}
In particular, $D_1(P\|Q)=\TV(P,Q)$.
If $p\le e^\varepsilon q$ almost everywhere on a measurable set $G$, then
$D_{e^\varepsilon}(P\|Q)\le P(G^c)$.
We apply this bound separately in each order of a privacy
comparison~\cite[Section~2.1]{IKM25}.

\subsection{Privacy tools}\label{sec:privacy-facts}

\paragraph{Privacy on fixed inputs.}
Write \(P_x=\Law(M(x))\) for the output distribution on a fixed dataset \(x\).
Definition~\ref{main:def:dp} is equivalent to
$D_{e^\varepsilon}(P_x\|P_{x'})\le\delta$ for each ordered
pair of neighboring datasets. In particular, for such a pair,
$\TV(P_x,P_{x'})\le e^\varepsilon-1+\delta$.
The distribution \(\Law(M(X))\) in Definition~\ref{main:def:sampling} includes
the draw $X\sim P^{\otimes n}$ as well as the mechanism's randomness.

\paragraph{Postprocessing.}
A fixed randomized postprocessing of a private output preserves its
privacy guarantee~\cite[Proposition~2.1]{DworkRoth2014}.

\paragraph{Privacy under perturbation.}
Suppose $M$ is $(\varepsilon,\delta)$-private and
$\TV(\Law(M(x)),\Law(\widetilde M(x)))\le\gamma$ for every fixed $x$.
Then $\widetilde M$ satisfies the same multiplicative privacy bound
with additive error at most $\delta+(1+e^\varepsilon)\gamma$
by Lemma~\ref{lem:prelim-privacy-comparison}(2).

\begin{fact}[Clipped Laplace noise]
\label{fact:prelim-clipped-laplace}
Let $g$ have sensitivity at most $\Delta_g$ on neighboring datasets.
For \(s>0\), let \(Z\sim\operatorname{Lap}(s)\) have density
\(e^{-|z|/s}/(2s)\). The release \(g(x)+Z\) is $(\Delta_g/s,0)$-private and
$\Prob\{|Z|>\ell\}=e^{-\ell/s}$ for $\ell\ge0$
\cite[Section~3.3]{DworkRoth2014}.
With $\operatorname{clip}_\ell(z)=\max\{-\ell,\min\{z,\ell\}\}$,
the release \(g(x)+\operatorname{clip}_\ell(Z)\) has multiplicative
privacy parameter \(\Delta_g/s\) and additive privacy error at most
\((1+e^{\Delta_g/s})e^{-\ell/s}\).
See Lemma~\ref{lem:prelim-laplace} for the proof.
\end{fact}

\begin{fact}[A mixture with a private branch]
\label{fact:prelim-private-mixture}
Suppose a mechanism chooses its private branch with a fixed
probability $w\in[0,1]$ on every input $x$. When $w>0$,
conditional on that choice its output has the distribution of
$M(x)$ for an $(\varepsilon,\delta_0)$-private mechanism $M$.
The other branch may use any input-dependent rule.
The mixture is $(\varepsilon,w\delta_0+1-w)$-private by
Lemma~\ref{lem:prelim-privacy-comparison}(3).
\end{fact}

\section{Private Gaussian multi-sampling}
\label{sec:upper-bound}\label{main:upper}\label{main:ub:upper}
\label{sec:ub-result}

We construct a sampler that is private on every dataset and has an
exact product distribution on Gaussian input. The proof starts with
conditional resampling from the empirical mean and scatter.

The following theorem gives the explicit form of
Theorem~\ref{main:intro:upper}.

\begin{theorem}[Private Gaussian sampling with independent outputs]
\label{thm:full-main}\label{thm:upper}
Let $d,m\ge1$ be integers, $0<\alpha,\delta<1$, and
$0<\varepsilon\le1$.
There is an $(\varepsilon,\delta)$-differentially private mechanism
$M$ that takes $n$ records in $\R^d$ and releases $m$ vectors in
$\R^d$ with the following guarantee. For every
$P=\mathcal N(\mu,\Sigma)$ with $\Sigma\succ0$, when
$X\sim P^{\otimes n}$, the output satisfies
\begin{equation}\label{eq:ub-product}
 \Law(M(X))=Q_P^{\otimes m},\qquad
 Q_P=(1-\eta)P+\eta\mathcal N(0,I_d),\quad
 \eta=\frac{\alpha}{2m}.
\end{equation}
In particular, $\TV(Q_P^{\otimes m},P^{\otimes m})\le\alpha/2$.
For a sufficiently large universal constant $c$, it suffices to use
\begin{equation}\label{eq:general-iid-positive}
 n=\left\lceil c\left[
 mL+\frac{(d+L)(\sqrt{mL}+L)}{\varepsilon}
 \right]\right\rceil,
 \qquad L=1+\log\frac{dm}{\alpha\delta\varepsilon}.
\end{equation}
The mechanism uses $O(n^3)$ operations.
\end{theorem}

As discussed in Step~3 of Section~\ref{main:technical}, weighting
changes the distribution of accepted batches, and a single rejection
affects all output positions. Both can introduce dependence. We use
two branches to recover exact independence on Gaussian input. The
private branch runs the private sampler $\mathcal A$, and the
correction branch supplies the rest of the target distribution.

Write $\star$ for a blank and $\star^m$ for the all-blank batch.
After exact resampling or private acceptance, leave each position
blank independently with probability $\eta=\alpha/(2m)$. A private
rejection gives $\star^m$. Under exact resampling with omissions,
this same outcome has probability $\eta^m$. At the end, fill every
blank independently from $R=\mathcal N(0,I_d)$.

Let $T=T(x)$ encode the input's empirical mean $\bar x$ and scatter
$S$, as in \eqref{eq:augmented-statistic}. For fixed $n$, it encodes
the same statistics as $\tau=(\bar x,S)$ in Step~3. On Gaussian
input, conditional on $T$, let $F_T$ be the distribution of exact
resampling with omissions, and let $G_T$ and $H_T$ be the private
and correction branch distributions before filling the blanks.
Section~\ref{sec:exact-correction} constructs the selector so that
\[
 F_T=wG_T+(1-w)H_T.
\]

Let $T_0=\operatorname{diag}(I_d,n)$ encode empirical mean zero and
scatter $I_d$. We draw a public proposal $\zeta\sim F_{T_0}$.
Using $\zeta$ and public randomness before accessing $x$, the
selector returns $\mathsf{Priv}=1$ to replace the proposal or
$\mathsf{Priv}=0$ to keep it. The decision depends on $\zeta$ so
that the proposals selected for replacement have distribution
$G_{T_0}$. The overall replacement probability is
$w=(1-e^{-\lambda/8}/n)^n$, where $\lambda$ is the base pruning
threshold. The keep probability $1-w$ bounds the additive privacy
cost of the correction branch.

For $S\succ0$, write $\Phi_T(\zeta)$ for the batch obtained by
mapping each retained vector $y$ to $\bar x+S^{1/2}y$ and leaving
blanks in place. Conditional on keeping the proposal,
$\Phi_T(\zeta)$ has distribution $H_T$.

\begin{algorithm}[H]
\caption{Preparing the output batch}
\label{alg:upper-summary}
\begin{algorithmic}[1]
\Require Input dataset $x$ and public parameters.
\State Draw $\zeta\sim F_{T_0}$.
\State Apply Algorithm~\ref{alg:main-selector} to $\zeta$ to obtain
       $\mathsf{Priv}$.
\If{$\mathsf{Priv}=0$} \Comment{Keep the proposal}
    \State Compute the empirical mean $\bar x$ and scatter $S$ of $x$.
    \State Set the batch to $\Phi_T(\zeta)$ if $S\succ0$, and to
           $\star^m$ otherwise.
\Else \Comment{Replace the proposal}
    \State Discard $\zeta$ and draw $Y\gets\mathcal A(x)$ with
           independent randomness (Algorithm~\ref{alg:base}).
    \State On acceptance, set the batch to $Y$ with independent
           omissions of probability $\eta$. On rejection, set it to
           $\star^m$.
\EndIf
\State Fill every blank independently from $R$ and return the batch.
\end{algorithmic}
\end{algorithm}

Section~\ref{sec:resampling} builds the exact and weighted samplers
and gives the stability test. Section~\ref{sec:joint-privacy} proves
privacy of the private sampler. Section~\ref{sec:exact-correction}
constructs the selector, derives the branch probabilities, and shows
that on Gaussian input the two branches recover exact resampling with
independent omissions. Section~\ref{sec:complete-sampler} gives the
full mechanism in Algorithm~\ref{alg:complete} and proves the theorem.

\subsection{Conditional resampling and stable weights}\label{sec:resampling}

We begin with a resampling rule that uses only the empirical mean
and scatter of the input. On Gaussian input, averaging over these
statistics gives a batch of independent draws from the original
Gaussian. Changing one record can substantially change the mean and scatter.
To limit this effect, we extend the resampling rule to weighted
statistics and choose weights that change little between neighboring
datasets. Setting every weight to one recovers the original rule.
We then use a private stability test to decide whether to draw a
batch from the weighted rule or reject. We call a batch drawn from the weighted rule a candidate batch.
We derive its joint density to compare candidate batches on
neighboring datasets and to construct the exact correction.

\subsubsection{Exact conditional resampling}\label{sec:exact}

We need an exact resampling rule that uses only the empirical mean
and scatter. For Gaussian input, fixing these statistics leaves a
uniform orientation that can be drawn without knowing the population
parameters.

View a dataset $x$ as an $n\times d$ matrix with rows $x_i^T$.
Write $\bar x$ for its mean, $S$ for its unnormalized scatter,
and $\widetilde x_i$ for the record vector.
\[
 \bar x=\frac1n\sum_i x_i,\qquad
 S=\sum_i(x_i-\bar x)(x_i-\bar x)^T,\qquad
 \widetilde x_i=\begin{pmatrix}x_i\\1\end{pmatrix}.
\]
The unweighted statistic $T$ below encodes the empirical mean, scatter, and sample size.
\begin{equation}\label{eq:augmented-statistic}
 T=\sum_i\widetilde x_i\widetilde x_i^T
 =\begin{pmatrix}S+n\bar x\bar x^T&n\bar x\\
                  n\bar x^T&n\end{pmatrix}.
\end{equation}
The extra coordinate lets a single matrix comparison control changes
in both the mean and scatter.

Step~1 of Section~\ref{main:technical} gives a formula for resampling
the first $m$ rows. We first identify the conditional distribution of
the full dataset, which we also use to generate reference datasets
for the correction. On Gaussian input with positive definite
covariance, $n>d$ ensures that the empirical scatter is positive
definite almost surely.

\begin{lemma}[Exact conditional resampling]\label{lem:full-conditioning}
Let $n>\max\{m,d\}$ and $X\sim P^{\otimes n}$ for a nonsingular
Gaussian $P$. Conditional on $(\bar X,S(X))$,
\[
 X\overset d=\one_n\bar X^T+US(X)^{1/2},
\]
where $U\in\R^{n\times d}$ is a uniform centered frame satisfying
$U^TU=I_d$ and $\one_n^TU=0$, independent of the statistics.
Consequently, if $U$ is drawn independently of $X$, the first $m$
rows of the reconstructed dataset $\one_n\bar X^T+US(X)^{1/2}$
have distribution $P^{\otimes m}$ after averaging over $X$.
\end{lemma}

\begin{proof}
Choose an $n\times(n-1)$ matrix $E$ whose columns form an orthonormal
basis for $\one_n^\perp$. By
Lemma~\ref{lem:prelim-gaussian-components}, the rows of
$G=E^TX$ are independent $\mathcal N(0,\Sigma)$ vectors,
independent of $\bar X$, and
$G^TG=S(X)\succ0$ almost surely. To identify the remaining randomness, write $Q=G(G^\top G)^{-1/2}$.
Independently of the input, draw a uniform $(n-1)\times(n-1)$
orthogonal matrix $O$. For fixed $G$, the columns of $Q$ are
orthonormal, so they can be extended to an orthonormal basis of
$\R^{n-1}$. Right invariance of $O$ implies
that $OQ$ has the distribution of the first $d$ columns of $O$.
In particular, the conditional distribution of $OQ$ given
$(G,\bar X)$ is uniform and constant in $(G,\bar X)$.

The density of $G$ depends only on $G^TG$, so for each fixed
orthogonal $O$, $OG$ has the same distribution as $G$. Left multiplication
by $O$ preserves $G^TG$ and sends the normalized frame $Q$ to $OQ$.
Since $\bar X$ is independent of $(O,G)$, the equality below holds
conditional on $O$ and hence also unconditionally.
\[
 (OQ,G^TG,\bar X)\overset d=(Q,G^TG,\bar X).
\]
Thus $Q$ is uniform and independent of $(\bar X,S(X))$. Reconstruct
$X$ as $\one_n\bar X^T+EG$, with $G=QS(X)^{1/2}$. By the centered
frame part of Lemma~\ref{lem:prelim-frame-projection}, $U=EQ$ is a
uniform centered frame. This proves the conditional representation.
Resampling from that representation and averaging over the
statistics recovers the original input distribution, whose first
$m$ rows have distribution $P^{\otimes m}$.
\end{proof}

With $\bar x$ and $S$ fixed, draw an independent uniform centered
frame $U$ and keep the first $m$ rows of
$\one_n\bar x^T+US^{1/2}$. The construction uses only
the empirical statistics and public randomness.
Lemma~\ref{lem:full-conditioning} gives
\begin{equation}\label{eq:exact-batch}
 X\sim P^{\otimes n},\qquad
 Y=\bigl(\one_n\bar X^T+US(X)^{1/2}\bigr)_{[m]}
 \quad\Longrightarrow\quad \Law(Y)=P^{\otimes m},
\end{equation}
where $U$ is independent of $X$ and $[m]$ selects the first $m$ rows.

Step~1 of Section~\ref{main:technical} draws these $m$ rows directly.
By Lemma~\ref{lem:prelim-frame-projection},
\[
 U_{[m]}\overset d=
 \left(I_m-\frac{\one_m\one_m^T}{n}\right)^{1/2}Z,
\]
where $Z$ consists of the first $m$ rows of a uniform
$(n-1)\times d$ frame. Substituting this identity in the resampling
rule above gives the batch formula in Step~1.

Replacing one record can change the empirical mean and scatter
arbitrarily. Since these statistics determine the resampling
distribution, neighboring datasets can produce very different batch
distributions. To control this change for privacy, we first extend
the resampling rule to weighted statistics and then choose weights
that limit the influence of individual records.

\subsubsection{Weighted resampling}\label{sec:ub-private-base}

Downweighting an influential record changes the empirical statistics. We let the weights change
the statistics in the resampling formula.
Assign weights $\rho_i\in[0,1]$ and put $b=\sum_i\rho_i$.
For $b>0$, $\theta=b^{-1}\sum_i\rho_i x_i$ is the weighted
mean and $C=\sum_i\rho_i(x_i-\theta)(x_i-\theta)^T$ is the
unnormalized weighted scatter. When $b=0$, use $\theta=0$ and
$C=0$ in the block representation. We replace $T$ (\ref{eq:augmented-statistic}) by
\begin{equation}\label{eq:weighted-statistic}
 B=\sum_i\rho_i\widetilde x_i\widetilde x_i^T
 =\begin{pmatrix}C+b\theta\theta^T&b\theta\\b\theta^T&b\end{pmatrix},
 \qquad 0\le\rho_i\le1,\qquad b=\sum_i\rho_i.
\end{equation}
The comparison below treats $B$ directly. For any positive definite
$B$ in the block form \eqref{eq:weighted-statistic} with $b>m$,
we use the mean $\theta$ and scatter $C$ encoded by its blocks.
The condition $b>m$ makes the row factor $\Gamma_b$ below positive
definite, and $n>\max\{m,d\}$ gives enough frame rows and columns.

Let $Z$ be the first $m$ rows of a uniform $(n-1)\times d$ frame.
For $B\succ0$ and $b>m$, generate the candidate batch by
\begin{equation}\label{eq:full-candidate}
 Y=\one_m\theta^T+\Gamma_b^{1/2}ZC^{1/2},\qquad
 \Gamma_b=I_m-\frac1b\one_m\one_m^T.
\end{equation}

To see that setting every weight to one recovers exact resampling,
let \(E\in\R^{n\times(n-1)}\) have orthonormal columns spanning
\(\one_n^\perp\). Let $Q$ be a uniform $(n-1)\times d$ frame, so
that $U=EQ$ is a uniform centered frame. Then
$E_{[m]}E_{[m]}^T=\Gamma_n$, and rotational invariance gives
$U_{[m]}\overset d=\Gamma_n^{1/2}Z$. Hence, when every weight equals one, $B=T$ and
\eqref{eq:full-candidate} agrees with the exact resampling rule in
\eqref{eq:exact-batch}.

Algorithm~\ref{alg:upper-weighted} draws an $(n-1)\times d$ frame,
with the number of rows determined by the original sample size
$n$. The total weight $b$ enters through $\Gamma_b$. We assume $n\ge m+d+2$
so the determinant exponents in the candidate density are
nonnegative.

\begin{algorithm}[H]
\caption{\textsc{WeightedResample}$(n,m,b,\theta,C)$}
\label{alg:upper-weighted}
\begin{algorithmic}[1]
\Require Original sample size $n\ge m+d+2$, batch size $m$,
         total weight $b>m$, weighted mean $\theta\in\mathbb R^d$,
         and unnormalized weighted scatter $C\succ0$.
\Ensure A full batch of $m$ vectors.

\State Draw $H\in\mathbb R^{(n-1)\times d}$ uniformly among
       matrices satisfying $H^TH=I_d$.
       \Comment{A uniform frame with orthonormal columns}

\State Let $Z$ be the first $m$ rows of $H$.

\State Form
       \[
       Y=
       \mathbf1_m\theta^T+
       \left(I_m-\frac1b\mathbf1_m\mathbf1_m^T\right)^{1/2}
       ZC^{1/2}.
       \]

\State \Return $(y_1,\ldots,y_m)$, where the $j$th row of $Y$
       is $y_j^T$.
\end{algorithmic}
\end{algorithm}

For privacy, we compare two candidate densities after applying
the same invertible affine map to every vector in both batches.
Using the weighted mean $\theta$ and scatter $C$ from the first
statistic, we apply $y\mapsto C^{-1/2}(y-\theta)$. This puts the
first statistic at mean zero and scatter $I_d$.

Weighting and rejection can introduce dependence among the outputs,
so we use a correction to restore exact independence on Gaussian
input. For this correction, we need to carry a batch prepared at
public reference statistics to the empirical statistics of the
input. Both steps use the following transformation property.

\begin{lemma}[Affine equivariance of weighted resampling]
\label{lem:weighted-resampling-equivariance}
Let $Y$ be drawn by \eqref{eq:full-candidate} with statistics
$b,\theta,C$. Fix an invertible $F\in\R^{d\times d}$ and $v\in\R^d$.
Let $Y'$ be drawn by the same rule with the statistics obtained by
transforming each record by $x_i\mapsto Fx_i+v$ and keeping the
weights fixed. Then
\[
 YF^T+\one_m v^T\overset d=Y'.
\]
\end{lemma}

\begin{proof}
The transformed statistics are $\theta'=F\theta+v$, $C'=FCF^T$,
and $b'=b$, so the row factor $\Gamma_b$ is the same for both
draws. Set $O=C^{1/2}F^T(C')^{-1/2}$. Since
$(C^{1/2}F^T)^T(C^{1/2}F^T)=C'$, the matrix $O$ is orthogonal.
Thus $C^{1/2}F^T=O(C')^{1/2}$, and right invariance of the
uniform frame gives $ZO\overset d=Z$. Substitution in
\eqref{eq:full-candidate} gives
\[
 YF^T+\one_m v^T
 =\one_m(\theta')^T+\Gamma_b^{1/2}ZO(C')^{1/2}
 \overset d=Y'.
\]
\end{proof}

When the same invertible map is applied to two candidate batches,
the change of variables contributes the same Jacobian factor to
both densities. Their density ratio is therefore preserved.

\subsubsection{Stable weights by averaging survivor sets}\label{sec:ub-stability-proof}
To control the weighted statistic on neighboring datasets, we adapt
the survivor-set averaging construction of Brown, Hopkins, and
Smith \cite[Section~3]{BHS23}, also used in
Gaussian single-sampling~\cite{IKM25}. Pruning at a single threshold
can change the survivors abruptly because deleting one record can
increase the influence of others. Averaging the survivor indicators
at nearby thresholds smooths this change. We apply the rule to the vectors $\widetilde x_i$ to control changes in the mean and scatter together.

The bound on how much the weighted statistic changes between
neighboring datasets applies when pruning can meet a threshold
after excluding few records on both inputs. We use a private test
to certify that this bound applies to the input and each of its
neighbors before drawing a candidate batch. The test uses a
stability score computed from the survivor counts.

For $I\subseteq[n]$, let
$T_I=\sum_{i\in I}\widetilde x_i\widetilde x_i^T$.
When $T_I$ is nonsingular, we measure the influence of record $i$
relative to $T_I$ by the quadratic form
$\widetilde x_i^\top T_I^{-1}\widetilde x_i$. At threshold $u$, start with
$I=[n]$ and repeatedly delete the smallest $i\in I$ satisfying
\[
 n\widetilde x_i^TT_I^{-1}\widetilde x_i>u.
\]
Return the survivors when no violation remains. If $T_I$ is singular
at any point, return the empty set. The factor $n$ is kept fixed
throughout pruning.

We choose consecutive thresholds so that a set meeting one threshold
still meets the next after deleting a record. Suppose every $i\in I$
satisfies $\widetilde x_i^T T_I^{-1}\widetilde x_i\le u/n<1$.
Removing one row leaves a statistic at least $(1-u/n)T_I$ in the
positive semidefinite order. Taking inverses reverses this order
for positive definite matrices by Lemma~\ref{lem:prelim-order-trace}.
Thus the inverse of the remaining statistic is at most
$(1-u/n)^{-1}T_I^{-1}$, so the quadratic form for each remaining
row is at most $(u/n)/(1-u/n)=u/(n-u)$. Multiplying by $n$ gives the bound $u/(1-u/n)$. Starting at $u_0=\lambda$, use this bound to
choose $u_{j+1}$ from $u_j$ for $0\le j<2k$.
\begin{equation}\label{eq:thresholds}
 u_j=\frac{\lambda}{1-j\lambda/n},\qquad 0\le j\le2k,
 \qquad u_{j+1}=\frac{u_j}{1-u_j/n},
\end{equation}
where $k,\lambda\ge1$ are integers and $n\ge16k\lambda$.
For $0\le j\le2k$, this condition gives
$1-j\lambda/n\ge7/8$, so every threshold is defined and lies
between $\lambda$ and $8\lambda/7$. Thus averaging over nearby
levels preserves the same influence scale.
For each $j$, run the pruning procedure from the full dataset at
threshold $u_j$, and denote its survivor set by $I_j$.

The lower thresholds determine the stability score $g$ used by
the private test in Section~\ref{sec:rejection-design}. The high thresholds define the weights by averaging.
The intermediate level $k$ records the set of deletions used in the
correction. Since the survivor sets are nested, every record surviving
at this level receives weight one. Define
\begin{equation}\label{eq:weights}
 \begin{split}
 g&=\min\bigl(\{j\in\{0,\ldots,k-1\}:|I_j|\ge n-j\}
                   \cup\{k\}\bigr),\\
 \rho_i&=\frac1k\sum_{j=k+1}^{2k}\one_{\{i\in I_j\}},\qquad
 B=\sum_i\rho_i\widetilde x_i\widetilde x_i^T.
 \end{split}
\end{equation}
Thus $g$ is the first level $j<k$ at which pruning at threshold
$u_j$ leaves out at most $j$ records. If no such level exists, set
$g=k$. When $g<k$, record the deleted set $A=[n]\setminus I_k$.

The next lemma gives the properties used by the test and correction.
Sensitivity of $g$ permits a private test, and the relative matrix
bound controls the candidate comparison on neighboring inputs.
The survivors at level $k$ ensure that the candidate is well defined
and give the correction its deletion set on acceptance. The count at
the base threshold will be used on rejection.

\begin{lemma}[Properties of the weights]\label{lem:full-stable}
The routine above has the following properties. Primes denote its
output on a neighboring dataset.
\begin{enumerate}
\item $0\le g\le k$ and $|g-g'|\le1$.
\item If $g<k$, then the recorded set $A$ consists of fewer than
      $k$ rows deleted at a threshold at least $\lambda$, and
      \[
      \rho_i=1\quad(i\notin A),\qquad
      0\prec T_{A^c}\preceq B\preceq T,\qquad
      n-|A|\le b\le n.
      \]
\item If $g,g'<k$, then
      \[
      \|B^{-1/2}B'B^{-1/2}-I_{d+1}\|_*\le16\lambda/n.
      \]
\item Pruning at threshold $\lambda$ returns at most $n-g$ survivors.
\end{enumerate}
An invertible affine change $x_i\mapsto Fx_i+v$ preserves the
sets $I_j$, score $g$, and weights $\rho_i$. It sends $B$ to
$\widetilde F B\widetilde F^\top$, where
$\widetilde F=\left(\begin{smallmatrix}F&v\\0&1\end{smallmatrix}\right)$.
\end{lemma}

\begin{proof}[Proof sketch]
Call a set valid at threshold $u$ if its statistic is positive
definite and all its rows meet the threshold relative to that
statistic. If a valid set $J$ is contained in the current set $I$,
then $T_I\succeq T_J$, so each record of $J$ still meets the threshold
relative to $T_I$. Pruning therefore preserves every valid subset.
When it returns a nonempty set, that set is the largest valid one.
Because the thresholds increase, the survivor sets are nested.

Let $i_*$ be the changed record index. The threshold calculation
above shows that $I_j\setminus\{i_*\}$ is valid at level $j+1$ on
the neighboring dataset whenever $I_j$ is nonempty. Thus
$I_j\setminus\{i_*\}\subseteq I'_{j+1}$, and exchanging the inputs
gives $I'_j\setminus\{i_*\}\subseteq I_{j+1}$. If $g<k-1$, the
first inclusion at $j=g$ gives
$|I'_{g+1}|\ge |I_g|-1\ge n-(g+1)$, so $g'\le g+1$.
The bound $g'\le k$ gives the same conclusion for $g\ge k-1$. Exchanging the inputs
yields $|g-g'|\le1$.

When $g<k$, nesting gives $|I_k|\ge n-g$, and each record in
$I_k$ has weight one. The other weights lie in $[0,1]$, so
$0\prec T_{I_k}\preceq B\preceq T$ and $|I_k|\le b\le n$.
For the base-threshold count, put
$a=n-|I_0|$. If $a<k$, then $|I_a|\ge n-a$ by nesting, so $g\le a$.
If $a\ge k$, then $g\le k\le a$.

For an unchanged record, the inclusions at adjacent levels shift
the summands in its weight by one level. Only an endpoint can
be lost, so its weight changes by at most $1/k$. When both scores
are below $k$, a decrease is possible only outside $I'_k$ and an
increase only outside $I_k$. Each complement has fewer than $k$
records, so the total weight change in each direction is at most
one on the unchanged records. The changed record adds at most
one to each total.

Every record of positive weight belongs to $I_{2k}$. Compared
with $T_{I_{2k}}$, the statistic $T_{I_k}$ omits fewer than $k$
records, each satisfying $\widetilde x_i^\top T_{I_{2k}}^{-1}\widetilde x_i\le u_{2k}/n$. Thus
$B\succeq T_{I_k}\succeq (1-ku_{2k}/n)T_{I_{2k}}$, which bounds
$\widetilde x_i^\top B^{-1}\widetilde x_i$ for every record $i$ of positive weight by
$2\lambda/n$. For each unchanged record, put $\widetilde x_i\widetilde x_i^\top$ in $E$ with coefficient $\rho_i-\rho_i'$ if its weight
decreases, and in $E'$ with coefficient $\rho_i'-\rho_i$ if its
weight increases. The changed record contributes its old weighted
outer product to $E$ and its new weighted outer product to $E'$.
Thus $E,E'\succeq0$ and $B'-B=E'-E$. The bounds on the weights and quadratic forms give
$\tr(B^{-1}E),\tr(B'^{-1}E')\le4\lambda/n$.
In particular, $E'\preceq B'/2$, so $B\succeq B'/2$ and
$\tr(B^{-1}E')\le8\lambda/n$. The nuclear-norm triangle
inequality bounds the relative change by the sum of the traces
of $B^{-1/2}EB^{-1/2}$ and $B^{-1/2}E'B^{-1/2}$, at most
$12\lambda/n$. Under the affine change in the lemma, $\widetilde x_i$ becomes
$\widetilde F\widetilde x_i$ and $T_I$ becomes
$\widetilde F T_I\widetilde F^\top$. Since $\widetilde F$ is invertible,
this preserves the rank of $T_I$. When $T_I$ is nonsingular,
\[
 (\widetilde F\widetilde x_i)^\top
 (\widetilde F T_I\widetilde F^\top)^{-1}
 (\widetilde F\widetilde x_i)
 =\widetilde x_i^\top T_I^{-1}\widetilde x_i.
\]
Thus the pruning decisions are unchanged.
Appendix~\ref{app:stable-weight-details} gives the full proof.
\end{proof}

The relative matrix bound in Lemma~\ref{lem:full-stable} requires
both scores to be below $k$. The test ensures this whenever the
sampler draws a candidate.

\subsubsection{The stability test}\label{sec:rejection-design}

Acceptance must guarantee that the input and every neighbor have
scores below \(k\), so the joint privacy comparison applies. The
branch selector also applies the test to a separate reference
dataset with \(n\) records, empirical mean zero, and scatter \(I_d\).
On this dataset, rejection must certify at least \(a_0\) deletions
at the base threshold, for a positive integer \(a_0\).
Section~\ref{sec:exact-correction} uses these deletions to show that
the selector's proposed replacement probability is at most one when
the simulated test rejects.

Adding Laplace noise to the sensitivity-one score lets us test it
privately. To make the conclusions on acceptance and rejection
deterministic, clip the noise to \([-\ell,\ell]\), replacing draws
below \(-\ell\) or above \(\ell\) by the corresponding endpoint.
Specifically, for \(\ell\ge1\) draw \(V\sim\operatorname{Lap}(2/\varepsilon)\)
and set \(W=\max\{-\ell,\min\{V,\ell\}\}\).
Fact~\ref{fact:prelim-clipped-laplace} bounds the additional privacy
error from clipping. Reject when \(g+W\ge a_0+\ell\), and set
\(k=a_0+\lceil2\ell\rceil+1\). The parameter choice in
Appendix~\ref{app:upper-calculations} ensures \(n-k>\max\{m,d\}\).
The private sampler $\mathcal A$ in Algorithm~\ref{alg:base}
tests the score first and draws a candidate with independent
randomness on acceptance.

\begin{algorithm}[H]
\caption{Sampler $\mathcal A$}
\label{alg:upper-base}\label{alg:base}
\begin{algorithmic}[1]
\Require Dataset $x=(x_1,\ldots,x_n)$, batch size $m$, and calibrated
         public parameters $\lambda,k,a_0,\ell,\varepsilon$.
\Ensure A full batch of $m$ vectors, or $\bot$.

\State Compute the weights $\rho_1,\ldots,\rho_n$ and stability
       score $g$ using \eqref{eq:weights}.

\State Draw $V\sim\operatorname{Lap}(2/\varepsilon)$.
\State Set $W\gets\max\{-\ell,\min\{V,\ell\}\}$.
       \Comment{Clip the Laplace draw}

\If{$g+W\ge a_0+\ell$}
    \State \Return $\bot$.
           \Comment{Reject the entire batch}
\EndIf

\State Set $b\gets\sum_i\rho_i$.
\State Set $\theta\gets b^{-1}\sum_i\rho_i x_i$.
\State Set $C\gets\sum_i\rho_i(x_i-\theta)(x_i-\theta)^T$.
       \Comment{Form the weighted statistics after acceptance}

\State \Return
       $\Call{WeightedResample}{n,m,b,\theta,C}$
       using independent randomness.
       \Comment{Algorithm~\ref{alg:upper-weighted}}
\end{algorithmic}
\end{algorithm}

If \(g<a_0\), then \(g+W<a_0+\ell\) and the test accepts.
If \(g\ge k-1\), then \(g+W\ge k-1-\ell\ge a_0+\ell\) and
the test rejects. Thus acceptance gives \(g<k-1\). Since
\(|g-g'|\le1\), all neighboring scores are below \(k\).
Lemma~\ref{lem:full-stable} gives \(B\succ0\) and
\(b\ge|I_k|>n-k>m\), so the weighted draw is well defined.

On rejection, \(W\le\ell\) gives \(g\ge a_0\), so
\(|I_0|\le n-a_0\) by Lemma~\ref{lem:full-stable}. On a
reference dataset every \(d+1\) rows are affinely independent
almost surely. Since \(a_0<k\) and \(n-k>d\), pruning cannot
encounter a singular statistic before the first \(a_0\) deletions.
Thus the correction can use those actual deletions.
Appendix~\ref{app:stable-weight-details} gives the details.

The weights and test are affine invariant. Together with the
equivariance of the weighted draw, this means an invertible affine
change of the input transforms the accepted vectors in the same
way and preserves the rejection probability. In
Section~\ref{sec:exact-correction}, this lets us construct the
correction at fixed public reference statistics and then transfer
it to the empirical statistics of the input.

\subsubsection{The density needed for privacy and correction}
\label{sec:candidate-density}

To prove privacy, we combine the privacy of the test with a
comparison of candidate distributions on neighboring datasets.
The candidate comparison bounds the density ratio on a
high-probability event where both densities are positive. In Section~\ref{sec:exact-correction}, we use the same density
formula to compare weighted resampling at $B$ with exact resampling
at $T$, the unweighted statistic of the same reference dataset.
That comparison uses only the retained output positions. We now
derive the density formula used in both arguments.

Take a weighted statistic $B\succ0$ of the form
\eqref{eq:weighted-statistic} with total weight $b>m$.
Its blocks specify the weighted mean $\theta$ and scatter $C$,
so \eqref{eq:full-candidate} defines the corresponding candidate.
For a fixed set of $r$ positions, $1\le r\le m$, let $f_B^{(r)}$
be the joint density of the selected rows in increasing position
order. Permuting the rows of the candidate batch leaves its
distribution unchanged. Thus any choice of $r$ positions gives
the same joint density for the selected rows.
When we select zero positions, the only outcome is the empty tuple,
which has probability one. We write $f_B^{(0)}=1$ for this case.

For an $r\times d$ matrix $Y$ of selected output rows, append a
column of ones to obtain $\widetilde Y$, in the same coordinates
as the weighted statistic $B$. This lets the lemma express the
support by testing whether $B-\widetilde Y^T\widetilde Y$ is
positive definite. The determinant factor $R_B(Y)$ and exponent
$\kappa_r$ describe the part of the density that varies with
$Y$ within that support.
\[
 \widetilde Y=[Y,\one_r],\qquad
 R_B(Y)=\det(I_r-\widetilde YB^{-1}\widetilde Y^T),\qquad
 \kappa_r=\frac{n-r-d-2}{2}.
\]

The lemma gives the support and separates the normalization from
the output-dependent term. This form lets us track their changes
together when comparing two statistics.

\begin{lemma}[Candidate density]\label{lem:full-density}
Let $B\succ0$ have total weight $b>m$, and suppose $n\ge m+d+2$.
For $1\le r\le m$, the density $f_B^{(r)}$ is zero outside
$B-\widetilde Y^T\widetilde Y\succ0$. On this set it is
\begin{equation}\label{eq:full-density}
 f_B^{(r)}(Y)=c_{n-1,r,d}A_r(b)(\det B)^{-r/2}
                  R_B(Y)^{\kappa_r},\qquad
 A_r(b)=\frac{b^{(n-2)/2}}{(b-r)^{(n-r-2)/2}},
\end{equation}
where $c_{N,r,d}$ is given in Lemma~\ref{ub-bg:haar-block-density}.
\end{lemma}

\begin{proof}
Let $Q$ be the full uniform $(n-1)\times d$ frame whose first $m$
rows form $Z$ in \eqref{eq:full-candidate}. Let $L$ consist of the
selected rows of $\Gamma_b^{1/2}$, padded on the right with
$n-1-m$ zero columns. Then $LQ$ is the selected block of
$\Gamma_b^{1/2}Z$. The row Gram matrix $LL^T$ is the corresponding
principal submatrix of $\Gamma_b$, namely
$\Gamma_{b,r}=I_r-\one_r\one_r^T/b$. It is positive definite
because $b>m\ge r$, so $L$ has full row rank. The projection part
of Lemma~\ref{lem:prelim-frame-projection} states that $LQ$ has
the same distribution as $(LL^T)^{1/2}Z_r$, where $Z_r$ consists
of the first $r$ rows of a uniform $(n-1)\times d$ frame. Thus
$LQ$ has the same distribution as $\Gamma_{b,r}^{1/2}Z_r$.
Restoring the mean and scatter factors gives the marginal representation
\[
Y=\one_r\theta^T+\Gamma_{b,r}^{1/2}Z_rC^{1/2},
\]
where $Z_r$ is the first $r$ rows of a uniform $(n-1)\times d$
frame. Lemma~\ref{ub-bg:haar-block-density}, with $N=n-1$ and
$p=d$, gives determinant exponent $(n-r-d-2)/2=\kappa_r$
and constant $c_{n-1,r,d}$. To express the change of variables
in terms of $B$, first note that the identity $\det(I-UV)=\det(I-VU)$
gives $R_B(Y)=\det(B-\widetilde Y^T\widetilde Y)/\det B$. With
$Y_c=Y-\one_r\theta^T$, the determinant and Schur-complement
identities in Lemma~\ref{lem:prelim-matrix-identities} give
\[
 \begin{gathered}
 \det B=b\det C,\qquad \det\Gamma_{b,r}=(b-r)/b,\\
 \det(B-\widetilde Y^T\widetilde Y)
 =(b-r)\det(C-Y_c^T\Gamma_{b,r}^{-1}Y_c).
 \end{gathered}
\]
Lemma~\ref{lem:prelim-affine-density}, applied with left factor
$\Gamma_{b,r}^{1/2}$ and right factor $C^{1/2}$, gives inverse
Jacobian determinant
$(\det\Gamma_{b,r})^{-d/2}(\det C)^{-r/2}$. Moreover,
\[
 \det(I_d-C^{-1/2}Y_c^T\Gamma_{b,r}^{-1}Y_cC^{-1/2})
 =\frac{b}{b-r}R_B(Y).
\]
Substituting this into the frame-block density gives the factor
\[
 (\det C)^{-r/2}
 \left(\frac b{b-r}\right)^{d/2+\kappa_r}
 =(\det B)^{-r/2}
 \frac{b^{(n-2)/2}}{(b-r)^{(n-r-2)/2}},
\]
which gives $A_r(b)(\det B)^{-r/2}$ in \eqref{eq:full-density}.
The constant $c_{n-1,r,d}$ comes from the normalized frame density,
so the change of variables preserves normalization. Since $b-r>0$,
the Schur complement gives
\[
 B-\widetilde Y^T\widetilde Y\succ0
 \quad\Longleftrightarrow\quad
 C-Y_c^T\Gamma_{b,r}^{-1}Y_c\succ0.
\]
Multiplying on the left and right by $C^{-1/2}$ preserves positive
definiteness and gives
\[
 C^{-1/2}(C-Y_c^\top\Gamma_{b,r}^{-1}Y_c)C^{-1/2}
 =I_d-Z_r^\top Z_r.
\]
The condition $I_d-Z_r^\top Z_r\succ0$ is exactly the support condition
for the frame block.
\end{proof}

The correction will use this formula for the retained positions.
We first use \(r=m\) to compare full candidates on neighboring
inputs. By Lemma~\ref{lem:full-density}, the densities under \(B\)
and \(B'\) are both positive at \(Y\) exactly when
\[
 B-\widetilde Y^\top\widetilde Y\succ0
 \quad\text{and}\quad
 B'-\widetilde Y^\top\widetilde Y\succ0.
\]
On this common region, the normalization and the output-dependent
determinant factor determine the density ratio.

\subsection{Joint privacy of $\mathcal{A}$}\label{sec:joint-privacy}\label{sec:private-joint}
For neighboring datasets whose scores are both below $k$,
Lemma~\ref{lem:full-stable} bounds the nuclear norm of the relative
change in the weighted statistic by $16\lambda/n$. For privacy, we need to turn that
matrix bound into a comparison of whole candidate batches. Because
the support depends on the statistic, the comparison must also
bound the probability of a batch outside the neighboring support.
We establish the comparison first, then combine it with the
clipped Laplace test.

The next lemma compares the joint distributions of candidate batches
drawn from two weighted statistics $B$ and $B'$. It bounds the privacy
exponent by $O(\Delta(\sqrt{mt}+t))$, where $\Delta$ is the nuclear norm
of the relative change $B^{-1/2}(B'-B)B^{-1/2}$ and $t$ controls the
exceptional probability. The $\sqrt{mt}$ term gives the square-root
dependence on batch size in the privacy term of the sample bound.
Lemma~\ref{lem:full-base-privacy} combines this bound for the batch
draw with the privacy guarantee of the clipped Laplace test to
establish privacy of $\mathcal A$.

\begin{lemma}[Comparing two candidate distributions]
\label{lem:full-candidate-privacy}
There are universal $c_*,C_*>0$ with the following property.
Let $t\ge1$ and $n\ge C_*(k+m+d+t)$. Let
$B,B'\in\R^{(d+1)\times(d+1)}$ be positive definite weighted statistics with total weights $b,b'\in[n-k,n]$, as in
\eqref{eq:weighted-statistic}. The sample-size condition gives
$n-k>m$, so \eqref{eq:full-candidate} defines both draws.
Generate $Y$ and $Y'$ using $B$ and $B'$, respectively. Define
\[
 \Delta=\|B^{-1/2}B'B^{-1/2}-I_{d+1}\|_*,\qquad
 s=\sqrt{mt}+t.
\]
If $\Delta s\le c_*$, then, for every measurable $E$,
\[
 \Prob\{Y\in E\}\le e^{C_*\Delta s}\Prob\{Y'\in E\}+C_*(d+1)e^{-t}.
\]
The same statement holds with $B,B'$ reversed, with the original
$\Delta$ and the same constants.
\end{lemma}

\paragraph{Proof roadmap.}
\emph{Cancellation in one dimension} shows why the first-order
privacy loss has mean zero when only the scatter changes.
\emph{The linear term in arbitrary dimension} establishes the same
cancellation when the mean, scatter, and total weight change together.

\emph{Concentration in a fixed direction} gives two estimates in
each eigenvector direction of the relative matrix change. The centered
estimate controls the fluctuation of the linear term. The uncentered
estimate is used in \emph{Support and remainder bounds} to show that
the sampled batch has positive density under the neighboring statistic
and to bound the higher-order terms. This completes the comparison
of the candidate distributions.

\emph{Privacy of the private sampler} combines the candidate comparison
with the clipped Laplace test to establish privacy for the sampler
that either releases a batch or rejects.

\paragraph{Cancellation in one dimension.}
The scalar case isolates the cancellation behind the $\sqrt m$
dependence. The matrix argument below proves the comparison for
arbitrary $d$.

Take $d=1$, fix the mean $\theta$ and total weight $b$, and let
$v=C>0$ be the weighted sum of squared deviations. To isolate its effect,
remove the common mean and row transformation. This gives
\[
 y=\Gamma_b^{-1/2}(Y-\theta\one_m)=\sqrt v\,Z.
\]
Here $N=n-1$ and $Z$ consists of the first $m$ coordinates of a
uniform direction in $\R^N$. Suppose $N>m+2$, and put
$\kappa=(N-m-2)/2$. Let $f_v$ denote the density of $y$ when the
weighted scatter is $v$.
The frame density and affine change of variables in
Lemmas~\ref{ub-bg:haar-block-density}
and~\ref{lem:prelim-affine-density} give
\[
 f_v(y)=c_{N,m,1}v^{-m/2}(1-\|y\|^2/v)^\kappa
 \qquad\text{for }\|y\|^2<v.
\]
The density is zero elsewhere, and $c_{N,m,1}$ is independent of $v$.
Rescaling both batches by $v^{-1/2}$ reduces the comparison to
values $1$ and $1+h$, where $h$ is the relative change in $v$.
Fix $y$ with $a=\|y\|^2<1$. For sufficiently small $|h|$, both
densities are positive and
\[
 \log\frac{f_1(y)}{f_{1+h}(y)}
 =\frac m2\log(1+h)+
  \kappa\left[\log(1-a)-\log\left(1-\frac{a}{1+h}\right)\right].
\]
Differentiating at $h=0$ gives the linear coefficient.
\[
 \left.\frac{d}{dh}\log\frac{f_1(y)}{f_{1+h}(y)}\right|_{h=0}
 =\frac m2-\kappa\frac{\|y\|^2}{1-\|y\|^2}.
\]
To see how these terms cancel in expectation, draw $y$ with density
$f_1$, so $y=Z$. By Lemma~\ref{lem:prelim-gaussian-direction},
dividing a standard Gaussian vector in $\R^N$ by its length gives a
uniform unit vector. We can therefore generate $y$ by normalizing
such a vector and keeping its first $m$ coordinates.
Splitting its first $m$ and remaining $N-m$ coordinates gives
\[
 \frac{\|y\|^2}{1-\|y\|^2}\overset d=
 \frac{\chi_m^2}{\chi_{N-m}^2},
\]
with independent numerator and denominator. Since $N-m>2$, the
mean formula in Lemma~\ref{lem:ub-scalar-facts} gives
$m/(N-m-2)$ for this ratio. Thus
\[
 \E_{y\sim f_1}\left[\frac m2-\kappa\frac{\|y\|^2}{1-\|y\|^2}\right]
 =\frac m2-\frac{N-m-2}{2}\frac{m}{N-m-2}=0.
\]
The linear term is centered. For $t\ge1$ and $N\ge32(m+t)$,
we have $N-m\ge16(m+t)$, so Lemma~\ref{lem:ub-scalar-facts} gives
$\left|\frac m2-\kappa\frac{\|y\|^2}{1-\|y\|^2}\right|
\le C(\sqrt{mt}+t)$ except with probability $4e^{-t}$. We next show that, in any dimension, the first-order term in the
joint log-density ratio has mean zero even when the weighted mean,
scatter, and total weight change together.
\paragraph{The linear term in arbitrary dimension.}
\label{sec:ub-comparison-proof}
Fix $B,B'$ and draw $Y$ by \eqref{eq:full-candidate} using $B$.
When the mean, scatter, and total weight can all change, centering
requires the combined first-order change of the normalization and
the factor that depends on $Y$. In Lemma~\ref{lem:full-density},
the corresponding log-density terms are
$\log A_m(b)-\frac m2\log\det B$ and $\kappa_m\log R_B(Y)$.
The factor $A_m$ in \eqref{eq:full-density} depends only on the
total weight once $n,m$ are fixed. Write
\[
 \phi(z)=\log A_m(z)
 =\frac{n-2}{2}\log z-\frac{n-m-2}{2}\log(z-m),\qquad z>m.
\]

Apply the same map $y\mapsto C^{-1/2}(y-\theta)$ to each vector in
both candidate batches, using $\theta$ and $C$ from $B$.
Lemma~\ref{lem:weighted-resampling-equivariance} identifies the
transformed candidate distributions, and the common Jacobian
cancels in the density ratio. On a vector $(y^\top,1)^\top$, the affine map acts by left multiplication with
\[
 L=\begin{pmatrix}
 C^{-1/2}&-C^{-1/2}\theta\\
 0&1
 \end{pmatrix}.
\]
Let $\bar B=LBL^\top$ and $O=\bar B^{-1/2}LB^{1/2}$.
Since $OO^\top=I_{d+1}$, the relative change transforms to
$O(B^{-1/2}B'B^{-1/2}-I_{d+1})O^\top$, preserving its nuclear
norm $\Delta$. Keep $B,B'$ for the transformed statistics and
$Y,Y'$ for the transformed batches. Then
$B=\operatorname{diag}(I_d,b)$. In these coordinates, set
\[
 \begin{gathered}
 H=B^{-1/2}B'B^{-1/2}-I_{d+1},\qquad \kappa=\kappa_m,\\
 W=[Y,\one_m/\sqrt b]=\widetilde YB^{-1/2}
 \in\R^{m\times(d+1)}.
 \end{gathered}
\]
The support condition $B-\widetilde Y^\top\widetilde Y\succ0$
gives $I_{d+1}-W^\top W\succ0$ after multiplying on both sides by
$B^{-1/2}$. Thus the singular values of $W$ are below one, and
$I_m-WW^\top\succ0$. Define
\[
 V_Y=W^{\top}(I_m-WW^{\top})^{-1}W
   =(I_{d+1}-W^{\top}W)^{-1}-I_{d+1}.
\]
The second expression follows from
Lemma~\ref{lem:prelim-matrix-identities}(4) with matrix argument
$W^\top$.
To compute the linear term in $\log(f_B^{(m)}(Y)/f_{B'}^{(m)}(Y))$,
fix $Y$ in the support and vary $B$ along
$B_\xi=B^{1/2}(I_{d+1}+\xi H)B^{1/2}=B+\xi(B'-B)$.
The derivative is taken at $\xi=0$, where the density stays positive
for sufficiently small $|\xi|$. The
matrix inside the determinant $R_{B_\xi}(Y)$ is
$I_m-W(I_{d+1}+\xi H)^{-1}W^{\top}$. Its derivative at $\xi=0$ is
$WHW^{\top}$, so the log-determinant derivative is
$\tr((I_m-WW^{\top})^{-1}WHW^{\top})=\tr(HV_Y)$.
For the two normalization terms, we have
\[
 \left.\frac{d}{d\xi}\log\det(I_{d+1}+\xi H)\right|_{\xi=0}
 =\tr H,\qquad b_\xi=b+\xi(b'-b),
\]
where $b_\xi$ is the total weight of $B_\xi$.
To center the resulting linear term we need the mean of $V_Y$.
Let $e=e_{d+1}$ be the last standard basis vector in $\R^{d+1}$.
Lemma~\ref{lem:ub-directional-comparison} gives
\begin{equation}\label{eq:mean-score-matrix}
 \kappa\E V_Y=\frac m2I_{d+1}-b\phi'(b)ee^{\top}.
\end{equation}
Since $b'-b=bH_{d+1,d+1}$, the entire linear term is therefore
\begin{equation}\label{eq:centered-linear-term}
 \frac m2\tr H-\kappa\tr(HV_Y)-(b'-b)\phi'(b)
 =-\kappa\tr\bigl(H(V_Y-\E V_Y)\bigr).
\end{equation}
The linear term is therefore centered even when all three
statistics change.

Write $H=\sum_{j=1}^{d+1}h_j u_j u_j^\top$, where the orthonormal
eigenvectors $u_j$ are fixed before $Y$ is drawn. The centered
linear term is
\[
 -\kappa\tr\bigl(H(V_Y-\E V_Y)\bigr)
 =-\sum_{j=1}^{d+1}h_j\kappa\,u_j^\top(V_Y-\E V_Y)u_j.
\]
Since $\sum_j|h_j|=\Delta$, bounding each centered term
$\kappa u_j^\top(V_Y-\E V_Y)u_j$ in absolute value by
$O(\sqrt{mt}+t)$ bounds the linear term by
$O(\Delta(\sqrt{mt}+t))$. We use the uncentered sum
$\sum_j|h_j|u_j^\top V_Yu_j$ to verify that $Y$ lies in the
support of the neighboring density and to bound the remainder.

\paragraph{Concentration in a fixed direction.}
\label{sec:ub-direction-proof}
The mean identity \eqref{eq:mean-score-matrix} gives the
cancellation in \eqref{eq:centered-linear-term}. To bound the
remaining fluctuation, we need a bound on
$u^\top(V_Y-\E V_Y)u$ in each fixed eigenvector direction of the
relative change. We also need a bound on $u^\top V_Yu$ to show
that the neighboring density is positive at $Y$ and to control
the Taylor remainder. The next lemma supplies the mean identity
and both bounds.

\begin{lemma}[Fluctuations in a fixed direction]
\label{lem:ub-directional-comparison}
Suppose $t\ge1$, $n\ge C(m+d+t)$, and $b\in[n/2,2n]$.
Then \eqref{eq:mean-score-matrix} holds. For each fixed unit vector
$u\in\R^{d+1}$, except with probability $Ce^{-t}$ under
the rule \eqref{eq:full-candidate} with $B=\operatorname{diag}(I_d,b)$,
\begin{equation}\label{eq:directional-bounds}
 |\kappa u^{\top}(V_Y-\E V_Y)u|\le C(\sqrt{mt}+t),\qquad
 u^{\top} V_Y u\le C(m+t)/n.
\end{equation}
\end{lemma}

To prove the lemma, we split $u$ into its first $d$ coordinates
and its last coordinate. The two diagonal contributions reduce
to ratios of squared projection lengths, which give their means
and deviations. For the mixed contribution, we use concentration
of a signed projection to obtain the square-root dependence on $m$.

\begin{proof}[Proof of Lemma~\ref{lem:ub-directional-comparison}]
Write the fixed unit vector as $u=(\rho v,\gamma)$, where
$\rho\ge0$ and $\rho^2+\gamma^2=1$. If $\rho=0$, choose any
fixed unit $v\in\R^d$. Otherwise $v$ is determined by the first
$d$ coordinates of $u$. In either case, $v$ is fixed before $Y$
is drawn. Set
\[
 \nu=n-m-d,\qquad \kappa=\frac{\nu-2}{2},\qquad
 s=\sqrt{mt}+t.
\]
The quantity $\nu$ is the denominator degrees of freedom in both
length ratios below. By taking the universal constant in the sample
assumption sufficiently large, we have
\[
 \nu\ge16(m+t),\qquad \nu\ge16(d+t),\qquad
 b-m\ge n/4.
\]
In particular, $\nu$, $\nu-2$, and $b-m$ are comparable to $n$.

\textit{Diagonal and mixed blocks.}
The quadratic form in $u$ has two diagonal terms and a mixed
term. Factoring out $\Gamma_b^{1/2}$ puts all three in terms
of the frame block $Z$ and the inverse of $I_m-ZZ^\top$.
Recall that $Y=\Gamma_b^{1/2}Z$, where $Z$ is the first $m$ rows
of a uniform $(n-1)\times d$ frame. Put
$G=(I_m-ZZ^{\top})^{-1}$. Since
$W=[Y,\one_m/\sqrt b]$ and
$\Gamma_b=I_m-\one_m\one_m^{\top}/b$,
\[
 \begin{aligned}
 I_m-WW^{\top}
 &=\Gamma_b-\Gamma_b^{1/2}ZZ^{\top}\Gamma_b^{1/2}
  =\Gamma_b^{1/2}(I_m-ZZ^{\top})\Gamma_b^{1/2},\\
 (I_m-WW^{\top})^{-1}
 &=\Gamma_b^{-1/2}G\Gamma_b^{-1/2}.
 \end{aligned}
\]
Since $\Gamma_b\one_m=(1-m/b)\one_m$, we have
$\Gamma_b^{-1/2}\one_m=\sqrt{b/(b-m)}\,\one_m$.
Substituting these identities into
$V_Y=W^{\top}(I_m-WW^{\top})^{-1}W$ gives
\begin{equation}\label{eq:directional-blocks}
 V_Y=
 \begin{pmatrix}
 Z^{\top}GZ&Z^{\top}G\one_m/\sqrt{b-m}\\
 \one_m^{\top}GZ/\sqrt{b-m}&\one_m^{\top}G\one_m/(b-m)
 \end{pmatrix}
 =\begin{pmatrix}V_{11}&V_{12}\\V_{12}^{\top}&V_{22}\end{pmatrix}.
\end{equation}
The three quantities for direction $u$ are
$v^{\top}V_{11}v$, $V_{22}$, and $v^{\top}V_{12}$.

\textit{The two length ratios.}
The two diagonal distributions supply the means for centering and
the tails for concentration. For $v^{\top}V_{11}v$, use
\eqref{eq:frame-ratio} in Lemma~\ref{lem:ub-frame-quadratic} with
$(r,p)=(m,d)$. For $V_{22}$, write $\widehat w=\one_m/\sqrt m$.
The inverse identity in Lemma~\ref{lem:prelim-matrix-identities}(4),
applied to $Z$, gives
\[
 G=I_m+Z(I_d-Z^{\top}Z)^{-1}Z^{\top}.
\]
It follows that
\[
 V_{22}=\frac m{b-m}
 \left(1+\widehat w^{\top}Z(I_d-Z^{\top}Z)^{-1}Z^{\top}\widehat w\right).
\]
The quadratic form in parentheses is covered by the same frame
lemma applied to $Z^{\top}$. Indeed, $Z^{\top}$ is a $d\times m$ block of the transpose of a
uniform $(n-1)\times(n-1)$ orthogonal matrix.
Transposition preserves that distribution. Exchanging $(r,p)$ from
$(m,d)$ to $(d,m)$ changes the numerator degrees of freedom but
leaves the denominator $n-r-p=\nu$ unchanged. Thus
\begin{equation}\label{eq:two-ratios}
 v^{\top}V_{11}v\overset d=\frac{\chi_m^2}{\chi_\nu^2},\qquad
 V_{22}\overset d=\frac m{b-m}
       \left(1+\frac{\chi_d^2}{\chi_\nu^2}\right).
\end{equation}
Each ratio has independent numerator and denominator. These marginal
identities suffice for the union bound below.

\textit{The mean of \(V_Y\).}
The mean is the quantity needed to cancel the normalization terms
in \eqref{eq:centered-linear-term}. By Lemma~\ref{lem:ub-scalar-facts},
the two chi-square ratios in \eqref{eq:two-ratios} have finite means.
Thus the diagonal entries of $V_Y$ have finite means. Since
$V_Y\succeq0$, we have
$|(V_Y)_{ij}|\le\tfrac12((V_Y)_{ii}+(V_Y)_{jj})$, so every
entry is integrable. For each unit $v$, the first ratio gives
$\E[v^\top V_{11}v]=m/(\nu-2)$, so $\E V_{11}=mI_d/(\nu-2)$.
Using the mean $d/(\nu-2)$ of the second ratio and
$2\kappa=\nu-2$ gives
\[
 \kappa\E V_{11}=\frac m2 I_d,\qquad
 \kappa\E V_{22}
 =\frac{m(\nu+d-2)}{2(b-m)}
 =\frac{m(n-m-2)}{2(b-m)}.
\]
The distribution is unchanged by $Z\mapsto-Z$. In
\eqref{eq:directional-blocks}, this transformation changes the sign
of $V_{12}$ and leaves the diagonal blocks unchanged. Consequently $\E V_{12}=0$, and
\[
 \kappa\E V_Y=
 \begin{pmatrix}
 (m/2)I_d&0\\
 0&\dfrac{m(n-m-2)}{2(b-m)}
 \end{pmatrix}.
\]
To compare the last entry with the weight-dependent normalization,
differentiate the two logarithms in $\phi$ to obtain
\[
 \phi'(z)=\frac{n-2}{2z}-\frac{n-m-2}{2(z-m)}
 =\frac{m(z-n+2)}{2z(z-m)}.
\]
At $z=b$, multiplying by $b$ and subtracting from $m/2$ gives
\[
 \frac m2-b\phi'(b)=\frac{m(n-m-2)}{2(b-m)},
\]
the last entry of the mean matrix. This proves the mean identity
\eqref{eq:mean-score-matrix}, including its last coordinate.

\textit{Diagonal fluctuations.}
The centered bounds control the diagonal contributions to the
linear term. The uncentered bounds also enter the mixed-term
estimate and the support and remainder bounds below. Apply
Lemma~\ref{lem:ub-scalar-facts} to each marginal in
\eqref{eq:two-ratios} and take a union bound. Except with probability
$Ce^{-t}$, we have
\begin{equation}\label{eq:directional-diagonal-bounds}
 \begin{aligned}
 |\kappa v^{\top}(V_{11}-\E V_{11})v|
   &\le Cs,\\
 |\kappa(V_{22}-\E V_{22})|
   &\le\frac{Cm}{b-m}(\sqrt{dt}+t)\le Cs,\\
 v^{\top}V_{11}v&\le C(m+t)/n,\qquad
 V_{22}\le Cm/n.
 \end{aligned}
\end{equation}
For the second inequality, the required comparison follows from
\[
 \frac m n\sqrt{dt}
 =\sqrt{mt}\,\frac{\sqrt{md}}n\le C\sqrt{mt},
 \qquad \frac m n t\le Ct.
\]
For the last bound, the scalar ratio estimate gives
$V_{22}\le Cm(1+(d+t)/n)/n\le Cm/n$. It remains to bound the
mixed term, for which the diagonal estimates alone are too large.

\textit{The signed mixed term.}
A direction with both components also needs a bound on
$v^\top V_{12}$. Positive semidefiniteness gives
$|v^\top V_{12}|^2\le(v^\top V_{11}v)V_{22}$, which with the
diagonal bounds yields order $\sqrt{m(m+t)}$ for
$\kappa|v^\top V_{12}|$. The signed projection bound supplies a
factor of order $\sqrt{t/m}$, giving the required scale.
The direction bound \eqref{eq:frame-cross} in
Lemma~\ref{lem:ub-frame-quadratic}, with $w=\one_m$, gives,
except with an additional probability $2e^{-t}$,
\[
 |v^{\top}Z^{\top}G\one_m|
 \le\sqrt{\frac{2t}{m}(\one_m^{\top}G\one_m)
       (v^{\top}Z^{\top}GZv)(1+v^{\top}Z^{\top}GZv)}.
\]
Divide by $\sqrt{b-m}$ and use
\eqref{eq:directional-blocks} to obtain
\[
 \kappa|v^{\top}V_{12}|
 \le\kappa\sqrt{\frac{2t}{m}V_{22}
                  (v^{\top}V_{11}v)(1+v^{\top}V_{11}v)}.
\]
On the event in \eqref{eq:directional-diagonal-bounds}, we have
$\kappa\le Cn$,
$v^{\top}V_{11}v\le C(m+t)/n$, and $V_{22}\le Cm/n$.
Also $1+v^{\top}V_{11}v\le C$, because $n\ge C(m+t)$.
Substitution now yields
\begin{equation}\label{eq:directional-mixed-bound}
 \kappa|v^{\top}V_{12}|
 \le Cn\sqrt{\frac{t}{m}\frac{m}{n}\frac{m+t}{n}}
 =C\sqrt{t(m+t)}\le C(\sqrt{mt}+t)=Cs.
\end{equation}
The mixed term has mean zero, so this is already its centered bound.

\textit{Combining the estimates in direction \(u\).}
The comparison needs both bounds in the eigenvector directions
of $H$. To get the bounds for the fixed direction $u=(\rho v,\gamma)$,
combine the three block estimates. Expanding the centered quadratic
form gives
\[
 \begin{aligned}
 \kappa u^{\top}(V_Y-\E V_Y)u
 &=\rho^2\kappa v^{\top}(V_{11}-\E V_{11})v
   +2\rho\gamma\kappa v^{\top}V_{12}
   +\gamma^2\kappa(V_{22}-\E V_{22}).
 \end{aligned}
\]
Equations~\eqref{eq:directional-diagonal-bounds} and
\eqref{eq:directional-mixed-bound}, together with
$|2\rho\gamma|\le1$, bound its absolute value by $Cs$.
For the uncentered quadratic form, restrict $V_Y\succeq0$ to the
two orthonormal directions $(v,0)$ and $(0,1)$. The resulting
$2\times2$ matrix is positive semidefinite, so its largest
eigenvalue is at most its trace. Thus
\[
 u^{\top}V_Yu\le v^{\top}V_{11}v+V_{22}\le C(m+t)/n.
\]
Both conclusions hold outside a set of probability $Ce^{-t}$ after
combining the diagonal and mixed-term exceptional events. These are
exactly the two bounds in \eqref{eq:directional-bounds}.
\end{proof}

\paragraph{Support and remainder bounds.}
We have a centered linear term and two bounds in each fixed
direction. Applying them in the eigenvector directions of $H$
gives one event on which we can control the linear term, verify
that the neighboring density is positive, and bound the remainder.

\begin{proof}[Completion of the proof of Lemma~\ref{lem:full-candidate-privacy}]
Use the normalized coordinates $B=\operatorname{diag}(I_d,b)$
and the matrices $H,W,V_Y$ defined above. Recall that
$\Delta=\|H\|_*$. Let $P_B$ and $P_{B'}$ denote the candidate
distributions from $B$ and $B'$, respectively. The output $Y$
in this part of the proof is drawn from $P_B$.

The sample-size condition ensures $b\ge n-k\ge n/2$ and lets
us apply Lemma~\ref{lem:ub-directional-comparison}. Diagonalize
$H=\sum_jh_ju_ju_j^{\top}$ and put
$|H|=\sum_j|h_j|u_ju_j^{\top}$. Apply the directional lemma to these
$d+1$ fixed directions. Let $\mathcal E_B$ be the event that $Y$
lies in the support under $B$ and both directional bounds hold
in all these directions. The union bound gives
$P_B(\mathcal E_B^c)\le C(d+1)e^{-t}$. On $\mathcal E_B$,
summing the estimates using $\sum_j|h_j|=\Delta$ gives
\begin{equation}\label{eq:comparison-event}
 |\kappa\tr(H(V_Y-\E V_Y))|\le C\Delta s,\qquad
 R_H:=\tr(|H|V_Y)\le C\Delta(m+t)/n,
 \quad s=\sqrt{mt}+t.
\end{equation}
The quantity $R_H$ is the sum of the uncentered quadratic forms
weighted by $|h_j|$. We use it first for support and then for
the remainder.

For support, put $K=I_{d+1}-W^\top W\succ0$, so $K^{-1}=I_{d+1}+V_Y$.
Expanding $K^{-1/2}HK^{-1/2}$ as the sum of
$h_j(K^{-1/2}u_j)(K^{-1/2}u_j)^\top$ gives
\[
 \|K^{-1/2}HK^{-1/2}\|_{\op}
 \le\sum_j|h_j|u_j^\top K^{-1}u_j
 =\Delta+R_H<1.
\]
The last inequality uses $\Delta+R_H\le\Delta[1+C(m+t)/n]$ and
$\Delta\le\Delta s\le c_*$ for small enough $c_*$.
Consequently $K+H=I_{d+1}+H-W^\top W\succ0$.
This is equivalent to the neighboring support condition
$B'-\widetilde Y^{\top}\widetilde Y\succ0$, so the neighboring density
is positive at $Y$.

To bound the remainder, factor out $R_B(Y)$ before expanding
the determinant. This will let us use the bound on $R_H$.
Put $Q=I_m-WW^{\top}\succ0$ and $G_H=H(I+H)^{-1}$. The identity
$I-(I+H)^{-1}=G_H$ yields
\[
 \begin{aligned}
 R_{B'}(Y)
 &=\det(Q+WG_HW^{\top})\\
 &=\det Q\,\det(I_{d+1}+G_HW^{\top}Q^{-1}W)\\
 &=R_B(Y)\det(I_{d+1}+G_HV_Y).
 \end{aligned}
\]
The middle equality is the determinant identity
$\det(I+UV)=\det(I+VU)$. Apply that identity once more, using
the symmetric positive semidefinite square root of $V_Y$, to obtain
\[
 R_{B'}(Y)=R_B(Y)\det(I_{d+1}+M),\qquad
 M=V_Y^{1/2}H(I+H)^{-1}V_Y^{1/2}.
\]
The matrix $M$ is symmetric because $H$ commutes with $(I+H)^{-1}$.
The factorization uses only a square root of $V_Y$, so it also holds
when $V_Y$ is singular.
Consequently the exact log-density ratio is
\begin{equation}\label{eq:exact-privacy-loss}
 \log\frac{f_B^{(m)}(Y)}{f_{B'}^{(m)}(Y)}
 =\frac m2\log\det(I+H)-\kappa\log\det(I+M)
        -\phi(b')+\phi(b).
\end{equation}
The potentially large coefficient $\kappa=O(n)$ now multiplies
the log determinant of $I+M$. We control $M$ using $R_H$,
whose bound in \eqref{eq:comparison-event} has a factor $1/n$.

We next verify the smallness needed to expand both logarithms.
First, $\|H\|_{\op}\le\Delta\le c_*\le1/2$. Since $H$ and
$(I+H)^{-1}$ have the same eigenvectors, set $a_j=V_Y^{1/2}u_j$.
Then
\[
 M=\sum_j\frac{h_j}{1+h_j}a_ja_j^\top,\qquad
 R_H=\sum_j|h_j|\|a_j\|^2.
\]
Use $\|a_ja_j^\top\|_*=\|a_j\|^2$ for the norm bound and
$h_j/(1+h_j)-h_j=-h_j^2/(1+h_j)$ for the trace error.
Since $|h_j|\le\Delta$, this gives
\[
 \|M\|_*\le\frac{R_H}{1-\Delta}\le2R_H,\qquad
 |\tr M-\tr(HV_Y)|\le\frac{\Delta R_H}{1-\Delta}
 \le2\Delta R_H.
\]
In particular, \eqref{eq:comparison-event} implies
\[
 \|M\|_{\op}\le\|M\|_*\le C\Delta(m+t)/n\le1/2
\]
for sufficiently small $c_*$, because $n\ge C_*(m+d+t)$.
Lemma~\ref{lem:prelim-matrix-perturbation}, together with
$\|\cdot\|_F\le\|\cdot\|_*$, therefore bounds the logarithmic
remainders for both $H$ and $M$ by $C\|H\|_*^2$ and
$C\|M\|_*^2$, respectively.
The scalar remainder
$\phi(b')-\phi(b)-(b'-b)\phi'(b)$ is also quadratic.
Differentiating the two terms in $\phi'(z)$ and grouping them gives
\[
 \phi''(z)=\frac{n-m-2}{2}
 \left(\frac{1}{(z-m)^2}-\frac{1}{z^2}\right)-\frac{m}{2z^2},
 \qquad
 \frac{1}{(z-m)^2}-\frac{1}{z^2}
 =\frac{m(2z-m)}{z^2(z-m)^2}.
\]
For $z\in[n-k,n]$, both $z$ and $z-m$ are of order $n$.
The difference of inverse squares is $O(m/n^3)$, so
$|\phi''(z)|\le Cm/n^2$. Taylor's theorem with
$|b'-b|=b|H_{d+1,d+1}|\le n\Delta$ gives the scalar bound below.
The four errors relative to the
centered linear term \eqref{eq:centered-linear-term} are therefore
\[
 \begin{aligned}
 \frac m2\bigl|\log\det(I+H)-\tr H\bigr|
     &\le Cm\Delta^2,\\
 \kappa\bigl|\tr M-\tr(HV_Y)\bigr|
     &\le Cn\Delta R_H,\\
 \kappa\bigl|\log\det(I+M)-\tr M\bigr|
     &\le CnR_H^2,\\
 \bigl|\phi(b')-\phi(b)-(b'-b)\phi'(b)\bigr|
     &\le Cm\Delta^2.
 \end{aligned}
\]
Thus $m\Delta^2$ comes from the two normalization terms,
$n\Delta R_H$ from replacing $H(I+H)^{-1}$ by $H$, and
$nR_H^2$ from the quadratic residual log-determinant. Substitution
of \eqref{eq:comparison-event} gives
\[
 \begin{aligned}
 m\Delta^2+n\Delta R_H+nR_H^2
 &\le C\Delta^2\left[m+(m+t)+\frac{(m+t)^2}{n}\right]\\
 &\le C\Delta^2(m+t)\le C\Delta s.
 \end{aligned}
\]
Here $n\ge C_*(m+d+t)$ bounds the fraction, while
$m+t\le s^2$ and $\Delta s\le c_*$ give the final inequality.
Together with \eqref{eq:comparison-event}, this bounds the absolute
log-density ratio by $C\Delta s$ on $\mathcal E_B$. For any measurable
$E$, integrating over $E\cap\mathcal E_B$ gives
\[
 P_B(E)
 \le e^{C\Delta s}P_{B'}(E\cap\mathcal E_B)+P_B(\mathcal E_B^c)
 \le e^{C\Delta s}P_{B'}(E)+C(d+1)e^{-t}.
\]
For the reverse comparison we need an event of high probability
under $P_{B'}$. Put $D=B'^{-1/2}B^{1/2}$. The reversed relative
matrix before subtracting the identity is $DD^\top$, whose
eigenvalues are those of $D^\top D=(I_{d+1}+H)^{-1}$. Thus the
eigenvalues of the reverse change are $-h_j/(1+h_j)$, and their
absolute values sum to at most $\Delta/(1-\Delta)\le2\Delta$. Starting with a sufficiently small
universal $c_*$ permits the same estimates in reverse; enlarging
$C_*$ gives the claim with the original $\Delta$.
\end{proof}

\paragraph{Privacy of the private sampler.}
The candidate comparison controls the batch draw when both scores
are below $k$. To obtain privacy of $\mathcal A$, we also need to
account for the decision to release the batch. The clipped test makes
both executions reject when either score equals $k$. On the other
pairs, we use half the multiplicative budget for the test and half
for candidate generation.

\begin{lemma}[Privacy of $\mathcal A$]\label{lem:full-base-privacy}
For a sufficiently large universal constant $C_{\rm priv}$, take
\[
 \ell=1+\frac2\varepsilon\log\frac{32}{\delta},\qquad
 t=\log\frac{C_{\rm priv}(d+1)}{\delta},\qquad
 k=a_0+\lceil2\ell\rceil+1.
\]
If
\begin{equation}\label{eq:base-size}
 n\ge C_{\rm priv}\left[
 k\lambda+m+d+t+
 \frac{\lambda(\sqrt{mt}+t)}{\varepsilon}\right],
\end{equation}
then $\mathcal A$ is $(\varepsilon,\delta/4)$-DP.
\end{lemma}

\begin{proof}
Use the theorem's privacy parameter range $0<\varepsilon\le1$
and $0<\delta<1$, and fix neighboring datasets. The size condition
\eqref{eq:base-size} gives $n\ge C_{\rm priv}k\lambda\ge16k\lambda$,
so Lemma~\ref{lem:full-stable} applies. If either score equals $k$,
sensitivity one gives that both are at least $k-1$. Recall that
$W$ here is the clipped Laplace draw, with $W\ge-\ell$. In both
executions,
\[
 g+W\ge k-1-\ell=a_0+\lceil2\ell\rceil-\ell\ge a_0+\ell,
\]
so both samplers return $\bot$. Otherwise both scores are below
$k$. By Lemma~\ref{lem:full-stable}, both weighted statistics are
positive definite and their total weights exceed $n-k>m$.
The candidates therefore exist on both inputs of this pair. For
the privacy analysis, generate the candidate independently of the
test even on rejection, and discard it if the test rejects. This
gives the same released output as the algorithm and permits
composition of the two steps.

First bound the privacy cost of the test. The score has sensitivity
one and the Laplace noise scale is $2/\varepsilon$, so
Fact~\ref{fact:prelim-clipped-laplace} gives $1/(2/\varepsilon)
=\varepsilon/2$ privacy before clipping. The threshold test is
postprocessing. Let $P_x$ and $\widetilde P_x$ be the decision
distributions on input $x$ with unclipped and clipped noise,
respectively. Coupling them with the same Laplace draw bounds the
probability of different decisions by
$\tau_{\rm clip}=e^{-\varepsilon\ell/2}$. For each decision event $E$
and neighboring inputs $x,x'$,
\[
 \begin{aligned}
 \widetilde P_x(E)
 &\le P_x(E)+\tau_{\rm clip}\\
 &\le e^{\varepsilon/2}P_{x'}(E)+\tau_{\rm clip}\\
 &\le e^{\varepsilon/2}\widetilde P_{x'}(E)
       +(1+e^{\varepsilon/2})\tau_{\rm clip}.
 \end{aligned}
\]
The clipped test therefore has additive privacy error at most
\[
 (1+e^{\varepsilon/2})\tau_{\rm clip}
 =\frac{\delta}{32}(1+e^{-\varepsilon/2})\le\delta/16.
\]
For the candidate, set $s=\sqrt{mt}+t$ and let $\Delta$ be the relative matrix
change in Lemma~\ref{lem:full-candidate-privacy}.
Stability and \eqref{eq:base-size} give
\[
 \Delta s\le\frac{16\lambda(\sqrt{mt}+t)}n
 \le\frac{16\varepsilon}{C_{\rm priv}}\le c_*
\]
for sufficiently large $C_{\rm priv}$, since $\varepsilon\le1$.
Since $\lambda\ge1$, \eqref{eq:base-size} gives
$n\ge C_{\rm priv}(k+m+d+t)$. Taking $C_{\rm priv}\ge32C_*$ both
ensures the comparison lemma's sample-size requirement and gives
$C_*\Delta s\le16C_*\varepsilon/C_{\rm priv}\le\varepsilon/2$.
Also,
$C_*(d+1)e^{-t}=C_*\delta/C_{\rm priv}\le\delta/16$.
Thus the candidate is $(\varepsilon/2,\delta/16)$-private on
this fixed neighboring pair, in both directions.
For each fixed test decision, the second stage draws from the
candidate distribution on the given input with independent randomness.
The candidate privacy bound therefore applies for each decision.
By Lemma~\ref{lem:prelim-privacy-rules}, the joint decision and
candidate are $(\varepsilon,\delta/8)$-private on this pair.
Returning the candidate on acceptance and $\bot$ on rejection
is postprocessing.
Together with the deterministic-rejection case above, this proves
the stronger $(\varepsilon,\delta/8)$ guarantee on every neighboring
pair, and hence the stated bound.
\end{proof}

The private sampler supplies the private branch. Applying an
invertible affine map to the input transforms accepted vectors by
the same map, and using the same Laplace draw preserves the test
decision. The correction uses this property to transfer the batch
distribution from public reference statistics to the input
statistics, including the rejection outcome.

\subsection{Exact correction of the batch distribution}
\label{sec:exact-correction}\label{sec:correction-principle}

The private sampler $\mathcal A$ downweights records when forming
the batch, and a single rejection affects all $m$ output positions.
Both changes can introduce dependence. To recover exact independence
on Gaussian input, we use a mixture. The private branch runs
$\mathcal A$ with high probability, and the correction branch
supplies the rest of the target distribution. Keeping the
correction branch rare limits its additive privacy cost.

For this mixture to reproduce the target, the target must allow
the rejection outcome. We represent rejection by leaving all $m$
positions blank. After exact resampling and after private acceptance,
we leave each position blank independently with probability $\eta$.
The target therefore assigns probability $\eta^m$ to the all-blank
batch. At the end, we fill each blank independently from $R$.

We draw a proposal and decide whether to keep it using only public
randomness. This keeps the overall branch probabilities fixed
across input datasets. To make this decision without knowing the
input distribution, we work at reference statistics with empirical
mean zero and scatter $I_d$. Conditional Gaussian resampling and
affine equivariance let us transfer the construction from these
reference statistics to the empirical statistics of the input.
If the selector keeps the proposal, we shift and rescale its
retained vectors using the input statistics. If it selects the
proposal for replacement, we discard it and run $\mathcal A$ on
the input. Algorithm~\ref{alg:complete} gives the full mechanism.

\begin{table}[!htbp]
\centering
\small
\caption{Notation for the exact correction.}
\label{tab:correction-notation}
\begin{tabularx}{\linewidth}{@{}lX@{}}
\toprule
Notation & Meaning \\
\midrule
\(x\) & Observed input dataset. \\
\(T(x)\) & Matrix encoding the empirical mean and scatter of \(x\). \\
\(T_0\) & Public reference statistic for mean zero and scatter \(I_d\). \\
\addlinespace
\(\star\) & Blank symbol. \\
\(J\) & Positions that contain vectors. \\
\(z\) & Matrix whose rows are the retained vectors in increasing position order. \\
\((J,z)\) & Partial batch of \(m\) positions with the rows of \(z\) at \(J\) and \(\star\) elsewhere. \\
\addlinespace
\(F_T\) & Target distribution from exact resampling at \(T\) with independent omissions. \\
\(G_T\) & Private branch's partial-batch distribution on Gaussian input conditional on \(T\), including rejection. \\
\(H_T\) & Distribution of a kept proposal after transport to \(T\). \\
\(w\) & Input-independent mixture weight of the private branch in \(F_T=wG_T+(1-w)H_T\). \\
\addlinespace
\(\zeta\) & Public proposal sampled from \(F_{T_0}\). \\
\(A\) & Deletion set guessed before sampling \(U\). \\
\(U\) & Separate simulated reference dataset sampled using the reweighting for \(A\). \\
\(V\) & Independent Laplace noise for the simulated stability test. \\
\(A_*(U,V)\) & Deletion set recorded by the simulated run. It must match \(A\) for replacement. \\
\(\pi(A)\) & Probability that the selector guesses \(A\). \\
\(r_A(U)\) & Density at \(U\) of the reweighted reference distribution relative to the uniform one. \\
\(h_{U,V}(\zeta)\) & Ratio of the simulated partial-batch density to the proposal density at \(\zeta\). \\
\(\mathsf{Priv}\) & Branch decision. \(1\) replaces the proposal with a fresh private call on \(x\). \(0\) keeps and transports it. \\
\bottomrule
\end{tabularx}
\end{table}

\paragraph{Proof roadmap.}
\emph{Partial batches and the target distribution} formalizes the
mixture. \emph{Replacing selected proposals} shows which proposals
must be selected for replacement so that the two branches recover
the target distribution.

\emph{The branch probability for a given proposal} identifies the
required probability and explains why a single reference simulation
can produce a score above one. We then use determinant bounds to
reweight the simulation, explain why the deletion set is guessed
before drawing the reference dataset, and give a direct sampler
for the reweighted distribution. The final two paragraphs show
that the selector uses valid probabilities and that the selected
proposals have the required distribution.

\paragraph{Partial batches and the target distribution.}
A \emph{partial batch} \((J,z)\) records the retained positions
\(J\subseteq[m]\) and their vectors \(z\in\R^{|J|\times d}\) in
increasing position order. The other positions contain
\(\star\notin\R^d\). For example, the batch \((y_1,\star,y_3)\)
has \(J=\{1,3\}\), and the rows of \(z\) are \(y_1^\top\) and
\(y_3^\top\), in that order.
The private branch applies the same independent omissions after
acceptance and uses \(J=\varnothing\) after rejection.
We index the distributions from Step~3 by the unweighted statistic
\(T\), which encodes \(\tau=(\bar x,S)\) for fixed \(n\).
Let \(F_T\) denote the distribution of exact resampling with
omissions at statistic \(T\). Let \(G_T\) be the distribution obtained
by running \(\mathcal A\) on Gaussian input conditional on \(T\) and
applying the omission and rejection rule. Let \(H_T\) be the distribution
of the transformed proposal conditional on choosing the correction
branch. We seek
a selector whose overall probability \(w\) of taking the private
branch is close to one, and for which the two branches give
\begin{equation}\label{eq:partial-batch-mixture}
 F_T=wG_T+(1-w)H_T.
\end{equation}
This identity requires \(wG_T(E)\le F_T(E)\) for every partial-batch
event \(E\), leaving the mass \(F_T-wG_T\) for the correction branch.
In particular, the weighted private probability of the all-blank batch
must fit within the target probability \(\eta^m\). The correction
branch may use the empirical mean and scatter, so
its probability \(1-w\) must fit within the additive privacy budget.

The selector makes one branch decision for the batch. Averaging
\(F_{T(X)}\) over the empirical statistics gives independent draws
from \(P\) with independent omissions. Filling those blanks
independently from \(R\) gives \(Q_P^{\otimes m}\).

\paragraph{Replacing selected proposals.}
We can prepare the correction without knowing \(P\). By
Lemma~\ref{lem:full-conditioning}, the conditional input at any
nonsingular \(T\) is an affine image of a uniform centered frame.
Since the private sampler is equivariant under this transformation,
we work at empirical mean zero and scatter \(I_d\), so the reference statistic is \(T_0=\operatorname{diag}(I_d,n)\). Centering
and rescaling a standard Gaussian matrix gives a uniform centered
frame with these statistics.

Draw the public proposal \(\zeta=(J,z)\) from \(F_{T_0}\) by
taking the first \(m\) rows of one such reference dataset and applying
the independent omissions. To transport a proposal to statistic
\(T\) with mean \(\bar x\) and scatter \(S\), keep \(J\) fixed and
transform its retained rows by
\[
 \Phi_T(J,z)=(J,\one_{|J|}\bar x^\top+zS^{1/2}).
\]
The resulting proposal \(\Phi_T(\zeta)\) has distribution \(F_T\).

The private reference distribution \(G_{T_0}\) is the distribution
of the partial batch obtained by applying \(\mathcal A\) and the
omission and rejection rule to a uniform centered reference dataset.

The next lemma shows why selecting proposals with the private
reference distribution at \(T_0\) suffices to recover the target
distribution at the input statistics.

\begin{lemma}[Correction by replacing proposals]
\label{lem:ub-correction-principle}
Let \((\zeta,\mathsf{Priv})\) be drawn independently of the input,
with \(\zeta\sim F_{T_0}\) and \(\mathsf{Priv}\in\{0,1\}\).
Suppose that for a fixed \(w\in[0,1]\) and every measurable event \(E\),
\begin{equation}\label{eq:marked-target}
 \Prob\{\zeta\in E,\ \mathsf{Priv}=1\}=wG_{T_0}(E).
\end{equation}
Use this pair in the branch steps of Algorithm~\ref{alg:upper-summary}.
Then
\begin{enumerate}
\item \(\Prob\{\mathsf{Priv}=1\}=w\).
\item On Gaussian input, conditional on \(T(X)=T\), the batch
      before filling the blanks has distribution \(F_T\).
\end{enumerate}
\end{lemma}

\begin{proof}
Taking \(E\) to be the whole partial-batch space in
\eqref{eq:marked-target} gives \(\Prob\{\mathsf{Priv}=1\}=w\).
For Gaussian input, condition on \(T(X)=T\). The decision \(\mathsf{Priv}\) is independent
of the input, and the private call uses independent randomness, so
the private branch contributes \(wG_T(E)\) to the probability of an
event \(E\). By Lemma~\ref{lem:full-conditioning}, the conditional
dataset is obtained by shifting and rescaling the rows of a uniform
centered reference dataset. Affine equivariance of \(\mathcal A\)
and the omission rule then give
\(G_T(E)=G_{T_0}(\Phi_T^{-1}(E))\).

Let \(v\) be the partial output before filling the blanks. Apply
\eqref{eq:marked-target} to \(\Phi_T^{-1}(E)\) to identify the
contribution of the private branch with the transported proposals
that were selected for replacement.
\begin{equation}\label{eq:correction-identity}
 \begin{aligned}
 \Prob\{v\in E\mid T(X)=T\}
 &=wG_T(E)
   +\Prob\{\Phi_T(\zeta)\in E,\ \mathsf{Priv}=0\}\\
 &=\Prob\{\Phi_T(\zeta)\in E,\ \mathsf{Priv}=1\}
   +\Prob\{\Phi_T(\zeta)\in E,\ \mathsf{Priv}=0\}\\
 &=\Prob\{\Phi_T(\zeta)\in E\}.
 \end{aligned}
\end{equation}
For \(0<w<1\), \(H_T\) in \eqref{eq:partial-batch-mixture} is
the distribution of \(\Phi_T(\zeta)\) conditional on
\(\mathsf{Priv}=0\).
\end{proof}

\paragraph{The branch probability for a given proposal.}
\label{sec:mark-construction}
Fix the proposal \(\zeta\). To satisfy \eqref{eq:marked-target},
the probability of replacing \(\zeta\) must equal \(w\) times the
density of \(G_{T_0}\) relative to \(F_{T_0}\) at \(\zeta\).
The private reference distribution \(G_{T_0}\) averages over the
reference dataset and the stability-test noise. To express this
average, let \(U\) have the ordinary uniform reference distribution
on \(n\)-row datasets and let \(V\sim\operatorname{Lap}(2/\varepsilon)\).
In this average, \(U\) and \(V\) are independent of each other and
of \(\zeta\). The rows of \(U\) determine the weights and stability
score, and \(W=\operatorname{clip}_\ell(V)\) gives the noise used in
the test. In the selector below, the simulated dataset is drawn
from a reweighted reference distribution.

Given \(U,V\), we can evaluate the conditional output density at
the existing proposal. On acceptance, use the weighted statistic
\(B(U)\) and the omission rule. On rejection, the conditional
output is the all-blank batch. We compute the density ratio
relative to \(F_{T_0}\) because its average over ordinary reference
simulations, multiplied by \(w\), gives the required replacement
probability for the proposal. The ratio is
\begin{equation}\label{eq:run-density}
 h_{U,V}(J,z)=
 \begin{cases}
 f_{B(U)}^{(|J|)}(z)/f_{T_0}^{(|J|)}(z),&\text{on acceptance},\\
 \eta^{-m}\one_{\{J=\varnothing\}},&\text{on rejection}.
 \end{cases}
\end{equation}
The first case uses \(f_B^{(0)}=1\). For a retained set of size \(r\),
the same factor \((1-\eta)^r\eta^{m-r}\) occurs in both distributions,
so it cancels in the ratio. On acceptance, \(B(U)\preceq T_0\) by
Lemma~\ref{lem:full-stable}, so the support condition in
Lemma~\ref{lem:full-density} puts all accepted candidates in the
proposal support. On rejection, the all-blank batch has mass one in
the conditional distribution and mass \(\eta^m\) under \(F_{T_0}\).
Density ratios may be set to zero outside the proposal support.

For each fixed proposal, averaging over the uniform reference
frame and independent Laplace draw gives the density ratio of
\(G_{T_0}\) to \(F_{T_0}\). Hence for every event \(E\),
\[
 G_{T_0}(E)
 =\E_{\zeta}\!\left[\one_{\{\zeta\in E\}}
       \E_{U,V}[h_{U,V}(\zeta)]\right].
\]
Thus the required probability at \(\zeta\) is
\(w\E_{U,V}[h_{U,V}(\zeta)]\). The score from a single ordinary
simulation, \(w h_{U,V}(\zeta)\), has this expectation. If the
simulated test rejects and the proposal is all blank, the score
is \(w\eta^{-m}\), which exceeds one when \(w>\eta^m\).
Capping the score at one would change its expectation.

{\centering\itshape
How can we turn \(w h_{U,V}(\zeta)\) into a valid replacement
probability while preserving its expectation?\par}

For a fixed proposal \(\zeta\), we use \emph{importance sampling}
to change how we draw the separate reference dataset \(U\).
The stability-test noise keeps its original distribution.
If we double the sampling density at a dataset and halve its
score, its contribution to the average stays the same. In general,
dividing the score by the ratio of the new density to the original
one preserves the required expectation. We need to choose a density
that reduces the large scores enough to form valid probabilities.

\paragraph{Choosing the reweighting.}
\label{sec:ub-det-proof}

We base the new density on how the determinant changes when we
delete records from the simulated reference dataset \(U\). The set
\(J\) indexes the retained positions of the separate proposal
\(\zeta\). Since we draw \(U\) independently of \(\zeta\), the
reweighting must control the score for every proposal.
For a dataset \(x\) with scatter \(S\), let \(A\) be a set of
\(a\) row indices. Measure the decrease in the determinant of the
unweighted statistic when we delete those rows by
\[
 D_A=\frac{\det T}{\det T_{A^c}}
     =\frac{n\det S}{(n-a)\det S(x_{A^c})}
\]
whenever \(T_{A^c}\succ0\). The second equality follows from the
Schur complement. Each scatter uses the mean of its own records.
For a uniform centered reference dataset \(U\), these complements
are nonsingular almost surely for \(a\le k\) under the size
assumptions below.

Let \(A_*(U,V)\) be the set of reference records whose deletions
we record on a simulated run. On acceptance, let \(A_*=[n]\setminus I_k\) be the set of all records
deleted by pruning at threshold \(u_k\). Every record in \(I_k\)
has weight one in \(B(U)\). Records in \(A_*\) can still have
positive weight because the weights average survivor sets at
higher thresholds. Setting their weights to zero gives
\(T_{A_*^c}(U)\), and setting them to one gives \(T_0\). Thus
\(0\prec T_{A_*^c}(U)\preceq B(U)\preceq T_0\) and
\(n-|A_*|\le b\le n\). Part 1 of the determinant lemma below
uses these bounds to show that the weighted density at the proposal
is at most \(D_{A_*}^{|J|}\le D_{A_*}^m\) times the target density.

On rejection, the all-blank proposal has density ratio
\(\eta^{-m}\). Here the recorded deletions give a lower bound
on the determinant ratio, which we use to offset this factor.
The test guarantees \(g\ge a_0\), so pruning at the base threshold
\(\lambda\) leaves at most \(n-a_0\) survivors by
Lemma~\ref{lem:full-stable}. Appendix~\ref{app:stable-weight-details}
shows that the first \(a_0\) deletions occur before the statistic
can become singular, almost surely. We take \(A_*\) to consist
of those first \(a_0\) deleted indices.

In both cases, \(|A_*|<k\). Each recorded deletion occurs at
a threshold of at least \(\lambda\). By part 2 of the determinant
lemma, each such deletion reduces the determinant by a factor
of at least \(e^{\lambda/n}\). The ratio of the original determinant
to the determinant after deletion therefore satisfies
\(D_{A_*}\ge e^{|A_*|\lambda/n}\).

For a fixed set \(A\), make simulations with larger \(D_A(U)\) more
likely by sampling \(U\) with density \(r_A\) relative to the
ordinary uniform reference distribution. Set
\[
 q=\lfloor n/3\rfloor,\qquad
 Z_a=\E_{\rm uniform}[D_A(U)^q],\qquad
 r_A(U)=D_A(U)^q/Z_a.
\]

Exchangeability makes \(Z_a\) depend only on \(a=|A|\), and
\(D_\varnothing=Z_0=1\). When \(A=A_*\), dividing by \(r_A(U)\)
multiplies the score by \(Z_a D_A^{-q}\). On acceptance, this reduces the
determinant factor in the bound from \(D_A^m\) to \(D_A^{m-q}\).
The parameters give \(q>m\), so a larger determinant ratio makes
this remaining factor smaller. On rejection, the \(a_0\) recorded
deletions give \(D_A^{-q}\le e^{-qa_0\lambda/n}\), which helps
offset the all-blank factor \(\eta^{-m}\). The choice of \(q\)
also leaves enough rows in the smaller frame used to sample
from \(r_A\).
\paragraph{Why we guess the deleted set.}

The determinant bounds apply to the set \(A_*(U,V)\) recorded
by the test, so we need the set used for reweighting to match it.
To draw \(U\) from \(r_A\), we must choose a projection using the
indices in \(A\). The test supplies \(A_*(U,V)\) only after \(U\)
and \(V\) have been drawn. We therefore draw a guess \(A\)
independently of the proposal and check it after the test. If
\(|A|<k\), we draw \(U\) using \(r_A\), draw \(V\), and compute
\(A_*(U,V)\). We keep the original proposal if \(|A|\ge k\),
if \(A_*(U,V)\) is undefined, or if \(A_*(U,V)\ne A\).
Write \(\pi(A)\) for the probability of drawing \(A\). On a match,
use the candidate replacement probability
\begin{equation}\label{eq:mark-design}
 \frac{w\,h_{U,V}(J,z)}{\pi(A)r_A(U)}.
\end{equation}

To see why both factors appear, fix one simulated pair \((U,V)\).
This pair can contribute through exactly one guess,
\(A=A_*(U,V)\). The guess contributes probability \(\pi(A)\), and
\(r_A(U)\) is the density of the reweighted pair relative to the
ordinary reference pair. The test noise has the same distribution
in both cases. Dividing by \(\pi(A)r_A(U)\) restores the contribution
\(w h_{U,V}(\zeta)\) to the average.
The proof sums over possible guesses. The selector draws one guess.
It remains to prove that on a match,
\[
 w h_{U,V}(\zeta)\le \pi(A)r_A(U).
\]

We include each index independently with probability
\(p=e^{-\lambda/8}/n\) and set \(w\) equal to the probability of
an empty guess. Thus
\[
 \pi(A)=p^{|A|}(1-p)^{n-|A|},\qquad w=\pi(\varnothing)=(1-p)^n .
\]
The small value of \(p\) keeps \(w\) close to one.
When \(A=A_*=\varnothing\), the test accepted, since rejection
records \(a_0\ge1\) deletions. All weights are one, \(B=T_0\),
and \(h_{U,V}=r_\varnothing=1\). This choice of \(w\) makes
\eqref{eq:mark-design} equal to one in this case. For a nonempty
matching set, the determinant growth must also cover the reciprocal
guess probability. The replacement identity below shows that
the overall probability of taking the private branch is \(w\).
\paragraph{Determinant bounds for the selector.}
Part 1 of the next lemma bounds the score on acceptance.
Part 2 supplies determinant growth on both test outcomes.
Part 3 controls \(Z_a\) and gives a direct sampler for \(r_A\).

\begin{lemma}[Determinant bounds for the correction]\label{lem:ub-det-tools}
Assume
\begin{equation}\label{eq:correction-size}
 \begin{gathered}
 n\ge256(m+d+1),\qquad n\ge16k\lambda,\\
 \lambda\ge16(d+2+\log n),\qquad
 a_0\lambda\ge16m\log(1/\eta).
 \end{gathered}
\end{equation}
Then the following hold.
\begin{enumerate}
\item If \(|A|=a\le k\), \(0\prec T_{A^c}\preceq B\preceq T\),
      and \(b\in[n-a,n]\), then
      \(f_B^{(r)}\le D_A^r f_T^{(r)}\) for \(0\le r\le m\).
\item If \(A\) consists of the first \(a\le k\) rows deleted at a
      threshold at least \(\lambda\), and the statistics before and
      after deletion are positive definite, then
      \(D_A\ge e^{a\lambda/n}\).
\item For a uniform centered frame and a fixed set \(A\) of \(a\le k\)
      rows, \(Z_a\le e^{2(d+1)a}\). The normalizing constant can be
      evaluated in \(O(qd)\) operations, and a frame with density
      \(D_A^q/Z_a\) can be sampled in \(O(n^3)\) operations.
\end{enumerate}
A frame drawn uniformly or with density \(r_A\) is in affine
general position almost surely.
\end{lemma}

\begin{proof}[Proof of parts 1 and 2 of Lemma~\ref{lem:ub-det-tools}]
\label{sec:det-density}
For part 1, we bound how much weighting can increase the density
of the $r$ retained output vectors. Since \(B\preceq T\), every
\(Y\) with \(f_B^{(r)}(Y)>0\) also satisfies \(f_T^{(r)}(Y)>0\).
On this support, Lemma~\ref{lem:full-density} gives
\[
 \frac{f_B^{(r)}(Y)}{f_T^{(r)}(Y)}
 =
 \frac{A_r(b)}{A_r(n)}
 \left(\frac{\det T}{\det B}\right)^{r/2}
 \left(\frac{R_B(Y)}{R_T(Y)}\right)^{\kappa_r}.
\]
The last factor is at most one because \(B^{-1}\succeq T^{-1}\)
gives \(R_B(Y)\le R_T(Y)\), and \(\kappa_r\ge0\).
For the middle factor, \(T_{A^c}\preceq B\) gives
\[
 \frac{\det T}{\det B}
 \le \frac{\det T}{\det T_{A^c}}=D_A.
\]
Thus the determinant ratio for deleting all records in \(A\)
bounds the change in this factor when those records are downweighted.
Combining the two estimates gives
\[
 \frac{f_B^{(r)}(Y)}{f_T^{(r)}(Y)}
 \le D_A^{r/2}\frac{A_r(b)}{A_r(n)}.
\]
The remaining factor \(A_r(b)/A_r(n)\) accounts for the change in
total weight from \(n\) to \(b\). To bound it, observe that, for any
direction \(u\), the scatter quadratic form of a subset is the
minimum over \(s\in\R\) of the sum of \((u^\top x_i-s)^2\)
over that subset. Removing observations cannot increase this
minimum, so \(S(x_{A^c})\preceq S\). Determinant monotonicity then
gives \(D_A\ge n/(n-a)\).
For \(b\in[n-a,n]\), the size assumptions give
\[
 \frac{d}{db}\log\bigl(b^{r/2}A_r(b)\bigr)
 =\frac{r(2b-n-r+2)}{2b(b-r)}\ge0.
\]
It follows that \(A_r(b)/A_r(n)\le(n/b)^{r/2}\le D_A^{r/2}\),
proving the first bound. For \(r=0\) the assertion follows from
\(f_B^{(0)}=f_T^{(0)}=1\).

\label{sec:det-growth}
For part 2, let \(v=\widetilde x_i\) and let \(H\) be the statistic just before record \(i\) is deleted. The deletion threshold is at least \(\lambda\), so \(v^\top H^{-1}v>\lambda/n\). The determinant
formula for a rank-one deletion gives
\[
 \frac{\det H}{\det(H-vv^\top)}
 =\frac1{1-v^\top H^{-1}v}\ge e^{\lambda/n}.
\]
The ratios telescope over the first \(a\) deletions, giving
\(D_A\ge e^{a\lambda/n}\).
\end{proof}
\paragraph{Sampling the reweighted reference dataset.}
Fix the guessed set \(A\). The determinant ratio depends only on
the component of the frame lost when we delete \(A\) and recenter
the survivors. We can therefore draw from \(r_A\) by sampling this
component under the reweighted distribution and then completing
the frame. For \(a>0\), let \(\mathcal H_A\) be the
\((n-a-1)\)-dimensional subspace of centered vectors supported
on \(A^c\). Let \(E_A\in\R^{n\times a}\) have orthonormal columns
spanning its orthogonal complement within \(\one_n^\perp\).
Write \(F_A=E_A^\top U\).

The projection onto \(\mathcal H_A\) replaces the rows in \(A\)
by zeros and subtracts the mean of the surviving rows from each
surviving row. The Gram matrix of this projection is therefore
the scatter of the surviving rows. Since \(U^\top U=I_d\) and
the removed component is \(E_AF_A\), that scatter is
\(I_d-F_A^\top F_A\).
The determinant expression for \(D_A\) therefore gives
\begin{equation}\label{eq:determinant-frame-block}
 D_A^{-1}=\frac{n-a}{n}\det(I_d-F_A^\top F_A).
\end{equation}
For a uniform centered \(U\), \(F_A\) has the distribution of the
first \(a\) rows of a uniform \((n-1)\times d\) frame by
Lemma~\ref{lem:prelim-frame-projection}. Multiplying its density
by \(D_A^q\) subtracts \(q\) from the exponent
\((n-a-d-2)/2\) of \(\det(I_d-F_A^\top F_A)\).
By Lemma~\ref{ub-bg:haar-block-density}, the resulting exponent
is that of a block of a uniform \((n-1-2q)\times d\) frame.
The size assumptions give \(n-1-2q\ge a+d\). Both densities have
support \(I_d-F_A^\top F_A\succ0\), so matching the exponent
identifies the normalized density.

Reweighting only uses \(F_A\), so the conditional distribution of
the projection onto \(\mathcal H_A\) given \(F_A\) is unchanged.
This projection has Gram matrix \(I_d-F_A^\top F_A\).
Right multiplication by \((I_d-F_A^\top F_A)^{-1/2}\) normalizes
this projection to a frame. Rotations within \(\mathcal H_A\) fix
\(F_A\) and preserve the conditional distribution of the normalized
projection. It therefore has the same uniform distribution in
\(\mathcal H_A\) for almost every \(F_A\). Thus draw \(F_A\) as the
first \(a\) rows of the smaller uniform frame and independently
draw a uniform \(d\)-column frame \(H_A\) in \(\mathcal H_A\).
Then set
\begin{equation}\label{eq:weighted-frame-recipe}
 U=H_A(I_d-F_A^\top F_A)^{1/2}+E_AF_A .
\end{equation}
For \(a=0\), draw a uniform centered frame. The two cases give
a direct sampler for \(r_A\). The uniform centered reference dataset
is in affine general position almost surely by
Appendix~\ref{app:stable-weight-details}. Absolute continuity
transfers this property to the reweighted draw. The evaluation and
bound of \(Z_a\) in part 3 are given in
Appendix~\ref{app:upper-calculations}, where
\eqref{eq:full-det-exact} is a product of \(O(qd)\) factors.
Generating the frames and forming their bases and matrix
products takes \(O(n^3)\) operations. This gives the draw needed
by the selector once it has chosen a set \(A\).

Algorithm~\ref{alg:main-selector} takes the existing proposal and
returns the decision \(\mathsf{Priv}\). The simulated test outcome on \(U\)
determines the recorded set and the formula for the density ratio.
For a matching guess, a simulated rejection can still select the
private branch when the proposal is empty. If \(\mathsf{Priv}=1\),
the mechanism makes an independent call to \(\mathcal A\) on the
observed dataset. On a match, substituting
\(\pi(A)=w(p/(1-p))^a\) and \(r_A(U)=D_A(U)^q/Z_a\) into
\eqref{eq:mark-design} gives the probability used by the selector.
\begin{equation}\label{eq:mark-probability}
 h_{U,V}(J,z)Z_a
 \left(\frac{1-p}{p}\right)^aD_A(U)^{-q}.
\end{equation}

\begin{algorithm}[H]
\caption{Choose whether to keep the proposal}
\label{alg:main-selector}
\begin{algorithmic}[1]
\Require Existing proposal \(\zeta=(J,z)\sim F_{T_0}\) and public parameters.
\Ensure A decision \(\mathsf{Priv}\in\{0,1\}\), with \(1\) to replace the
        proposal and \(0\) to keep it.
\State Form \(A\) by including each index of \([n]\) independently with
       probability \(p\), using randomness independent of \(\zeta\).
\If{\(|A|\ge k\)}
    \State \Return \(0\).
\EndIf
\State Set \(a\gets|A|\). Draw \(U\) with density \(r_A\) by
       \eqref{eq:weighted-frame-recipe} if \(a>0\), or draw a uniform
       centered frame if \(a=0\).
\State Draw \(V\sim\operatorname{Lap}(2/\varepsilon)\) independently of
       \(\zeta,A,U\).
\State Compute the weights and score \(g\) on \(U\). Retain \(I_k\)
       and the ordered deletion list at the base threshold \(\lambda\).
\If{\(g+\operatorname{clip}_\ell(V)\ge a_0+\ell\)}
    \State Record a simulated rejection. Set \(A_*(U,V)\) to the
           first \(a_0\) deleted indices at threshold \(\lambda\),
           if available.
\Else
    \State Record a simulated acceptance and set \(A_*(U,V)=[n]\setminus I_k\).
\EndIf
\If{\(A_*(U,V)\) is undefined or \(A\ne A_*(U,V)\)}
    \State \Return \(0\).
\EndIf
\State Evaluate \(h_{U,V}(J,z)\) at the proposal by \eqref{eq:run-density}.
\If{the density ratio is undefined}
    \State \Return \(0\).
\EndIf
\State \Return \(1\) with probability \eqref{eq:mark-probability}, and \(0\) otherwise.
\end{algorithmic}
\end{algorithm}

Algorithm~\ref{alg:public} returns the proposal and the selector's
decision \(\mathsf{Priv}\).

\begin{algorithm}[H]
\caption{Public preparation}
\label{alg:public}
\begin{algorithmic}[1]
\Require Public parameters.
\Ensure A proposal $\zeta$ and a decision $\mathsf{Priv}\in\{0,1\}$.
\State Draw $\zeta=(J,z)\sim F_{T_0}$.
\State Apply Algorithm~\ref{alg:main-selector} to $\zeta$ to obtain
       $\mathsf{Priv}$.
\State \Return $(\zeta,\mathsf{Priv})$.
\end{algorithmic}
\end{algorithm}

The selector uses the original proposal and at most one separate
simulated dataset. We now verify its two guarantees.

\paragraph{Validity of the replacement probability.}
On a match, \eqref{eq:mark-probability} has a factor
\(Z_a((1-p)/p)^a\) that depends only on the size of the guess.
The bound on \(Z_a\) and the choice of \(p\) give
\[
 Z_a\left(\frac{1-p}{p}\right)^a
 \le\exp\bigl(a(2d+2+\log n+\lambda/8)\bigr)
 \le e^{a\lambda/4}.
\]
On a match after simulated acceptance, the retained-row ratio in
\eqref{eq:run-density} is at most \(D_A^{|J|}\le D_A^m\), by
part 1 of Lemma~\ref{lem:ub-det-tools}. Part 2 gives
\(D_A\ge e^{a\lambda/n}\). Since \(q=\lfloor n/3\rfloor\) and
\(n\ge256(m+d+1)\), we have \(q-m\ge n/4\) and \(q\ge5n/16\).
After compensation, the remaining determinant factor \(D_A^{m-q}\)
offsets the cost of the guess and normalizer. The probability in
\eqref{eq:mark-probability} is at most
\[
 e^{a\lambda/4}D_A^{m-q}\le1.
\]
On a match after simulated rejection, \(a=a_0\), and the ratio is zero
unless \(J=\varnothing\). The \(a_0\) deletions guarantee enough
determinant growth to absorb the empty-proposal ratio \(\eta^{-m}\).
For \(J=\varnothing\), the probability is at most
\[
 \eta^{-m}e^{a_0\lambda/4}D_A^{-q}
 \le\eta^{-m}e^{-a_0\lambda/16}\le1.
\]
The last inequality uses \(a_0\lambda\ge16m\log(1/\eta)\).
These bounds cover both possible test outcomes on a match.
A nonmatching guess gives \(\mathsf{Priv}=0\).

\paragraph{Distribution of proposals selected for replacement.}
It remains to check the distribution of the proposals selected for replacement. Let \(\nu\) be the joint distribution before reweighting,
with a uniform centered frame \(U\) and independent Laplace noise \(V\).
Given a guess \(A\) with \(|A|<k\), the simulated pair has density
\(r_A(U)\) relative to \(\nu\), because the independent noise draw
has the same distribution in both experiments. For a fixed proposal
and each such guess, multiply
the probability of the guess and the density of the simulated
frame by the replacement probability to obtain
\begin{equation}\label{eq:mark-cancellation}
 \pi(A)r_A(U)\one_{\{A=A_*(U,V)\}}
 \frac{w h_{U,V}(\zeta)}{\pi(A)r_A(U)}
 =w h_{U,V}(\zeta)\one_{\{A=A_*(U,V)\}}.
\end{equation}
The sampling factors cancel. For \(\nu\)-almost every \((U,V)\),
the recorded set \(A_*(U,V)\) has fewer than \(k\) elements, so its
term is the only nonzero match indicator in the sum over guesses.
Summing and integrating with respect to \(\nu\) gives
\begin{equation}\label{eq:auxiliary-output-law}
 \begin{aligned}
 &\Prob\{\zeta\in E,\ \mathsf{Priv}=1\}\\
 &\quad=w\E_{\zeta}\!\left[\one_{\{\zeta\in E\}}
       \int h_{U,V}(\zeta)
       \left(\sum_{\substack{A\subseteq[n]\\|A|<k}}
        \one_{\{A=A_*(U,V)\}}\right)d\nu(U,V)\right]\\
 &\quad=wG_{T_0}(E).
 \end{aligned}
\end{equation}
This is \eqref{eq:marked-target}. The sum over guesses is only
used in the proof. Taking \(E\) to be the whole space gives
\(\Prob\{\mathsf{Priv}=1\}=w\), so the probability of using
the correction branch is bounded by
\begin{equation}\label{eq:correction-privacy-cost}
 1-w=1-(1-p)^n\le np=e^{-\lambda/8}.
\end{equation}
On Gaussian input, the replacement identity gives the conditional
partial-batch distribution \(F_T\) at the input statistics.
The bound on \(1-w\) will control the additive privacy cost of the
correction branch.

\subsection{The complete mechanism and its guarantees}
\label{sec:complete-sampler}\label{sec:upper-proof}
\label{sec:upper-at-glance}\label{sec:complete-algorithm}

Algorithm~\ref{alg:complete} puts the public preparation and the
input-dependent steps in execution order. The preparation chooses
the branch before accessing the input. In the private branch, the
mechanism draws a candidate only if the stability test accepts.
Rejection leaves all positions blank. In either branch, the final
step fills the blanks independently.

\begin{algorithm}[H]
\caption{Mechanism $M$}
\label{alg:upper-overview}\label{alg:complete}
\begin{algorithmic}[1]
\Require Dataset $x=(x_1,\ldots,x_n)$ and public parameters
         chosen in Section~\ref{sec:upper-calibration}.
\Ensure An ordered batch of $m$ vectors.

\State Run Algorithm~\ref{alg:public} once, obtaining
       $\zeta$ and $\mathsf{Priv}\in\{0,1\}$.
\State Write $\zeta=(\zeta_1,\ldots,\zeta_m)$, using $\star$ for blank positions.

\If{$\mathsf{Priv}=1$} \Comment{Private branch}
    \State Discard $\zeta$.
    \State Draw $Y\gets\mathcal A(x)$ using independent randomness.
           \Comment{Algorithm~\ref{alg:upper-base}}

    \If{$Y=\bot$}
        \State Set $v_1=\cdots=v_m=\star$.
               \Comment{The entire batch was rejected}
    \Else
        \State Write $Y=(y_1,\ldots,y_m)$.
        \State Independently for each $j\in[m]$, set $v_j=y_j$
               with probability $1-\eta$, and $v_j=\star$
               otherwise, using independent randomness.
    \EndIf

\Else \Comment{$\mathsf{Priv}=0$: Correction branch}
    \State Compute
           $\bar x=n^{-1}\sum_i x_i$ and
           $S=\sum_i(x_i-\bar x)(x_i-\bar x)^T$.
    \If{$S\succ0$}
        \State For every $j\in[m]$, set $v_j=\star$
               if $\zeta_j=\star$. Otherwise set
               $v_j=\bar x+S^{1/2}\zeta_j$.
    \Else
        \State Set $v_1=\cdots=v_m=\star$.
    \EndIf
\EndIf

\State Fill every blank position with an independent
       draw from $R=\mathcal N(0,I_d)$.
\State \Return $(v_1,\ldots,v_m)$.
\end{algorithmic}
\end{algorithm}

The guarantees below use the choices of \(\ell,t,k\) in
Lemma~\ref{lem:full-base-privacy}, the size conditions
\eqref{eq:base-size} and \eqref{eq:correction-size}, and
\(e^{-\lambda/8}\le\delta/4\). Appendix~\ref{app:upper-calculations}
verifies that the parameter choice satisfies these conditions
simultaneously.

\paragraph{Privacy on every dataset.}
Fix an input \(x\). By \eqref{eq:auxiliary-output-law}, the mechanism
takes the private branch with probability \(w\) independent of \(x\).
This branch runs Algorithm~\ref{alg:base} with randomness independent
of the branch choice, then applies the omission and rejection rules
as postprocessing. It therefore has privacy parameters
\((\varepsilon,\delta/4)\) by Lemma~\ref{lem:full-base-privacy}.
The correction branch can depend on the empirical mean and scatter. By
Fact~\ref{fact:prelim-private-mixture}, its probability \(1-w\)
adds at most that amount to the additive privacy parameter.
The partial output therefore has multiplicative parameter
\(\varepsilon\) and additive parameter bounded by
\[
 w\frac{\delta}{4}+(1-w)
 \le\frac{\delta}{4}+e^{-\lambda/8}
 \le\frac{\delta}{2}\le\delta.
\]
Filling the blanks is postprocessing, so \(M\) is
\((\varepsilon,\delta)\)-private.

\paragraph{Exact distribution and accuracy.}
Let \(X\sim P^{\otimes n}\), so \(T(X)\succ0\) almost surely.
By the replacement identity and Lemma~\ref{lem:ub-correction-principle},
conditional on \(T(X)\) the batch before filling is distributed as
exact resampling with independent omissions of probability \(\eta\).
Averaging over \(T(X)\) using \eqref{eq:exact-batch} and filling
the blanks independently from \(R\) gives
\[
 \Law(M(X))=\bigl((1-\eta)P+\eta R\bigr)^{\otimes m}
 =Q_P^{\otimes m}.
\]

To bound the distance to \(P^{\otimes m}\), draw one batch from
\(P^{\otimes m}\) and construct its coupled \(Q_P^{\otimes m}\)
batch by replacing each position independently with probability
\(\eta\) using a draw from \(R\). The batches agree whenever no
replacement is made, so the coupling inequality gives
\[
 \TV(Q_P^{\otimes m},P^{\otimes m})
 \le1-(1-\eta)^m\le m\eta=\alpha/2.
\]

\paragraph{Sample size.}
Two requirements determine the leading terms in the sample bound.
The privacy estimate contributes a term of order
\(\lambda(\sqrt{mt}+t)/\varepsilon\).
For the selector on a rejected simulation, the bound
\(a_0\lambda\ge16m\log(1/\eta)\), together with \(n\ge16k\lambda\)
and \(k>a_0\), accounts for the term linear in \(m\).
Appendix~\ref{app:upper-calculations} chooses
\(\lambda=\Theta(d+L)\), \(t=O(L)\), and the remaining parameters
in \eqref{eq:parameter-choice}. Substitution gives
\eqref{eq:general-iid-positive}.

\paragraph{Running time.}\label{sec:ub-runtime}
The selector uses the direct frame construction in
Section~\ref{sec:exact-correction} to draw at most one dataset with
density \(r_A\). Computing the weights requires pruning on that
dataset and, if the private branch is chosen, on the actual input.
We bound the cost of one pruning run and then sum over thresholds.

Write $D=d+1$. At one threshold, forming the statistic and, when
it is nonsingular, its inverse and all row scores costs
$O(nD^2+D^3)$. If pruning deletes record $i$ from a set with statistic $H\succ0$, set $v=\widetilde x_i$, $a=v^TH^{-1}v$, and $u=H^{-1}v$.
For $a<1$,
\[
 (H-vv^T)^{-1}=H^{-1}+\frac{uu^T}{1-a}.
\]
For each remaining record $j$, set $z=\widetilde x_j$. The updated quadratic form is
$z^\top(H-vv^\top)^{-1}z=z^\top H^{-1}z+(z^\top u)^2/(1-a)$. Updating the inverse and scores therefore costs
$O(D^2+nD)$ per deletion. A singular statistic, including a deletion
with $a=1$, ends that threshold's pruning run. At most $n$ deletions give $O(n^2D)$
operations per threshold.

There are $2k+1$ thresholds. Since $k\lambda\le n/16$ and
$\lambda\ge D$, pruning costs $O(kn^2D)=O(n^3)$. It is run at
most twice, once on the simulated dataset and once in the private call. All other
matrix operations and frame sampling cost $O(n^3)$. The frame density normalizing constant
cancels in \eqref{eq:run-density}; only the normalizer $Z_{|A|}$ corresponding to the guessed set must be
evaluated, in $O(qd)$ operations. The total running time is \(O(n^3)\).

This completes the proof of Theorem~\ref{thm:full-main} for
\(\varepsilon>0\).

\subsubsection{The full privacy range}\label{sec:ub-range}

Two constructions extend the bound to the full privacy range.
The first releases uniformly selected distinct input rows. The
second applies Theorem~\ref{thm:full-main} at a positive privacy
parameter and absorbs any extra multiplicative loss into $\delta$.
Taking the smaller sample bound gives the upper bound matching
Theorem~\ref{thm:lb}.

\begin{corollary}\label{cor:iid-full-range}
For integers $d,m\ge1$, $0<\alpha,\delta<1$, and $0\le\varepsilon\le1$,
there is an $(\varepsilon,\delta)$-DP mechanism using
\[
 \widetilde O\left(m+\min\left\{
 \frac{d\sqrt m}{\varepsilon+\delta},\frac m\delta
 \right\}\right)
\]
input samples and polynomially many operations. For every nonsingular
Gaussian $P$, its output distribution under $P^{\otimes n}$ is exactly
$Q_P^{\otimes m}$, with $Q_P$ as in Theorem~\ref{thm:full-main}.
Moreover, $\TV(Q_P^{\otimes m},P^{\otimes m})\le\alpha/2$.
\end{corollary}

\begin{proof}
Use $R=\mathcal N(0,I_d)$ and $\eta=\alpha/(2m)$ as in
Theorem~\ref{thm:full-main}.

\emph{The $m/\delta$ bound.}
Take $n=\lceil m/\delta\rceil$ and select $m$ distinct input indices
uniformly in random order. Independently replace each selected record
with a draw from $R$ with probability $\eta$. On neighboring inputs,
use the same indices and replacement randomness. The outputs can
differ only if the changed index is selected, which happens with
probability $m/n\le\delta$. This gives $(0,\delta)$-privacy.

For $X\sim P^{\otimes n}$, conditional on the chosen indices, the
selected records have distribution $P^{\otimes m}$ because the
indices are distinct and chosen independently of $X$. The independent
replacements therefore give $Q_P^{\otimes m}$.

\emph{The $d\sqrt m/(\varepsilon+\delta)$ term.}
To apply Theorem~\ref{thm:full-main} even when $\varepsilon=0$,
use the positive parameter $\varepsilon'=(\varepsilon+\delta)/2$.
Set $\delta'=\delta/2$, reserving the other half of the additive
budget for any excess of $\varepsilon'$ over $\varepsilon$.
The theorem gives the required output distribution $Q_P^{\otimes m}$.

Let $h=(\varepsilon'-\varepsilon)_+$. Then
$h\le\delta/2$ and $\varepsilon'\le\varepsilon+h$.
For any ordered pair of neighboring inputs, let $P_0,P_1$ be
the corresponding output distributions. For every measurable
event $E$, the theorem gives
$P_0(E)\le e^{\varepsilon+h}P_1(E)+\delta/2$.
Multiplying by $e^{-h}$ and adding $(1-e^{-h})P_0(E)$ gives
\[
 \begin{aligned}
 P_0(E)
 &\le e^\varepsilon P_1(E)
       +e^{-h}\delta/2+(1-e^{-h})P_0(E)\\
 &\le e^\varepsilon P_1(E)+\delta/2+h\\
 &\le e^\varepsilon P_1(E)+\delta,
 \end{aligned}
\]
where the second inequality uses $1-e^{-h}\le h$.
Since the ordered pair was arbitrary, this proves
$(\varepsilon,\delta)$-privacy.

In the theorem's sample bound, $1/\varepsilon'=2/(\varepsilon+\delta)$
and $\varepsilon'\delta'\ge\delta^2/4$. The logarithm is therefore
at most $1+\log(4dm/(\alpha\delta^2))$, so this construction uses
$\widetilde O(m+d\sqrt m/(\varepsilon+\delta))$ input samples.

Choose the construction with the smaller sample bound. The total
variation guarantee follows from the bound on $Q_P^{\otimes m}$
in Theorem~\ref{thm:full-main}.
\end{proof}

\section{Lower bounds for jointly accurate sampling}
\label{sec:lower-bound}\label{main:lower}

Ghazi et al.~\cite{GHKM23} derive single-sampling lower bounds
by constructing private estimators from repeated sampler calls.
For identity covariance, they run the sampler a constant number
of times on disjoint input blocks. Coordinatewise medians of the
independent outputs, clipped to the parameter range, give a mean
estimate with expected squared \(\ell_2\) error \(O(d)\). A
private mean-estimation lower bound then bounds the sampler's
sample complexity. For bounded unknown covariance, they generate
\(O(d^2)\) outputs for a Gaussian learner. To share input records
across these calls, they sample random subsets of a larger dataset
and account for privacy using subsampling and advanced composition.

Cheu and Nayak~\cite{CheuNayak2025} give a lower-bound argument for
weak multi-sampling of Gaussians with bounded mean and covariance
under zero-concentrated differential privacy. Weak multi-sampling
requires independent outputs from a common distribution close to the
target in total variation. Their argument uses \(d+1\) generated
outputs as public observations for a Gaussian learner that also receives
private observations. The learner's guarantee assumes that the public
observations come from a Gaussian close to the target in total variation.
With such outputs, an overly sample-efficient sampler would yield a
learner that violates a private Gaussian-learning lower bound.

Our proof asks what the batch itself reveals about the covariance.
We split each vector into two blocks and normalize their empirical
cross-moment to obtain a matrix direction \(T\). We measure its
\emph{alignment} with the unknown cross-covariance \(B\) by the
Frobenius inner product \(\langle B,T\rangle\). We choose a Gaussian
family for which this inner product is large and positive with high
probability over the Gaussian prior on \(B\) and an \emph{ideal batch}
of \(m\) independent draws from the target Gaussian. Joint total
variation accuracy transfers this alignment event to the released
batch, whose vectors may be dependent.

We prove the quantitative form of Theorem~\ref{main:intro:lower}.

\begin{theorem}\label{thm:lb}\label{main:lb:theorem}
There is a universal constant \(c>0\) with the following property.
Let \(d,m,n\ge1\) be integers, \(0\le\alpha<1\),
\(0\le\varepsilon\le1\), and \(0<\delta<1\). Suppose that a
measurable randomized mechanism
\(M:(\R^d)^n\to(\R^d)^m\) is
\((\varepsilon,\delta)\)-differentially private and satisfies
\[
\TV\bigl(\Law(M(X)),P^{\otimes m}\bigr)\le\alpha
\]
whenever \(X\sim P^{\otimes n}\) for a Gaussian distribution
\(P=\mathcal N(\mu,\Sigma)\) with \(\Sigma\succ0\). Define
$
\beta=1-\alpha,\,\,
 a_\beta=\sqrt{2\log(8/\beta)},\,\,
 L_\delta=\log(128/\delta).
$
Then
\begin{equation}\label{eq:lb:theorem}
n\ge c\beta^3\left[
m+\min\left\{
\frac{d\sqrt m}{a_\beta(\varepsilon+\delta L_\delta)},
\frac{m}{\delta L_\delta}
\right\}\right].
\end{equation}
In addition, \(n\ge\beta^3m/8\). The released vectors may be dependent,
and no restriction relating \(\delta\) to \(n\) is required.
\end{theorem}

Fix a mechanism satisfying the theorem's assumptions. Throughout the
proof, \(c,C>0\) denote universal constants whose values may change.
The dimension-dependent argument uses a Gaussian family with two blocks
and an unknown cross-covariance. Section~\ref{sec:batch-alignment}
chooses the noise level so that the direction from an ideal batch
has a large positive inner product with \(B\) even though each record
carries little information about \(B\). We weight the direction from the
released batch to make its inner product with \(B\) nonnegative
while preserving the direction on the success event.
Each record score is the gradient of its log-density with respect to
the cross-covariance. Gaussian integration by parts then lower-bounds
the \emph{total record agreement}, the sum of expected inner products
of the weighted direction with the scores scaled by the noise standard
deviation.

Section~\ref{sec:lower-privacy} bounds the same sum using privacy.
Conditional on the Gaussian parameter, replacing one record
makes the new output direction independent of the removed
record's centered score. This gives a zero-mean comparison.
The additive \(\delta\) term bounds the excess density mass in each
direction of the privacy comparison. This mass can carry large
agreements. We control their combined contribution by summing matrices
formed from the Gaussian record scores with data-dependent coefficients.
The bound adds terms of order \(d\sqrt{n\delta L_\delta}\) and
\(n\delta L_\delta\). The first
yields the \(m/(\delta L_\delta)\) branch in \eqref{eq:lb:theorem}.

The matrix argument uses a dimension cutoff, which we choose when
transferring ideal alignment to the released batch.
Section~\ref{sec:scalar-lower-bound} uses one unknown mean coordinate
to give the privacy-dependent bound below that cutoff and the
linear cost in \(m\) in every dimension.

\subsection{Accuracy forces agreement with input records}
\label{sec:batch-alignment}
We first construct a Gaussian family whose cross-covariance is
the unknown matrix. We extract a direction from the batch and
choose the noise level so that, for an ideal batch of independent
draws, this direction aligns with the unknown matrix. Joint
accuracy transfers the alignment event to the released batch.
We then use a Gaussian score identity to turn this alignment
into a lower bound on agreement with the input records.
Section~\ref{sec:lower-privacy} uses privacy to bound the same
agreement from above.

We build the family so that the matrix we want the batch to reveal
appears as a cross-covariance. For the matrix construction take
\(d\ge2\), set \(k=\lfloor d/2\rfloor\), and split the first
\(2k\) coordinates into two blocks of size \(k\). We generate the
second block by applying an unknown matrix \(B\in\R^{k\times k}\)
to the first block and adding Gaussian noise. With noise scale
\(\sigma>0\), one record is generated by
\begin{equation}\label{eq:lb:model}
z\sim\mathcal N(0,I_k),\qquad
w=Bz+\sigma g,\qquad g\sim\mathcal N(0,I_k),
\end{equation}
where \(z,g\) are independent. Thus the cross-covariance is
\(\E[wz^\top\mid B]=B\). In odd dimension, append an independent standard Gaussian coordinate.
Write \(P_B\) for the resulting Gaussian distribution. Its covariance on the two blocks is
\[
\begin{pmatrix}I_k&B^\top\\ B&BB^\top+\sigma^2I_k\end{pmatrix}.
\]
Its quadratic form at \((u,v)\) is
\(\|u+B^\top v\|^2+\sigma^2\|v\|^2\), which is strictly positive when
\((u,v)\ne0\). The mechanism must therefore be accurate for every
\(B\). For the analysis, draw the entries of \(B\) independently from
\(\mathcal N(0,1)\). Conditional on \(B\), draw the input
\(X=(X_1,\ldots,X_n)\sim P_B^{\otimes n}\). The mechanism sees only
\(X\). Probabilities and expectations in the matrix argument include
the prior on \(B\), the sample, and mechanism randomness, except
where we condition explicitly.

\paragraph{Extracting the direction.}
The released batch gives an empirical counterpart to
\(\E[wz^\top\mid B]\). Put the first \(k\) and next \(k\) coordinates
of released vector \(j\) in column \(j\) of \(Z\) and \(W\),
respectively, with \(Z,W\in\R^{k\times m}\). Normalize their
cross-moment to obtain a direction
\begin{equation}\label{eq:lb:statistic}
A=WZ^\top,\qquad
T=\begin{cases}A/\|A\|_F,&A\ne0,\\0,&A=0.\end{cases}
\end{equation}
The usual factor \(1/m\) cancels under normalization. For every
possible output, \(T\) is defined and \(\|T\|_F\le1\).

\paragraph{Choosing the noise level.}
The score identity \eqref{eq:lb:identity} below gives a factor of
\(\sigma\) in the lower bound on record agreement. We therefore increase \(\sigma\)
with \(m\) while keeping the ideal batch's alignment of order
\(\sqrt{k}\). Larger \(\sigma\) makes individual records less
informative about \(B\) and increases the noise in the batch
cross-moment.

Conditional on the same \(B\), draw an ideal batch of \(m\) independent
records from \(P_B\). Put the first \(k\) and next \(k\) coordinates
of ideal record \(j\) in column \(j\) of \(Z^\star\) and \(W^\star\),
respectively. Let \(G\in\R^{k\times m}\) contain the corresponding
standard Gaussian residuals \(g\) as columns. Thus \(G\) is independent
of \(B,Z^\star\).
Set \(A^\star=W^\star(Z^\star)^\top\), and obtain \(T^\star\) by the
same normalization as in \eqref{eq:lb:statistic}. For
\(A^\star\ne0\), the alignment is
\(\langle B,T^\star\rangle=\langle B,A^\star\rangle/\|A^\star\|_F\).
Write \(S=Z^{\star}(Z^{\star})^\top\) and \(t=\tr S\). Then
\(A^{\star}=BS+\sigma G(Z^{\star})^\top\). The first moment of
the numerator and the variance of its noise term show how
\(\sigma\) affects the signal and fluctuations. Given \(Z^{\star}\),
\[
\E[\langle B,A^{\star}\rangle\mid Z^{\star}]=kt,\qquad
\operatorname{Var}(\sigma\langle B,G(Z^{\star})^\top\rangle\mid Z^{\star})=k\sigma^2t.
\]
We must also control the normalization by \(\|A^{\star}\|_F\). Given \(Z^{\star}\),
\(\E[\|A^{\star}\|_F^2\mid Z^{\star}]=k\tr(S^2)+k\sigma^2t
\le kt^2+k\sigma^2t\).
With \(\sigma^2\) on the scale of \(t\), the numerator has mean
\(kt\), the standard deviation of its noise term is of order
\(\sqrt{k}\,t\), and the denominator has second moment \(O(kt^2)\).
These scales suggest an inner product of order \(\sqrt{k}\)
after normalization. Since \(t\) is typically of order \(km\),
we choose
\begin{equation}\label{eq:lb:noise}
\sigma^2=km.
\end{equation}

This choice lets the noise variance grow with the batch size.

\paragraph{Transferring alignment to the released batch.}
With this noise level, Lemma~\ref{lem:lb:direction} in
Appendix~\ref{app:alignment-concentration} shows that
\(\langle B,T^\star\rangle\ge\sqrt{k}/8\) with probability at least
\(1-26/k\) over \(B\) and the ideal draws.
For each fixed \(B\), joint total variation accuracy applied to
the event that the normalized cross-moment has inner product
at least \(\sqrt{k}/8\) with \(B\) gives
\[
\Pr\{\langle B,T\rangle\ge\sqrt{k}/8\mid B\}
\ge\Pr\{\langle B,T^\star\rangle\ge\sqrt{k}/8\mid B\}-\alpha.
\]
Averaging over \(B\) gives a lower bound of \(\beta-26/k\). To make
this a constant fraction of \(\beta\), assume for the rest of
the matrix argument that
\begin{equation}\label{eq:lb:large-d}
d\ge1024/\beta.
\end{equation}
Since \(k\ge d/3\), this ensures \(26/k\le\beta/8\) and yields
\begin{equation}\label{eq:lb:actual}
\Pr\{\langle B,T\rangle\ge\sqrt k/8\}\ge7\beta/8.
\end{equation}

\paragraph{Keeping positive alignment.}
We need a lower bound on expected alignment. Negative values of
\(\langle B,T\rangle\) outside the event in \eqref{eq:lb:actual}
could cancel the positive contribution from that event.

We weight the direction to keep its contribution on the success
event and to make its alignment nonnegative everywhere. With
\(T\) held fixed, a linear interpolation from zero to full weight
makes the weighted direction continuous in \(B\) with a bounded
derivative wherever it is differentiable. This is useful for
integration by parts. Define
\begin{equation}\label{eq:lb:weight}
\widetilde T
=T\min\{1,\max\{0,8\langle B,T\rangle/\sqrt k\}\}.
\end{equation}
The weight is used only in the analysis and may depend on the unknown
\(B\). The weighted direction satisfies \(\|\widetilde T\|_F\le1\) and
\(\langle B,\widetilde T\rangle\ge0\), and equals \(T\) on the
success event in \eqref{eq:lb:actual}. Therefore
\begin{equation}\label{eq:lb:weighted}
\E\langle B,\widetilde T\rangle\ge\frac{7\beta\sqrt k}{64}.
\end{equation}

\paragraph{Relating alignment to input records.}
The privacy comparison will change one input record at a time, so we
express the expected alignment in terms of scores of individual records.
For the first two blocks \((z_i,w_i)\) of record \(X_i\), write
\(g_i=(w_i-Bz_i)/\sigma\). If \(p_B\) is the density of \(P_B\),
differentiating with the record held fixed gives
\(\nabla_B\log p_B(X_i)=g_i z_i^\top/\sigma\).
This gradient is the score of record \(i\). Its inner product
with the weighted direction is
\begin{equation}\label{eq:lb:record}
\langle\widetilde T,\nabla_B\log p_B(X_i)\rangle
=\frac1\sigma\langle\widetilde T,g_i z_i^\top\rangle.
\end{equation}
Thus the total record agreement is
\(\sum_i\E\langle\widetilde T,g_i z_i^\top\rangle\).

We use integration by parts in \(B\), as in Gaussian fingerprinting
and score arguments~\cite{KMS22,CWZ2023}. A related posterior argument
appears in Narayanan's covariance-estimation lower bound~\cite{Narayanan25}.
The posterior score contains two terms. The Gaussian prior
contributes \(-B\), and the likelihood contributes the sum of the
record scores. Integrating by parts relates these two terms, with
a correction from differentiating the weight. When differentiating
\(\widetilde T\), we hold \(X\) and the unweighted direction \(T\) fixed and
differentiate only the weight in \eqref{eq:lb:weight}.

Fix an input \(x=(x_1,\ldots,x_n)\). The posterior density of \(B\)
given \(X=x\) is
\[
q_x(B)\propto\exp\left(-\frac12\|B\|_F^2
-\frac1{2\sigma^2}\sum_i\|w_i-Bz_i\|^2\right),
\qquad
\nabla_B\log q_x(B)=-B+\frac1\sigma\sum_i g_i z_i^\top.
\]
This posterior is a nondegenerate Gaussian. Given the input, the distribution
of the unweighted output direction \(T\) does not depend on \(B\), so
revealing \(T\) leaves this posterior unchanged.
With \(T\) held fixed, write \(\operatorname{div}_B\) for the
divergence with respect to the entries of \(B\), so
\(\operatorname{div}_B\widetilde T
=\sum_{r,s}\partial(\widetilde T)_{rs}/\partial B_{rs}\).
Integrate \(\operatorname{div}_B(\widetilde Tq_x(B))\) over \(B\).
When \(T=0\), the vector field is identically zero. Otherwise,
the boundedness of \(\widetilde T\) and the Gaussian decay of the
posterior make the boundary term at infinity vanish. The boundary
terms at \(\langle B,T\rangle\in\{0,\sqrt k/8\}\) cancel because
\(\widetilde T\) is continuous across the two breakpoints. Thus
the integral is zero.
Expanding the divergence gives
\[
0=\E\!\left[
\operatorname{div}_B\widetilde T-\langle B,\widetilde T\rangle
+\frac1\sigma\sum_i\langle\widetilde T,g_i z_i^\top\rangle
\,\middle|\,X=x,T\right].
\]

Averaging over \(X,T\) gives the identity
\begin{equation}\label{eq:lb:identity}
\frac1\sigma\sum_i\E\langle\widetilde T,g_i z_i^\top\rangle
=\E\langle B,\widetilde T\rangle
-\E\operatorname{div}_B\widetilde T.
\end{equation}

The expectations are finite because \(\widetilde T\) and its
divergence are bounded and \(B\) and \(g_i z_i^\top\) have finite first
moments. With \(T\) fixed, differentiating the weight gives, almost
everywhere in \(B\),
\[
\operatorname{div}_B\widetilde T
=\frac{8\|T\|_F^2}{\sqrt k}
\one_{\{0<\langle B,T\rangle<\sqrt k/8\}}
\le\frac8{\sqrt k}.
\]
The cutoff \eqref{eq:lb:large-d} now serves a second purpose.
It gives \(\beta k\ge1024/3\), so the retained alignment in
\eqref{eq:lb:weighted} dominates the derivative term. Combining
the two bounds in \eqref{eq:lb:identity} yields
\begin{equation}\label{eq:lb:lower}
\sum_i\E\langle\widetilde T,g_i z_i^\top\rangle
\ge\sigma\left(\frac{7\beta\sqrt k}{64}-\frac8{\sqrt k}\right)
\ge\frac{\beta k\sqrt m}{16}.
\end{equation}

The record score contains a factor \(1/\sigma\).
Solving \eqref{eq:lb:identity} for total record agreement therefore
multiplies the alignment by \(\sigma=\sqrt{km}\), producing the
\(k\sqrt m\) scale in \eqref{eq:lb:lower}.

\subsection{Privacy bounds record agreement}\label{sec:lower-privacy}

To bound total record agreement from above, for each input record we
compare the output directions before and after replacing that record
by an independent draw. We measure both directions against the
original record. After replacement this agreement has
mean zero and expected absolute value at most one. The multiplicative
part of the density comparison contributes at most
\(e^\varepsilon-1\) times this expected absolute value for each
record. Summing over the records gives a total contribution of at
most \(n(e^\varepsilon-1)\le 2n\varepsilon\).
In each direction of the privacy comparison, the additive
allowance \(\delta\) bounds the excess density mass. We control
the contribution from all these remainders to obtain
\begin{equation}\label{eq:lb:upper-target}
\sum_i\E\langle\widetilde T,g_i z_i^\top\rangle
\le C\left(n\varepsilon+k\sqrt q+q\right),
\qquad q=n\delta L_\delta.
\end{equation}

\paragraph{Comparing with a replaced record.}
Conditional on \(B\), draw an independent copy \(X'\) of \(X\),
and let \(X^{(i)}\) be \(X\) with \(X_i\) replaced by \(X_i'\). Write
\(\widetilde T^{(i)}\) for the weighted direction obtained by running
\(M\) on \(X^{(i)}\) and using \eqref{eq:lb:weight} with the same \(B\).
The agreement is \(\langle\widetilde T,g_i z_i^\top\rangle\) before
replacement and \(\langle\widetilde T^{(i)},g_i z_i^\top\rangle\)
after replacement. Both inner products use the matrix from the
original record. Conditional on \(B\), this centered matrix is
independent of \(\widetilde T^{(i)}\), so the second inner product has
mean zero. The agreement can have either sign, so we bound the
absolute difference of the two joint densities. We set up the
comparison so that swapping the records brings both additive
remainders under the joint distribution with the original output.

Write \((z_i',w_i')\) for the replacement blocks and set
\(g_i'=(w_i'-Bz_i')/\sigma\).
Let \(\mathsf P\) and \(\mathsf Q_i\) be two distributions on
\((B,X,X',v)\). Under \(\mathsf P\), use \(v=\widetilde T\).
Under \(\mathsf Q_i\), use \(v=\widetilde T^{(i)}\).
Let \(S_i\) exchange \(X_i\) and \(X_i'\), fixing all other
coordinates of the tuple. Conditional on \(B\), the distribution
of \((X,X')\) is invariant under this swap. Under \(\mathsf P\),
the dataset used to generate \(v\) is the original \(X\). In
the swapped coordinates, that dataset is obtained by replacing
the \(i\)th record of the new \(X\) by the \(i\)th record of the
new \(X'\). Thus the swap sends \(\mathsf P\) to \(\mathsf Q_i\),
and applying it again sends \(\mathsf Q_i\) to \(\mathsf P\).

For fixed \(B,X,X'\), the inputs \(X\) and \(X^{(i)}\) differ
in at most one record. Computing the weighted direction with
the same \(B\) is the same deterministic postprocessing under
either input. For a measurable event \(E\) on the joint space,
apply privacy to \(\{v:(B,X,X',v)\in E\}\) at each fixed
\(B,X,X'\) and then average over their common distribution.
This gives
\(\mathsf P(E)\le e^\varepsilon\mathsf Q_i(E)+\delta\).
Applying the same inequality to the swapped event \(S_i(E)\) gives
\(\mathsf Q_i(E)\le e^\varepsilon\mathsf P(E)+\delta\).
Thus the functions below may depend on \(B\), even though the mechanism
does not know it.

Under both distributions, use the agreement function
\(f_i(B,X,X',v)=\langle v,g_i z_i^\top\rangle\).
Under \(\mathsf P\), this is the original record agreement.
Under \(\mathsf Q_i\), conditional on \(B,v\), the removed vectors
\(g_i,z_i\) remain independent standard Gaussians. For a fixed matrix
\(H\), the bilinear form \(\langle H,g_i z_i^\top\rangle\) has mean
zero and second moment \(\|H\|_F^2\). Since \(\|v\|_F\le1\),
\begin{equation}\label{eq:lb:fresh}
\E_{\mathsf Q_i}f_i=0,\qquad
\E_{\mathsf Q_i}|f_i|\le1.
\end{equation}

\paragraph{Keeping track of the additive privacy error.}
To bound \(\E_{\mathsf P}f_i-\E_{\mathsf Q_i}f_i\) for the
signed, possibly unbounded function \(f_i\), we separate the density
remainder permitted by \(\delta\).
Let \(p_i,q_i\) be the densities of \(\mathsf P,\mathsf Q_i\)
with respect to \(\lambda_i=\mathsf P+\mathsf Q_i\), and put
\(r_i=(p_i-e^\varepsilon q_i)_+.\)
By \eqref{eq:prelim-hockey-stick}, \(\int r_i\,d\lambda_i\le\delta\).
To express this part of the density as a weight under
\(\mathsf P\), define \(b_i=r_i/p_i\) on \(\{p_i>0\}\)
and \(b_i=0\) elsewhere. Then
\(0\le b_i\le1,\,\, \E_{\mathsf P}b_i\le\delta.\)
Write \(r_i\circ S_i\) for \(r_i\) evaluated after swapping
\(X_i,X_i'\).
The swap \(S_i\) preserves \(\lambda_i\) and exchanges its two
densities, so
\(r_i\circ S_i=(q_i-e^\varepsilon p_i)_+
\quad\text{almost everywhere}.\)
We bound the absolute density difference by a multiplicative
term and the two additive remainders. The bound holds
\(\lambda_i\)-almost everywhere.
\[
|p_i-q_i|\le(e^\varepsilon-1)q_i+r_i+r_i\circ S_i.
\]
When \(p_i\ge q_i\), the first two terms bound \(p_i-q_i\).
When \(q_i>p_i\), use
\(q_i-p_i\le(e^\varepsilon-1)p_i+(q_i-e^\varepsilon p_i)_+\)
and \(p_i\le q_i\).

The expectations under both joint distributions are finite because
\(|f_i|\le\|g_i\|\|z_i\|\) and the marginal distribution of
\((B,X,X')\) is the same under both.
Since \(\E_{\mathsf Q_i}f_i=0\), the agreement satisfies
\[
\E\langle\widetilde T,g_i z_i^\top\rangle
=\int f_i(p_i-q_i)\,d\lambda_i
\le\int |f_i|\,|p_i-q_i|\,d\lambda_i.
\]
The \(q_i\) term contributes at most \(e^\varepsilon-1\) by
\eqref{eq:lb:fresh}. The forward remainder \(r_i=b_i p_i\)
gives an expectation under \(\mathsf P\). We express the reverse
remainder under the same distribution by changing variables
using \(S_i\), which is its own inverse and preserves
\(\lambda_i\). This gives
\[
\begin{split}
\int |f_i|(r_i\circ S_i)\,d\lambda_i
&=\int |f_i\circ S_i|r_i\,d\lambda_i\\
&=\E_{\mathsf P}[b_i|f_i\circ S_i|].
\end{split}
\]
Here \((f_i\circ S_i)(B,X,X',v)
=\langle v,g_i'(z_i')^\top\rangle\), since \(S_i\) swaps the two
records while leaving \(v\) fixed.
Together with \eqref{eq:lb:fresh}, this gives
\[
\E\langle\widetilde T,g_i z_i^\top\rangle
\le e^\varepsilon-1
+\E_{\mathsf P}\bigl[b_i(|f_i|+|f_i\circ S_i|)\bigr].
\]
Both remaining terms now use the original weighted direction
\(\widetilde T\).
We regard all \(b_i\) as functions of the same tuple
\((B,X,X',\widetilde T)\) under \(\mathsf P\).
Summing over \(i\) and using
\(e^\varepsilon-1\le2\varepsilon\), gives
\begin{equation}\label{eq:lb:exceptions}
\begin{split}
\sum_i\E\langle\widetilde T,g_i z_i^\top\rangle
\le 2n\varepsilon+\E\sum_i b_i\bigl(&
|\langle\widetilde T,g_i z_i^\top\rangle|\\
&+|\langle\widetilde T,g_i'(z_i')^\top\rangle|\bigr).
\end{split}
\end{equation}

\paragraph{Combining the contributions permitted by \(\delta\).}
The weights \(b_i\) may depend on the data and the released
direction, so they can favor large agreements. The swap is useful
because both absolute inner products in \eqref{eq:lb:exceptions}
now use the same direction \(\widetilde T\). By choosing signs
that absorb the absolute values, we can pair this direction with
one sum of matrices. The signed matrices have nonnegative inner
products with \(\widetilde T\), while their components orthogonal
to \(\widetilde T\) can still cancel in the sum. Set
\[
\begin{split}
c_i&=b_i\operatorname{sgn}\langle\widetilde T,g_i z_i^\top\rangle,\\
c_i'&=b_i\operatorname{sgn}\langle\widetilde T,g_i'(z_i')^\top\rangle,
\end{split}
\]
with \(\operatorname{sgn}(0)=0\). With these signs,
\(c_i\langle\widetilde T,g_i z_i^\top\rangle
=b_i|\langle\widetilde T,g_i z_i^\top\rangle|\), and similarly for
the replacement records. Since \(\|\widetilde T\|_F\le1\),
\[
\begin{split}
\sum_i b_i\bigl(&|\langle\widetilde T,g_i z_i^\top\rangle|
+|\langle\widetilde T,g_i'(z_i')^\top\rangle|\bigr)\\
&=\left\langle\widetilde T,
\sum_i\bigl(c_i g_i z_i^\top+c_i'g_i'(z_i')^\top\bigr)\right\rangle\\
&\le\left\|\sum_i\bigl(c_i g_i z_i^\top+c_i'g_i'(z_i')^\top\bigr)\right\|_F.
\end{split}
\]
At each fixed \(B\), the \(2n\) Gaussian pairs are independent
and each pair consists of two independent standard Gaussian
vectors. This product distribution is the same for every \(B\),
so the pairs remain independent after averaging over \(B\).
The coefficients satisfy
\(|c_i|,|c_i'|\le1\) and
\(\E|c_i|,\E|c_i'|\le\E b_i\le\delta\), but may depend on all
the pairs. Conditioning on the coefficients can change the
distribution of the Gaussian pairs. The following lemma bounds
the expected norm while allowing that dependence.

\begin{lemma}[Gaussian sums with data-dependent coefficients]\label{lem:lb:coefficients}
Let \((g_j,z_j)\), \(1\le j\le N\), be mutually independent pairs
of independent standard Gaussian vectors in \(\R^k\). Suppose random
coefficients \(c_j\in[-1,1]\) satisfy \(\E|c_j|\le\delta\), where
\(0<\delta<1\). The coefficients may depend arbitrarily on the vectors
and on other randomness. With \(Q=N\delta\log(128/\delta)\),
\begin{equation}\label{eq:lb:coefficients}
\E\left\|\sum_jc_jg_jz_j^\top\right\|_F
\le C\left(k\sqrt Q+Q\right).
\end{equation}
\end{lemma}

\begin{proof}
We round the coefficients to a signed selection whose mean,
conditional on the Gaussian pairs and the original coefficients,
is the original coefficient vector. This reduces the norm
estimate to a sum over sign patterns. For each fixed pattern, we use a bound
on its contribution on the event that rounding selects it.

Adjoin independent variables \(U_1,\ldots,U_N\), uniform on
\([0,1]\) and independent of all the original randomness. Set
\(V_i=\operatorname{sgn}(c_i)\mathbf1_{\{U_i\le|c_i|\}}\),
with \(\operatorname{sgn}(0)=0\).
Conditional on the original randomness, \(\E_U V_i=c_i\).
Convexity of the Frobenius norm therefore gives
\begin{equation}\label{eq:lb:rounding}
\E\left\|\sum_i c_i g_i z_i^\top\right\|_F
\le\E\left\|\sum_i V_i g_i z_i^\top\right\|_F.
\end{equation}
Write \(V=(V_1,\ldots,V_N)\), \(J=|\{i:V_i\ne0\}|\), and
\(K=\E J\). Since \(p_i=\Pr(V_i\ne0)=\E|c_i|\le\delta\),
\begin{equation}\label{eq:lb:selection-size}
K=\sum_i p_i\le N\delta.
\end{equation}

Lemma~\ref{lem:lb:fixed-selection} in
Appendix~\ref{app:fixed-selection-tails} bounds the contribution of
a deterministic pattern on any event of probability \(p>0\). The event
may depend on the Gaussian pairs and the rounding randomness.
Conditioning on such an event multiplies a tail probability by at
most \(1/p\). Shifting the exponential tail parameter by
\(\log(1/p)\) accounts for that factor.
Averaging these logarithmic costs over the selected patterns
gives the entropy term, as in information-theoretic bounds
for adaptive selection~\cite{RZ16}. The small probabilities
that individual \(V_i\) are nonzero bound this entropy.
For a finite-valued random variable \(Y\), write
\(H(Y)=\sum_y\Pr(Y=y)\log(1/\Pr(Y=y))\), with
zero-probability summands interpreted as zero.
Entropy subadditivity, which does not require independence, gives
\(H(V)\le\sum_i H(V_i)\).
First recording whether \(V_i\) is zero and then, if nonzero, its
sign gives
\[
H(V_i)\le h(p_i)+p_i\log2,\qquad
h(p)=-p\log p-(1-p)\log(1-p).
\]
Since \(-(1-p)\log(1-p)\le p\),
\(h(p)+p\log2\le p\log(2e/p)\).
The last expression is increasing on \((0,1]\), with derivative
\(\log(2/p)>0\), and extends continuously to zero. Consequently,
\begin{equation}\label{eq:lb:selection-entropy}
H(V)\le N\delta\log(2e/\delta).
\end{equation}

We now average over the patterns selected by rounding. For each
\(v\in\{-1,0,1\}^N\), write \(p_v=\Pr(V=v)\),
\(j=|\{i:v_i\ne0\}|\), and \(Y_v=\|\sum_i v_i g_i z_i^\top\|_F\).
On \(\{V=v\}\), the norm on the right of \eqref{eq:lb:rounding}
equals \(Y_v\). The zero pattern contributes zero. For each
nonzero pattern with \(p_v>0\), apply \eqref{eq:lb:event-selection}
to \(A=\{V=v\}\) to get
\begin{align*}
\E\left\|\sum_i V_i g_i z_i^\top\right\|_F
&=\sum_{v\ne0}\E[Y_v\mathbf1_{\{V=v\}}]\\
&\le C\sum_{v\ne0}p_v
 \bigl[k\sqrt{j+\log(1/p_v)}+j+\log(1/p_v)\bigr],
\end{align*}
Terms with \(p_v=0\) are omitted. We have
\[
\sum_{v\ne0}p_v[j+\log(1/p_v)]\le K+H(V).
\]
By Cauchy--Schwarz and \(\sum_{v\ne0}p_v\le1\),
\begin{align*}
\sum_{v\ne0}p_v\sqrt{j+\log(1/p_v)}
&\le\left(\sum_{v\ne0}p_v\right)^{1/2}
 \left(\sum_{v\ne0}p_v[j+\log(1/p_v)]\right)^{1/2}\\
&\le\sqrt{K+H(V)}.
\end{align*}
Therefore,
\begin{equation}\label{eq:lb:entropy-selection}
\E\left\|\sum_i V_i g_i z_i^\top\right\|_F
\le C\bigl[k\sqrt{K+H(V)}+K+H(V)\bigr].
\end{equation}
Finally, \eqref{eq:lb:selection-size} and
\eqref{eq:lb:selection-entropy} imply
\[
K+H(V)
\le N\delta[1+\log(2e/\delta)]
=N\delta\log(2e^2/\delta)
\le N\delta\log(128/\delta)=Q.
\]
Combining this with \eqref{eq:lb:rounding} and
\eqref{eq:lb:entropy-selection} proves the result.
The zero selection contributes zero throughout, so no assumption
\(N\delta\ge1\) is needed.
\end{proof}

Apply Lemma~\ref{lem:lb:coefficients} with \(N=2n\), so \(Q=2q\),
to the matrix sum of original and replacement records above.
Substituting its bound into \eqref{eq:lb:exceptions} proves
\eqref{eq:lb:upper-target}.

\paragraph{The two privacy-dependent alternatives.}
Comparing \eqref{eq:lb:lower} with \eqref{eq:lb:upper-target} gives
\begin{equation}\label{eq:lb:master}
\beta k\sqrt m
\le C\left[n(\varepsilon+\delta L_\delta)
+k\sqrt{n\delta L_\delta}\right].
\end{equation}
At least one of the two terms in brackets is at least a constant
multiple of the left side. The linear term then gives
\(n\ge c\beta k\sqrt m/(\varepsilon+\delta L_\delta)\),
whereas the square-root term gives
\(n\ge c\beta^2m/(\delta L_\delta)\).
Since \(k\ge d/3\), in either case
\[
n\ge c\min\left\{
\frac{\beta d\sqrt m}{\varepsilon+\delta L_\delta},
\frac{\beta^2m}{\delta L_\delta}\right\}.
\]
As \(\beta\le1\) and \(a_\beta\ge1\), this implies the privacy
term in \eqref{eq:lb:theorem}.

The square-root term in \eqref{eq:lb:master} gives the
\(m/(\delta L_\delta)\) alternative for fixed \(\alpha\).
To see what a separate bound for each record gives, set
\(Y=\|g_i z_i^\top\|_F\) and let \(U\) be independent and uniform on
\([0,1]\). The event \(\{U\le b_i\}\) has probability
\(p=\E b_i\le\delta\), and \(\E[Y\one_{\{U\le b_i\}}]=\E[b_iY]\).
For \(p>0\), apply \eqref{eq:lb:event-selection} with one selected
index. The bound is \(Cp[k\sqrt{1+\log(1/p)}+1+\log(1/p)]\).
Since \(k\ge1\) and \(p[1+\log(1/p)]\) is increasing on \((0,1]\),
this is \(O(k\delta[1+\log(1/\delta)])\). The contribution is zero
when \(p=0\), and the same bound holds for the replacement records.
Summing these separate estimates in \eqref{eq:lb:exceptions} bounds
the additive contribution by \(O(nk\delta L_\delta)\). If it accounts
for a constant fraction of the required \(\beta k\sqrt m\) agreement,
it gives \(n=\widetilde\Omega(\beta\sqrt m/\delta)\).

\subsection{Unknown-mean bounds and conclusion}
\label{sec:scalar-lower-bound}\label{sec:lb:scalar}

We now use a family with identity covariance and one unknown
mean coordinate. The ideal batch average has standard deviation
\(m^{-1/2}\), and joint accuracy gives the released batch
probability at least \(3\beta/4\) of locating the mean within
\(a_\beta/\sqrt m\). The posterior density given the input
limits this probability and gives the linear cost in \(m\)
from accuracy alone. For the privacy-dependent bound, we compare
the probability of decoding among several nearby means with
the differences between their output distributions that privacy
permits.

\begin{lemma}[Lower bounds from an unknown mean]\label{lem:lb:scalar}
Under the assumptions of Theorem~\ref{thm:lb},
\begin{equation}\label{eq:lb:scalar}
n\ge\frac{\beta^3m}{8},\qquad
n(\varepsilon+\delta)\ge
\frac{\beta^2\sqrt m}{16a_\beta}.
\end{equation}
\end{lemma}

\begin{proof}
Consider \(P_\theta=\mathcal N(\theta e_1,I_d)\) and input
\(X\sim P_\theta^{\otimes n}\). Let \(U\) be the mean of the first
coordinates in the released batch \(M(X)\), and put
\(r=a_\beta/\sqrt m\). The ideal average satisfies
\(U^\star\sim\mathcal N(\theta,1/m)\). The Gaussian tail bound and
total variation accuracy therefore give, for every fixed \(\theta\),
\begin{equation}\label{eq:lb:localize}
\Pr\{|U-\theta|\le r\mid\theta\}
\ge1-2e^{-a_\beta^2/2}-\alpha=3\beta/4.
\end{equation}

\paragraph{The input must locate the mean.}
To bound the success probability using accuracy alone, draw
\(\theta\sim\mathcal N(0,\tau^2)\) with \(\tau>0\), then draw
\(X\sim P_\theta^{\otimes n}\). The posterior of \(\theta\) given
\(X\) is Gaussian with variance \((n+\tau^{-2})^{-1}\).
Conditional on \(X\), the released average \(U\) is independent
of \(\theta\), so the posterior given \((X,U)\) is the same.
Its density is at most
\(\sqrt{(n+\tau^{-2})/(2\pi)}\) everywhere. Averaging the
conditional interval probability over the input and output gives
\[
\begin{split}
\Pr\{|U-\theta|\le r\}
&=\E_{X,U}\!\left[
\Pr\{\theta\in[U-r,U+r]\mid X,U\}\right]\\
&\le2r\sqrt{\frac{n+\tau^{-2}}{2\pi}}.
\end{split}
\]
Taking expectation of~\eqref{eq:lb:localize} over the prior on
\(\theta\) and comparing with this upper bound yields
\[
\frac{3\beta}{4}\le
2\frac{a_\beta}{\sqrt m}\sqrt{\frac{n+\tau^{-2}}{2\pi}}.
\]
Letting \(\tau\to\infty\) and rearranging gives
\[
n\ge\frac{9\pi\beta^2m}{32a_\beta^2}
\ge\frac{\beta^3m}{8}.
\]
The second inequality uses
\(\beta a_\beta^2=2\beta\log(8/\beta)\le2\log8<9\pi/4\).

\paragraph{Privacy limits changes between nearby means.}
We now compare output distributions at fixed means. A coupling
lets us use the privacy guarantee for one record to control the
difference between the two output distributions. For fixed
neighboring inputs, privacy bounds the total variation
distance between the output distributions by
\(e^\varepsilon-1+\delta\le2\varepsilon+\delta\).
For any distributions \(P,Q\) on \(\R^d\), draw each input-record
pair from a maximal coupling of \(P,Q\), independently across record
indices. This couples \(X_P\sim P^{\otimes n}\) with
\(X_Q\sim Q^{\otimes n}\), and each pair disagrees with probability
\(\TV(P,Q)\). Conditional on the coupled datasets, replace the
disagreeing records one at a time and use the triangle inequality. The expected number of
differing records is \(n\TV(P,Q)\). Taking expectation over the coupled
datasets and using convexity of total variation therefore gives
\begin{equation}\label{eq:lb:coupling}
\TV\bigl(\Law(M(X_P)),\Law(M(X_Q))\bigr)
\le n(2\varepsilon+\delta)\TV(P,Q).
\end{equation}
For the scalar Gaussian family, the relative entropy is
\((\theta-\theta')^2/2\), so Pinsker's inequality gives
\(\TV(P_\theta,P_{\theta'})\le|\theta-\theta'|/2\)
(see also Lemma~\ref{lem:prelim-gaussian-tv}).

The success probability \(3\beta/4\) can be small, so we use
\(K=\lceil4/\beta\rceil\) candidate means \(\theta_j=2rj\) for
\(0\le j<K\). Decode the candidate closest to \(U\), breaking ties
deterministically, and let \(E_j\) be the event that the decoder
returns \(j\). If the output had the same distribution for all
\(K\) candidates, the average decoding success over a uniform choice
of candidate would be \(1/K\le\beta/4\).
Write \(Q_j=\Law(M(X))\) when \(X\sim P_{\theta_j}^{\otimes n}\).
For an ideal batch from \(P_{\theta_j}\), the decoder succeeds
whenever \(|U^\star-\theta_j|<r\). The ideal average is continuous,
so ties have probability zero. The released average may have ties,
so we apply accuracy to the decoder event \(E_j\) itself to obtain
\(Q_j(E_j)\ge3\beta/4\).
For each \(j\),
\(Q_j(E_j)\le Q_0(E_j)+\TV(Q_j,Q_0)\).
The events \(E_j\) partition the output space, so
\(\sum_jQ_0(E_j)=1\). Summing the inequalities and dividing by \(K\) gives
\[
\frac{3\beta}{4}
\le\frac1K\sum_jQ_j(E_j)
\le\frac1K+\max_j\TV(Q_j,Q_0).
\]
Since \(1/K\le\beta/4\), some output distribution must satisfy
\(\TV(Q_j,Q_0)\ge\beta/2\). On the other hand,
\(j\le K-1<4/\beta\), so \eqref{eq:lb:coupling} and the
Gaussian comparison give
\[
\max_j\TV(Q_j,Q_0)
\le n(2\varepsilon+\delta)r\frac4\beta
\le\frac{8a_\beta n(\varepsilon+\delta)}{\beta\sqrt m}.
\]
Comparing the two bounds proves the second inequality in
\eqref{eq:lb:scalar}.
\end{proof}

\paragraph{Completing the proof.}
The matrix argument gives the privacy-dependent term for
\(d\ge1024/\beta\). For \(d<1024/\beta\),
Lemma~\ref{lem:lb:scalar} gives
\[
n\ge\frac{\beta^2\sqrt m}{16a_\beta(\varepsilon+\delta)}
\ge\frac{\beta^3d\sqrt m}
{16384a_\beta(\varepsilon+\delta L_\delta)}.
\]
Since the minimum in \eqref{eq:lb:theorem} is at most its first
term, this lower bound on \(n\) implies the required privacy term. Finally,
Lemma~\ref{lem:lb:scalar} gives \(n\ge\beta^3m/8\) in every
dimension. The maximum of these two lower bounds is at least half their
sum. This proves \eqref{eq:lb:theorem} and the additional assertion.
The same argument covers \(\varepsilon=0\), since \(\delta>0\)
keeps the denominators positive.

For a direct comparison with the upper-bound rate, note that
\(\varepsilon+\delta L_\delta\le L_\delta(\varepsilon+\delta)\).
The theorem therefore implies
\begin{equation}\label{eq:lb:log-form}
n\ge\frac{c\beta^3}{a_\beta L_\delta}
\left[m+\min\left\{
\frac{d\sqrt m}{\varepsilon+\delta},\frac m\delta
\right\}\right].
\end{equation}
For every fixed \(0\le\alpha<1\), this is
\(\widetilde\Omega_\alpha(m+\min\{d\sqrt m/(\varepsilon+\delta),
m/\delta\})\). For fixed \(0<\alpha<1\), the rate matches the
upper bound in Corollary~\ref{cor:iid-full-range} up to logarithmic
factors.

\section{Conclusion}

We gave a polynomial-time algorithm for Gaussian multi-sampling
under approximate differential privacy. The algorithm uses shared
stable statistics across the output batch and corrects for weighting
and rejection to obtain exactly independent outputs. Our lower
bound applies even when the outputs are dependent. Thus, for fixed
joint error, requiring exact independence increases the optimal
sample complexity by at most logarithmic factors.

\section*{AI Methodology Statement}

We used GPT-5.5 for mathematical discussions that contributed to
deriving the lower bound, and GPT-5.6 for discussions that contributed
to developing the construction for exactly independent and
identically distributed outputs.
We used GPT-6.1 to improve the grammar, wording, and organization
of the manuscript.

\renewcommand{\bibsection}{
  \section*{\refname}
  \addcontentsline{toc}{section}{\refname}
}
\bibliographystyle{plainnat}
\bibliography{references}

\appendix

\section{Supporting estimates for the upper bound}
\label{app:upper-estimates}\label{app:upper}

We collect the distribution and privacy facts used in both proofs.
We then derive the frame densities and concentration bounds needed
for the joint privacy comparison, prove stability of the weighted
statistic, and verify the parameter choices for the sampler.

Two datasets are neighboring if they have the same number of records
and differ in one record, as in Definition~\ref{main:def:dp}.
An upper bound on the additive privacy parameter can always be
replaced by one when it exceeds one.

\subsection{Distribution and privacy facts}\label{app:preliminary-proofs}
\label{sec:prelim-gaussians}\label{sec:prelim-frames}

\paragraph{Total variation and Gaussian resampling.}
We use coupling to compare batches and the common-covariance
Gaussian bound to compare inputs. We then record the Gaussian
identities that determine the distribution of the resampled rows.

\begin{lemma}[Coupling and postprocessing]\label{lem:prelim-coupling}
Let \(P,Q\) be distributions.
Every coupling \(U\sim P\), \(V\sim Q\) satisfies
\(\TV(P,Q)\le\Prob\{U\ne V\}\), and some coupling attains equality.
Applying the same randomized rule to \(P\) and \(Q\) cannot increase
their total variation distance.
\end{lemma}

See \cite[Theorems~2.4 and~2.12]{denHollander2012} for the
coupling inequality and maximal coupling. The total variation norm
there is twice the distance used here. The postprocessing bound is
the data-processing inequality for total variation
\cite[Fact~36]{IKM25}.

\begin{lemma}[Gaussians with a common covariance]\label{lem:prelim-gaussian-tv}
For \(\Sigma\succ0\),
\[
 \TV\bigl(\Normal(\mu,\Sigma),\Normal(\nu,\Sigma)\bigr)
 \le \tfrac12\|\Sigma^{-1/2}(\mu-\nu)\|.
\]
\end{lemma}

\begin{proof}[Proof of Lemma~\ref{lem:prelim-gaussian-tv}]
Translate, whiten, and rotate to obtain \(\Normal(0,I_d)\) and
\(\Normal(se_1,I_d)\), where \(s=\|\Sigma^{-1/2}(\mu-\nu)\|\).
If \(s=0\), the distributions coincide. For \(s>0\), the log ratio
of the first density to the second at \(x\) is \(s^2/2-sx_1\).
Thus the first density is larger on \(x_1<s/2\), and total variation
is the difference of the two probabilities of this halfspace.
Writing \(\Phi\) for the standard Gaussian distribution function
gives
\[
 \TV\bigl(\Normal(0,I_d),\Normal(se_1,I_d)\bigr)
 =\Phi(s/2)-\Phi(-s/2)\le s/\sqrt{2\pi}\le s/2 .
\]
The first inequality bounds the probability of \([-s/2,s/2]\) by
its length times the maximum standard Gaussian density.
See also \cite[Section~2]{DevroyeMehrabianReddad2023TV}.
\end{proof}

\begin{lemma}[Orthonormal Gaussian combinations]
\label{lem:prelim-gaussian-combinations}
Let \(X\in\R^{n\times d}\) have independent \(\Normal(0,\Sigma)\) rows.
Let \(A\in\R^{k\times n}\) be fixed or drawn independently of \(X\),
with \(AA^\top=I_k\). Then \(AX\) has independent
\(\Normal(0,\Sigma)\) rows and is independent of \(A\).
\end{lemma}

\begin{proof}[Proof of Lemma~\ref{lem:prelim-gaussian-combinations}]
Conditional on \(A\), the rows of \(AX\) are jointly Gaussian
with mean zero. Writing \(X_\ell^\top\) for row \(\ell\) of \(X\),
the transpose of row \(i\) of \(AX\) is \(\sum_\ell A_{i\ell}X_\ell\).
By independence of the input rows, the covariance between the
transposes of rows \(i\) and \(j\) is
\(\sum_{\ell=1}^n A_{i\ell}A_{j\ell}\Sigma=(AA^\top)_{ij}\Sigma\).
Each row has covariance \(\Sigma\) and distinct rows have zero
cross-covariance. Joint Gaussianity gives independence of the rows.
Their conditional distribution is the same for every admissible
\(A\), so \(AX\) is independent of \(A\).
\end{proof}

\begin{lemma}[Mean and orthogonal component]\label{lem:prelim-gaussian-components}
Let \(X\) have independent \(\Normal(\mu,\Sigma)\) rows, and let
\(E\) be a fixed matrix with orthonormal columns spanning \(\one_n^\perp\).
The sample mean \(\bar X=X^\top\one_n/n\) is independent of
\(E^\top X\), whose rows are independent \(\Normal(0,\Sigma)\)
vectors. If \(\Sigma\succ0\), every subset of \(d+1\) observations
is affinely independent almost surely. In particular, every
subset of at least \(d+1\) observations has positive definite scatter.
\end{lemma}

\begin{proof}[Proof of Lemma~\ref{lem:prelim-gaussian-components}]
Apply Lemma~\ref{lem:prelim-gaussian-combinations} to
\(X-\one_n\mu^\top\) using the orthogonal matrix whose first row is
\(\one_n^\top/\sqrt n\) and whose remaining rows are \(E^\top\).
The first transformed row is \(\sqrt n(\bar X-\mu)^\top\).
The other rows are \(E^\top X\), since \(E^\top\one_n=0\).
Independence of these rows proves the decomposition.

For \(J_0\subseteq[n]\) of size \(d+1\), let \(X_{J_0}\) be the
matrix of its rows. These rows are affinely independent if
\(\det[X_{J_0},\one_{d+1}]\ne0\). The determinant is a nonzero
polynomial, as witnessed by the points \(0,e_1,\ldots,e_d\).
When \(\Sigma\succ0\), the entries of \(X_{J_0}\) have a joint
density, so the polynomial vanishes with probability zero.
A finite union gives affine independence for all such \(J_0\).

For any \(J\subseteq[n]\) of size at least \(d+1\), write \(\bar X_J\)
for its mean and \(S(X_J)\) for its scatter. For \(v\ne0\),
\[
 v^\top S(X_J)v=\sum_{i\in J}\bigl(v^\top(X_i-\bar X_J)\bigr)^2 .
\]
If the sum were zero, the rows in \(J\) would all belong to an
affine hyperplane with normal \(v\). This contradicts affine
independence of any \(d+1\) rows in \(J\), so \(S(X_J)\succ0\).
\end{proof}

\begin{lemma}[Gaussian length and direction]\label{lem:prelim-gaussian-direction}
If \(g\sim\Normal(0,I_D)\), then \(g/\|g\|\) is uniform on the
unit sphere and is independent of \(\|g\|\).
The squared length \(\|g\|^2\) has the chi-square distribution
\(\chi_D^2\).
\end{lemma}

\begin{proof}[Proof of Lemma~\ref{lem:prelim-gaussian-direction}]
The Gaussian distribution is rotationally invariant. Conditional
on its length, its direction therefore has the uniform distribution
on the sphere, independently of the length. The squared length is
a sum of \(D\) independent squared standard Gaussians.
\end{proof}

Uniform frames are invariant under left and right orthogonal
multiplication. We use the left invariance to identify the selected
rows of the centered frame.

\begin{lemma}[Sampling and projecting frames]\label{lem:prelim-frame-projection}
\leavevmode
\begin{enumerate}
\item \emph{Sampling.} Let \(N\ge p\ge1\), and let \(G\in\R^{N\times p}\)
      have independent standard Gaussian entries. Then
      \(Q=G(G^\top G)^{-1/2}\) is a uniform frame independent of
      \(G^\top G\). The same frame distribution is obtained by drawing
      each column uniformly on the unit sphere in the orthogonal
      complement of the preceding columns.
\item \emph{Projection.} If \(Q\) is a uniform \(N\times p\) frame and
      \(L\in\R^{r\times N}\) has full row rank, then
      \(LQ\overset{d}{=}(LL^\top)^{1/2}Z\), where \(Z\) consists of
      the first \(r\) rows of a uniform \(N\times p\) frame.
\item \emph{Centered frames.} For \(n-1\ge d\), take \(U=EQ\) with
      \(Q\) uniform of size \((n-1)\times d\) and \(E\) as in
      Lemma~\ref{lem:prelim-gaussian-components}. Then
      \(U^\top U=I_d\), \(\one_n^\top U=0\), and the distribution of
      \(U\) is independent of the basis \(E\). If \(E_J\) consists of
      the rows indexed by \(J\subseteq[n]\) and \(|J|=r\), then
      \(E_JE_J^\top=I_r-\one_r\one_r^\top/n\).
\end{enumerate}
\end{lemma}

\begin{proof}
\emph{(1) Sampling.}
The determinant of the first \(p\) rows of \(G\) is a nonzero
polynomial in their entries. Since these entries have a joint
density, \(G\) has full column rank almost surely. Put
\(Q(G)=G(G^\top G)^{-1/2}\), and let \(O\) be an independent uniform
orthogonal matrix of order \(N\). For fixed \(G\) of full rank,
complete \(Q(G)\) to an orthogonal matrix. By right invariance of
\(O\), the first \(p\) columns of its product with this completion
have the uniform frame distribution. These columns are \(OQ(G)\).
Thus \(OQ(G)\) is uniform conditional on \(G\), with a distribution
that is constant in \(G\). Gaussian rotation invariance and
\((OG)^\top(OG)=G^\top G\) give
\[
 (OQ(G),G^\top G)\overset{d}{=}(Q(G),G^\top G).
\]
This proves the uniform distribution and independence from the
Gram matrix \cite[see also Proposition~9]{Tropp2012}.

Successive uniform draws in orthogonal complements give a frame
distribution invariant under left orthogonal transformations.
Any such distribution is unchanged by multiplication by an
independent uniform orthogonal matrix. Conditional on the initial
frame, this multiplication gives a uniform frame by the same
orthogonal completion argument. This proves uniqueness and
justifies the successive draws. Right multiplication by a fixed
orthogonal matrix preserves left invariance, so uniqueness also
gives right invariance.

\emph{(2) Projection.}
Complete the orthonormal rows of \((LL^\top)^{-1/2}L\) to the
rows of an orthogonal matrix. Its product with \(Q\) has the
distribution of \(Q\), by left invariance. Taking the first
\(r\) rows and multiplying by \((LL^\top)^{1/2}\) gives the
claimed distribution of \(LQ\).

\emph{(3) Centered frames.}
The identities \(E^\top E=I_{n-1}\) and \(\one_n^\top E=0\)
give the two constraints on \(U\). Any other orthonormal basis of
\(\one_n^\perp\) has the form \(EO\) for an orthogonal matrix \(O\)
of order \(n-1\). Left invariance gives \(EOQ\overset{d}{=}EQ\).
Finally, \(EE^\top=I_n-\one_n\one_n^\top/n\) is the orthogonal
projector onto \(\one_n^\perp\). Its principal submatrix on \(J\)
is \(E_JE_J^\top\), giving the claimed identity.
\end{proof}

\paragraph{Privacy comparisons.}
The density identity below identifies the mass that contributes to
the additive privacy parameter. Using \(P+Q\) as the common measure
also lets us apply it in the lower bound when the output distributions
lack Lebesgue densities.

\begin{lemma}[Density form of the privacy comparison]
\label{lem:prelim-privacy-remainder}
Let \(P,Q\) have densities \(p,q\) with respect to a common measure
\(\rho\), for instance \(P+Q\). With \(a_+=\max\{a,0\}\),
\[
 \sup_E\bigl(P(E)-e^\varepsilon Q(E)\bigr)
 =\int (p-e^\varepsilon q)_+\,d\rho .
\]
\end{lemma}

\begin{proof}[Proof of Lemma~\ref{lem:prelim-privacy-remainder}]
For any event \(E\), \(P(E)-e^\varepsilon Q(E)\) is the integral of
\(p-e^\varepsilon q\) on \(E\). This is at most the integral of
its positive part, with equality on \(\{p>e^\varepsilon q\}\).
\end{proof}

\begin{lemma}[Postprocessing and composition]
\label{lem:prelim-privacy-rules}
A randomized rule that accesses the input only through a private
output preserves its privacy parameters. If \(M_1\) is
\((\varepsilon_1,\delta_1)\)-private and \(M_2(\cdot,y)\) is
\((\varepsilon_2,\delta_2)\)-private for every fixed \(y\), then
draw \(Y_1=M_1(x)\) and then \(M_2(x,Y_1)\) using randomness
independent of the first call. Their joint release is
\((\varepsilon_1+\varepsilon_2,\delta_1+\delta_2)\)-private.
For a fixed neighboring pair, the same conclusion holds on that
pair if the first-stage comparison holds in both directions and
the second-stage comparison holds in both directions for every
fixed first-stage output.
\end{lemma}

See \cite[Proposition~2.1 and Theorem~B.1]{DworkRoth2014} for
postprocessing and composition. Keeping the budgets separate in
the composition proof gives the stated sums.

The next lemma gives the event comparison used for the candidate
density, the perturbation bound used for the clipped test, and the
privacy bound for the mixture with the correction branch.

\begin{lemma}[Distribution comparisons and privacy error]
\label{lem:prelim-privacy-comparison}
\leavevmode
\begin{enumerate}
\item Let \(p,q\) be the densities of \(P,Q\) with respect to a common
      measure. If \(P(G^c)\le\delta\) and \(p\le e^\varepsilon q\)
      almost everywhere on \(G\), then
      \(P(E)\le e^\varepsilon Q(E)+\delta\) for every event \(E\).
      For privacy, check both orders of the pair.
\item If \(M\) is \((\varepsilon,\delta)\)-private and
      \(\TV(\Law(M(x)),\Law(\widetilde M(x)))\le\gamma\) for every
      \(x\), then \(\widetilde M\) is
      \((\varepsilon,\delta+(1+e^\varepsilon)\gamma)\)-private.
\item Suppose a mechanism chooses its private branch with a fixed
      probability \(w\in[0,1]\) on every input \(x\). When \(w>0\),
      conditional on choosing that branch, its output has the
      distribution of \(M(x)\) for an \((\varepsilon,\delta_0)\)-private
      mechanism \(M\). The other branch may use any input-dependent
      rule. The mixture is \((\varepsilon,w\delta_0+1-w)\)-private.
\end{enumerate}
\end{lemma}

\begin{proof}[Proof of Lemma~\ref{lem:prelim-privacy-comparison}]
For (1), the integrand in Lemma~\ref{lem:prelim-privacy-remainder}
vanishes on \(G\). Its integral over \(G^c\) is at most \(P(G^c)\).

For (2), fix neighboring \(x,x'\) and an event \(E\). Total variation
bounds the change in its probability on either input. Thus
\begin{align*}
 \Prob\{\widetilde M(x)\in E\}
 &\le \Prob\{M(x)\in E\}+\gamma\\
 &\le e^\varepsilon\Prob\{M(x')\in E\}+\delta+\gamma\\
 &\le e^\varepsilon\Prob\{\widetilde M(x')\in E\}
       +\delta+(1+e^\varepsilon)\gamma .
\end{align*}

For (3), write \(N\) for the mixture. On input \(x\), the private
branch contributes \(w\Prob\{M(x)\in E\}\) and the other branch
contributes at most \(1-w\). Thus
\begin{align*}
 \Prob\{N(x)\in E\}
 &\le w\Prob\{M(x)\in E\}+1-w\\
 &\le e^\varepsilon w\Prob\{M(x')\in E\}+w\delta_0+1-w\\
 &\le e^\varepsilon\Prob\{N(x')\in E\}+w\delta_0+1-w .
\end{align*}
The last step uses that the private branch contributes
\(w\Prob\{M(x')\in E\}\) to the full probability on \(x'\).
\end{proof}

\begin{lemma}[Clipped Laplace noise]\label{lem:prelim-laplace}
Suppose \(g\) has sensitivity at most \(\Delta_g\) on neighboring datasets.
Let \(Z\sim\operatorname{Lap}(s)\) have density \(e^{-|z|/s}/(2s)\)
for \(s>0\). The release \(g(x)+Z\) is \((\Delta_g/s,0)\)-private.
For \(\ell\ge0\), set
\(\operatorname{clip}_\ell(z)=\max\{-\ell,\min\{z,\ell\}\}\).
The release \(g(x)+\operatorname{clip}_\ell(Z)\) is
\(\bigl(\Delta_g/s,(1+e^{\Delta_g/s})e^{-\ell/s}\bigr)\)-private.
\end{lemma}

\begin{proof}[Proof of Lemma~\ref{lem:prelim-laplace}]
For neighboring datasets, the log-density ratio of the ordinary
releases is at most \(|g(x)-g(x')|/s\le\Delta_g/s\) by the triangle
inequality in the Laplace density \cite[Theorem~3.6]{DworkRoth2014}.
Couple the ordinary release and the release with clipped noise using
the same \(Z\). They can differ only when \(|Z|>\ell\), which has
probability \(e^{-\ell/s}\). The coupling bound and
Lemma~\ref{lem:prelim-privacy-comparison}(2) give the stated
additive privacy bound.
\end{proof}

\subsection{Matrix and frame estimates}\label{app:frame-projections}
\label{sec:prelim-matrices}\label{sec:prelim-change-variables}

\paragraph{Matrix identities and changes of variables.}
Use the record vectors \(\widetilde x_i=(x_i^\top,1)^\top\). For \(I\subseteq[n]\), let
\(T_I=\sum_{j\in I}\widetilde x_j\widetilde x_j^\top\).
Deleting a record \(i\in I\) subtracts its rank-one contribution.
\[
 T_{I\setminus\{i\}}=T_I-\widetilde x_i\widetilde x_i^\top .
\]
The inverse update gives the new quadratic forms for the surviving records.
The determinant update measures how much pruning changes the
statistic, which enters the density bound for the correction.

\begin{lemma}[Determinants and inverses]\label{lem:prelim-matrix-identities}
\leavevmode
\begin{enumerate}
\item For \(C\succ0\), the symmetric block matrix
      \(\left(\begin{smallmatrix}A&B\\ B^\top&C\end{smallmatrix}\right)\)
      is positive definite if and only if \(A-BC^{-1}B^\top\succ0\).
      Its determinant is \(\det C\,\det(A-BC^{-1}B^\top)\).
\item For \(U\in\R^{k\times\ell}\) and \(V\in\R^{\ell\times k}\),
      \(\det(I_k+UV)=\det(I_\ell+VU)\).
\item Let \(A\succ0\) and \(s=u^\top A^{-1}u\). Then
      \(A-uu^\top\succ0\) if and only if \(s<1\).
      Its determinant is \(\det(A)(1-s)\), and for \(s<1\),
      \[
       (A-uu^\top)^{-1}
       =A^{-1}+\frac{A^{-1}uu^\top A^{-1}}{1-s}.
      \]
\item If \(\|Z\|_\op<1\), then
      \((I-ZZ^\top)^{-1}=I+Z(I-Z^\top Z)^{-1}Z^\top\).
\end{enumerate}
\end{lemma}

\begin{proof}
For (1), the Schur-complement factorization
\cite[Appendix~A.5.5]{BoydVandenberghe2004}
\[
 \begin{pmatrix} A&B\\ B^\top&C\end{pmatrix}
 =
 \begin{pmatrix}I&BC^{-1}\\0&I\end{pmatrix}
 \begin{pmatrix}A-BC^{-1}B^\top&0\\0&C\end{pmatrix}
 \begin{pmatrix}I&0\\C^{-1}B^\top&I\end{pmatrix}
\]
preserves positive definiteness because the outer factors are
transposes of one another and are invertible. Their determinants
are one. For (2), take the determinant of
\(\left(\begin{smallmatrix}I_k&U\\ -V&I_\ell\end{smallmatrix}\right)\)
by eliminating either off-diagonal block.
For (3), put \(v=A^{-1/2}u\), so \(s=\|v\|^2\). The identity
\[
 A-uu^\top=A^{1/2}(I-vv^\top)A^{1/2}
\]
reduces positivity and the determinant to those of \(I-vv^\top\).
When \(v=0\), this is the identity. Otherwise the eigenvalue in
the direction of \(v\) is \(1-s\), and the eigenvalue on \(v^\perp\)
is one. Thus positivity holds exactly when \(s<1\), and the
determinant is \(\det A\,(1-s)\). For \(s<1\), direct
multiplication gives \((I-vv^\top)^{-1}=I+vv^\top/(1-s)\).
Multiplying this inverse on both sides by \(A^{-1/2}\) gives the
stated inverse of \(A-uu^\top\).
For (4), multiply the claimed inverse by \(I-ZZ^\top\) and use
\((I-Z^\top Z)^{-1}-Z^\top Z(I-Z^\top Z)^{-1}=I\).
\end{proof}

\begin{lemma}[Order and trace bounds]\label{lem:prelim-order-trace}
If \(0\prec A\preceq B\), then \(B^{-1}\preceq A^{-1}\) and
\(\det A\le\det B\). For a symmetric matrix \(H\), set
\(|H|=(H^2)^{1/2}\). Then
\[
 \|H\|_\op\le\|H\|_F\le\|H\|_*,\qquad
 |\tr(HG)|\le\|H\|_*\|G\|_\op,\qquad
 -|H|\preceq H\preceq |H|
\]
for square \(G\) of the same size. If \(A\succeq0\), then
\(\|A\|_*=\tr A\) and
\(\|A^{1/2}HA^{1/2}\|_*\le\tr(A|H|)\).
\end{lemma}

\begin{proof}
For the order inequalities, \(K=A^{-1/2}BA^{-1/2}\succeq I\)
has eigenvalues at least one. Thus \(K^{-1}\preceq I\) and
\(\det K\ge1\). Since \(B^{-1}=A^{-1/2}K^{-1}A^{-1/2}\) and
\(\det B=\det A\,\det K\), the two comparisons follow.
For the norm and trace bounds, write \(H=\sum_j h_j u_j u_j^\top\)
with orthonormal \(u_j\). The singular values are \(|h_j|\),
and \(-|h_j|\le h_j\le |h_j|\) gives the two matrix inequalities.
For the trace bound,
\(|\tr(HG)|\le\sum_j|h_j|\,|u_j^\top G u_j|
\le\|H\|_*\|G\|_\op\).
The triangle inequality applied to the rank-one terms of
\(A^{1/2}HA^{1/2}\) bounds its nuclear norm by
\(\sum_j |h_j|\|A^{1/2}u_j\|^2=\tr(A|H|)\).
\end{proof}

Deleting records can only decrease the unnormalized scatter.
Indeed, for weights \(w_i\ge0\) of positive total mass, let \(C_w\)
be the unnormalized scatter about the weighted mean. Then
\(v^\top C_w v=\min_{a\in\R}\sum_i w_i(v^\top x_i-a)^2\).
The minimizer is the projection of the weighted mean onto \(v\).
The objective increases pointwise when the weights increase, so
\(C_w\preceq C_{w'}\) whenever \(w_i\le w_i'\)
for every \(i\).

\begin{lemma}[Inverse and log determinant perturbations]
\label{lem:prelim-matrix-perturbation}
Let \(H\) be symmetric and \(\rho=\|H\|_\op<1\). Then \(I+H\succ0\)
and
\[
 \|(I+H)^{-1}-I\|_*\le\frac{\|H\|_*}{1-\rho},\qquad
 |\log\det(I+H)-\tr H|
 \le\frac{\|H\|_F^2}{2(1-\rho)} .
\]
\end{lemma}

\begin{proof}
For each eigenvalue \(h\) of \(H\), we have \(1+h\ge1-\rho>0\).
The eigenvalues of \(I+H\) are therefore positive, and
\(\lvert(1+h)^{-1}-1\rvert=|h|/(1+h)\le |h|/(1-\rho)\).
For the log determinant, integrate the derivative of
\(\log(1+th)-th\) from \(t=0\) to \(1\) to get
\[
 \log(1+h)-h=-h^2\int_0^1\frac{t}{1+th}\,dt .
\]
Since \(1+th\ge1-\rho\), the integral is at most
\(1/(2(1-\rho))\). Sum the first bound over the eigenvalues
for the inverse and the second for the log determinant.
\end{proof}

\begin{lemma}[Density under an affine matrix transformation]
\label{lem:prelim-affine-density}
Suppose \(Z\in\R^{r\times d}\) has Lebesgue density \(f_Z\).
Let \(Y=D+AZC\), with \(D\) fixed and
\(A\in\R^{r\times r}\), \(C\in\R^{d\times d}\) fixed and invertible.
Then
\[
 f_Y(y)=|\det A|^{-d}|\det C|^{-r}
        f_Z\bigl(A^{-1}(y-D)C^{-1}\bigr).
\]
\end{lemma}

\begin{proof}
Left multiplication by \(A\) acts separately on the \(d\) columns,
and right multiplication by \(C\) acts separately on the \(r\)
rows. Their absolute Jacobians are \(|\det A|^d\) and
\(|\det C|^r\), respectively. Translation by \(D\) has Jacobian one,
so the absolute Jacobian of the composition is their product.
The change of variables uses the inverse map
\(y\mapsto A^{-1}(y-D)C^{-1}\) and the reciprocal Jacobian.
\end{proof}

\paragraph{Frame block densities.}
The sampler and the determinant correction both use the density of
a block of a uniform frame. We obtain this density one column at a
time. Given the preceding columns, the next column is uniform on
the sphere in their orthogonal complement, so its retained entries
are a projection of a uniform direction. We then use this same
projection to control the inverse quadratic forms that enter the
joint privacy comparison.

\begin{lemma}[Density of a projected uniform direction]
\label{lem:prelim-projected-direction}
Let \(D>r\ge 1\), let \(\xi\) be uniform on the unit sphere in
\(\R^D\), and let \(L\in\R^{r\times D}\) have full row rank.
For \(M=LL^\top\), the density of \(L\xi\) is
\begin{equation}\label{eq:projected-sphere-density}
 \frac{\Gamma(D/2)}{\pi^{r/2}\Gamma((D-r)/2)}
 (\det M)^{-1/2}(1-z^\top M^{-1}z)^{(D-r-2)/2}
\end{equation}
on \(z^\top M^{-1}z<1\), and is zero elsewhere. Here \(\Gamma\) is
the gamma function.
\end{lemma}

\begin{proof}
Write \(\xi=g/\|g\|\) for a standard Gaussian vector \(g\), and split
\(g\) into its first \(r\) coordinates and the remaining \(D-r\)
coordinates. Their squared lengths \(U,V\) are independent chi-square
variables with \(r,D-r\) degrees of freedom. The squared length of the
first \(r\) coordinates of \(\xi\) is \(U/(U+V)\), independent of their
uniform direction. Put \(t=U/(U+V)\) and \(s=U+V\). The inverse
change of variables \(U=ts,\ V=(1-t)s\) has absolute Jacobian \(s\).
Integrating the resulting joint density over \(s>0\) gives the
beta density
\[
 \frac{\Gamma(D/2)}{\Gamma(r/2)\Gamma((D-r)/2)}
 t^{r/2-1}(1-t)^{(D-r)/2-1},\qquad 0<t<1.
\]
In polar coordinates, after integrating over the angles, the volume
element at squared radius \(t\) is
\(\pi^{r/2}t^{r/2-1}\,dt/\Gamma(r/2)\).
Dividing the beta density by this factor gives
\eqref{eq:projected-sphere-density} for \(M=I_r\).
By the projection part of Lemma~\ref{lem:prelim-frame-projection},
\(L\xi\overset{d}{=}M^{1/2}\xi_{[r]}\), where \(\xi_{[r]}\)
contains the first \(r\) coordinates of \(\xi\). The density
transformation contributes the factor \((\det M)^{-1/2}\)
and changes the squared radius to \(z^\top M^{-1}z\).
\end{proof}

\begin{lemma}[Density of selected frame rows]\label{ub-bg:haar-block-density}
Let \(r,p\ge1\) and \(N\ge r+p\). The first \(r\) rows \(Z\) of a
uniform \(N\times p\) frame have Lebesgue density
\begin{equation}\label{eq:frame-block-density}
 f(Z)=c_{N,r,p}\det(I_p-Z^\top Z)^{(N-r-p-1)/2},\qquad
 c_{N,r,p}=\pi^{-rp/2}\prod_{j=1}^p
 \frac{\Gamma((N-j+1)/2)}{\Gamma((N-r-j+1)/2)}
\end{equation}
on \(I_p-Z^\top Z\succ0\), and zero elsewhere.
\end{lemma}

\begin{proof}
For \(j=0,\ldots,p\), let \(Z_j\) be the top \(r\) rows of the
first \(j\) columns, and let \(D_j=\det(I_r-Z_jZ_j^\top)\), so
\(D_0=1\). For \(j\ge1\), write \(z_j\) for the first \(r\)
entries of column \(j\). We show by induction that
\(A_j=I_r-Z_{j-1}Z_{j-1}^\top\succ0\) almost surely, starting with
\(A_1=I_r\).

Given the first \(j-1\) full columns, column \(j\) is uniform on
the unit sphere of their \((N-j+1)\)-dimensional orthogonal
complement. The row Gram matrix of the projection map to the
first \(r\) coordinates is the top-left \(r\times r\) block
of the projector onto that complement. The projector subtracts
the outer products of the preceding full columns from the identity,
so this block is \(A_j\). By the
induction hypothesis the projection has full row rank, and
\(N-j+1>r\) because \(N\ge r+p\) and \(j\le p\). Thus
Lemma~\ref{lem:prelim-projected-direction} gives its density.
This density depends on the full preceding columns only through
\(Z_{j-1}\), so the same formula gives the conditional density
of \(z_j\) given \(Z_{j-1}\).

The support condition \(z_j^\top A_j^{-1}z_j<1\) is equivalent
to \(A_j-z_jz_j^\top\succ0\) by
Lemma~\ref{lem:prelim-matrix-identities}(3). This gives
\(A_{j+1}\succ0\) almost surely and completes the induction.
Conversely, \(I_r-ZZ^\top\succ0\) implies all preceding support
conditions because \(Z_jZ_j^\top\preceq ZZ^\top\).
On this support, the rank-one determinant formula gives
\(1-z_j^\top A_j^{-1}z_j=D_j/D_{j-1}\).
The projected-direction density contributes a factor \(D_{j-1}^{-1/2}\)
and the ratio \(D_j/D_{j-1}\) to the power \((N-j-r-1)/2\).
The resulting factors are
\[
 D_j^{(N-j-r-1)/2} D_{j-1}^{-(N-j-r)/2}.
\]
The negative power of \(D_{j-1}\) cancels its positive power
from column \(j-1\) whenever \(j\ge2\). Since \(D_0=1\),
multiplying the factors leaves \(D_p^{(N-r-p-1)/2}\). The conditional normalizing constants
multiply to \(c_{N,r,p}\). Sylvester's determinant identity gives
\(D_p=\det(I_p-Z^\top Z)\), and either support condition is
equivalent to \(\|Z\|_{\mathrm{op}}<1\).
\end{proof}

\paragraph{Concentration of projected lengths and directions.}
The joint privacy comparison uses quadratic forms of the inverse
matrix associated with the frame block. The diagonal forms reduce
to ratios of squared Gaussian lengths. A mixed form depends on a
signed projection as well as a length. The chi-square and
directional bounds below control these two contributions.

\begin{lemma}[Chi-square deviations]\label{lem:prelim-chi-square}
For \(X\sim\chi_h^2\), \(h\ge1\), and \(t>0\),
\[
 \Prob\{X-h>2\sqrt{ht}+2t\}\le e^{-t},
 \qquad
 \Prob\{h-X>2\sqrt{ht}\}\le e^{-t}.
\]
\end{lemma}

\begin{proof}
The chi-square moment generating function gives
\[
 \log \E e^{\lambda(X-h)}
 =\frac h2[-\log(1-2\lambda)-2\lambda]
 \le\frac{h\lambda^2}{1-2\lambda},\qquad 0\le\lambda<1/2.
\]
For the other sign,
\(\log \E e^{\lambda(h-X)}
 =\frac h2[2\lambda-\log(1+2\lambda)]\le h\lambda^2\)
for \(\lambda\ge0\). The bounds follow by integrating the
derivatives of the two bracketed expressions from zero.
Apply the exponential Markov inequality with
\(\lambda=\sqrt t/(\sqrt h+2\sqrt t)\) in the first bound
and \(\lambda=\sqrt{t/h}\) in the second.
\end{proof}

\begin{lemma}[Length ratios and directional tails]\label{lem:ub-scalar-facts}
Let \(U\sim\chi_r^2\) and \(V\sim\chi_q^2\) be independent, where
\(r,q\ge1\) are integers. If \(q>2\), then \(\E(U/V)=r/(q-2)\).
For \(t\ge1\) and \(q\ge16(r+t)\), with probability at least
\(1-4e^{-t}\),
\begin{equation}\label{eq:ratio-bounds}
 \left|\frac{q-2}{2}\frac UV-\frac r2\right|
 \le C(\sqrt{rt}+t),\qquad
 \frac UV\le C(r+t)/q,
\end{equation}
where one may take \(C=8\). If \(\xi\) is a uniform direction
in \(\R^D\), then for fixed \(w\in\R^D\) and \(t>0\),
\begin{equation}\label{eq:sphere-tail}
 \Prob\{|\xi^\top w|>\|w\|\sqrt{2t/D}\}\le 2e^{-t}.
\end{equation}
\end{lemma}

\begin{proof}
For the mean, independence gives \(\E(U/V)=\E U\,\E V^{-1}\).
Here \(\E U=r\). When \(q>2\), integrating \(v^{-1}\) against the
\(\chi_q^2\) density gives
\[
 \E V^{-1}
 =\frac{\Gamma(q/2-1)}{2\Gamma(q/2)}=\frac{1}{q-2}.
\]

For the deviation, Lemma~\ref{lem:prelim-chi-square} gives, with
probability at least \(1-4e^{-t}\),
\[
 |U-r|\le 2\sqrt{rt}+2t,\qquad
 |V-q|\le 2\sqrt{qt}+2t,\qquad
 V\ge q-2\sqrt{qt}\ge q/2 .
\]
The last inequality uses \(q\ge16t\). On this event,
\[
 \frac{q-2}{2}\frac UV-\frac r2
 =\frac{(q-2)(U-r)+r(q-2-V)}{2V}.
\]
Its absolute value is at most \(|U-r|+(r/q)(|V-q|+2)\).
Using the displayed deviations, \(r/q\le1/16\), and \(t\ge1\),
this is at most \(3(\sqrt{rt}+t)\). Also
\(U\le r+2\sqrt{rt}+2t\le3(r+t)\), so \(U/V\le6(r+t)/q\).

For the directional bound, represent a standard Gaussian vector
as \(g=\|g\|\xi\), with independent length and direction.
Divide the \(2j\)-th moment of \(g^\top w\) by the \(j\)-th
moment of \(\|g\|^2\sim\chi_D^2\). Independence of length and
direction then gives, for \(j\ge1\),
\[
 \E(\xi^\top w)^{2j}
 =\frac{\|w\|^{2j}(2j-1)!!}{D(D+2)\cdots(D+2j-2)}
 \le (\|w\|^2/D)^j(2j-1)!! .
\]
The double factorial is the product of the first \(j\) positive
odd integers. Odd moments vanish by symmetry. Expanding the
exponential gives \(\E e^{s\xi^\top w}\le
e^{s^2\|w\|^2/(2D)}\), which yields \eqref{eq:sphere-tail}
by applying the exponential Markov inequality to each sign
with \(s=\sqrt{2Dt}/\|w\|\) when \(w\ne0\). When \(w=0\),
the bound holds with probability one.
\end{proof}

\paragraph{Inverse quadratic forms.}
The density comparison uses the inverse of \(I_r-ZZ^\top\).
Isolating one frame column reduces its diagonal quadratic form
to a length ratio. Conditional on that length, the mixed form
is a linear functional of a uniform direction.

\begin{lemma}[Inverse quadratic forms of frame blocks]
\label{lem:ub-frame-quadratic}
Let \(r,p\ge1\) and let \(Z\in\R^{r\times p}\) be the first
\(r\) rows and \(p\) columns of a uniform orthogonal matrix
of order \(n-1\), with \(n-1\ge r+p\). Set
\(G=(I_r-ZZ^\top)^{-1}\), which exists almost surely by the
support condition in Lemma~\ref{ub-bg:haar-block-density}.
For each fixed unit \(v\in\R^p\),
\begin{equation}\label{eq:frame-ratio}
 v^\top Z^\top GZv\overset{d}{=}\frac{\chi_r^2}{\chi_{n-r-p}^2},
\end{equation}
with independent numerator and denominator. For fixed \(w\in\R^r\)
and \(t>0\), with probability at least \(1-2e^{-t}\),
\begin{equation}\label{eq:frame-cross}
 |v^\top Z^\top Gw|
 \le
 \sqrt{\frac{2t}{r}(w^\top Gw)
    (v^\top Z^\top GZv)(1+v^\top Z^\top GZv)}.
\end{equation}
The ratio identity also holds for \(Z^\top\), with \(r,p\) exchanged.
\end{lemma}

\begin{proof}
Choose an orthogonal matrix \(O\) with \(Oe_1=v\).
Right multiplication by \(O\) preserves the distribution of \(Z\)
and leaves \(ZZ^\top\) unchanged. The first column of \(ZO\) is
\(Zv\), so we may take \(v=e_1\) and write \(Z=[z,D]\) with
\(z\) the first column. Column permutation invariance lets us
expose the full columns \(2,\ldots,p\) before column \(1\).
Conditional on those columns, the first column is uniform on
the unit sphere of their orthogonal complement, of dimension
\((n-1)-(p-1)=n-p\). The projection map from this complement to
the first \(r\) coordinates has row Gram matrix \(A=I_r-DD^\top\).
This matrix is positive definite almost surely by
Lemma~\ref{ub-bg:haar-block-density}, and \(A=I_r\) if \(p=1\).
Thus \(x=A^{-1/2}z\) is distributed as the first \(r\) coordinates
of a uniform direction in \(\R^{n-p}\).

Since \(G^{-1}=A-zz^\top=A^{1/2}(I_r-xx^\top)A^{1/2}\),
we have \(\tau=\|x\|^2<1\) almost surely. The rank-one inverse
formula gives
\[
 G=A^{-1/2}\left(I_r+\frac{xx^\top}{1-\tau}\right)A^{-1/2},
 \qquad z^\top Gz=\frac{\tau}{1-\tau},\qquad
 z^\top Gw=\frac{x^\top A^{-1/2}w}{1-\tau}.
\]
As in the projected-direction proof, \(\tau\) has the distribution
of \(U/(U+V)\) for independent \(U\sim\chi_r^2\) and
\(V\sim\chi_{n-p-r}^2\). Thus \(\tau/(1-\tau)\) has the distribution
of \(U/V\). This conditional distribution is the same for every
choice of the preceding columns, proving \eqref{eq:frame-ratio}.

Conditional also on \(\tau\), write \(x=\sqrt{\tau}\,\xi\)
with \(\xi\) a uniform direction in \(\R^r\). Apply
\eqref{eq:sphere-tail} to the fixed vector \(A^{-1/2}w\).
With conditional probability at least \(1-2e^{-t}\),
\[
 |z^\top Gw|
 \le
 \sqrt{\frac{2t}{r}(w^\top A^{-1}w)
                 \frac{\tau}{(1-\tau)^2}} .
\]
The inverse formula gives \(G\succeq A^{-1}\), and
\(\tau/(1-\tau)^2=(z^\top Gz)(1+z^\top Gz)\).
Substitution and averaging over the conditioning give
\eqref{eq:frame-cross}. Since the transpose of a uniform
orthogonal matrix is uniform, the ratio for \(Z^\top\) follows
by exchanging \(r,p\).
\end{proof}

Equation~\eqref{eq:frame-ratio} supplies the means and diagonal
fluctuations in the joint privacy comparison. The factor \(1/r\)
in \eqref{eq:frame-cross} controls the mixed term on the same
square-root scale.

\subsection{Pruning and stability}\label{app:stable-weight-details}

The sampler in Section~\ref{sec:resampling} uses a score and a
weighted statistic. We define both, bound their changes on neighboring
datasets, and explain how the correction uses the deletion history. The rule adapts
the stable weights of Brown, Hopkins, and Smith
\cite[Section~3 of the full version]{BHS23} to the record vectors \(\widetilde x_i\).

\paragraph{The weighting rule.}
For \(I\subseteq[n]\), write \(\widetilde x_i=(x_i^\top,1)^\top\) and set
\(T_I=\sum_{i\in I}\widetilde x_i\widetilde x_i^\top\).
At a threshold \(u\), start with \(I=[n]\). At each iteration,
return the empty set if \(T_I\) is singular. Otherwise compute \(n\widetilde x_i^\top T_I^{-1}\widetilde x_i\) for each \(i\in I\). If all values are at most \(u\), return \(I\).
If any exceeds \(u\), delete the smallest such index from \(I\)
and repeat. The factor \(n\) is the original dataset size
throughout the run.

Fix integers \(k,\lambda\ge1\) and assume \(n\ge16k\lambda\).
For each \(j=0,\ldots,2k\), run pruning from all \(n\) records at
\(u_j=\lambda/(1-j\lambda/n)\), and call the returned set \(I_j\).
The thresholds are spaced so that removing one record from a set
satisfying threshold \(u_j\) leaves a set satisfying
\(u_{j+1}=u_j/(1-u_j/n)\), as shown below.

Levels \(0,\ldots,k-1\) determine the score, and levels
\(k+1,\ldots,2k\) determine the weights.
\[
 g=\min\bigl(\{j\in\{0,\ldots,k-1\}:|I_j|\ge n-j\}\cup\{k\}\bigr),
 \qquad
 \rho_i=\frac1k\sum_{j=k+1}^{2k}\one_{\{i\in I_j\}} .
\]
The weighted statistic is \(B=\sum_i\rho_i\widetilde x_i\widetilde x_i^\top\),
and its total weight is \(b=\sum_i\rho_i\).
Level \(k\) lies just below the averaging range. We show that
the sets are nested, so every record in \(I_k\) receives weight
one. When the test on a simulated reference dataset accepts,
the correction uses \([n]\setminus I_k\). Keep the order of
deletions at \(u_0=\lambda\) for use if that test rejects.

\paragraph{Stability bounds.}
For neighboring datasets, write \(g'\), \(I'_j\), \(\rho_i'\),
and \(B'\) for the corresponding quantities on the second input.
The score satisfies \(0\le g\le k\) and \(|g-g'|\le1\). If \(g<k\),
then \(|I_k|\ge n-g\), all records in \(I_k\) have weight one,
\(0\prec T_{I_k}\preceq B\preceq T_{[n]}\), and \(|I_k|\le b\le n\).
If \(g,g'<k\), then
\[
 \|B^{-1/2}B'B^{-1/2}-I_{d+1}\|_*\le16\lambda/n .
\]
On every input the base-threshold run returns \(|I_0|\le n-g\)
survivors. An invertible affine change \(x_i\mapsto Fx_i+v\) preserves the
pruning decisions, score, and weights. It sends \(B\) to
\(\widetilde F B\widetilde F^\top\), where
\(\widetilde F=\left(\begin{smallmatrix}F&v\\0&1\end{smallmatrix}\right)\).

\begin{proof}[Proof of the stability bounds]
\paragraph{Survivor sets at a fixed threshold.}
Call a set \(J\) valid at threshold \(u\) if \(T_J\succ0\) and \(n\widetilde x_i^\top T_J^{-1}\widetilde x_i\le u\) for every \(i\in J\).
If the current set \(I\) contains a valid set \(J\), then
\(T_I\succeq T_J\succ0\). Inverse monotonicity shows that every
row in \(J\) also meets the threshold relative to \(T_I\).
Thus pruning preserves every valid set. Its output is the unique
largest valid set if one exists, and the empty set otherwise.

A set valid at \(u_j\) is also valid at \(u_{j+1}\), since the
thresholds increase. The largest valid sets are therefore nested,
\(I_0\subseteq\cdots\subseteq I_{2k}\), with empty sets allowed.

\paragraph{Comparing neighboring runs.}
Fix neighboring datasets \(x,x'\) and let \(i_*\) be their
changed index. Write \(I_j'\) for the survivor set on \(x'\).
Fix \(0\le j<2k\) with \(I_j\ne\varnothing\).
To compare the survivor sets, remove \(i_*\) from \(I_j\) if
it is present. In that case put \(v=\widetilde x_{i_*}\).
The threshold condition gives \(v^\top T_{I_j}^{-1}v\le u_j/n<1\).
Cauchy--Schwarz in the inner product defined by \(T_{I_j}\)
gives \(vv^\top\preceq(v^\top T_{I_j}^{-1}v)T_{I_j}\).
Therefore the remaining statistic is at least
\((1-u_j/n)T_{I_j}\succ0\). Taking inverses and using the
threshold bound at \(u_j\) gives, for every \(i\in I_j\setminus\{i_*\}\),
\[
 n\widetilde x_i^\top T_{I_j\setminus\{i_*\}}^{-1}\widetilde x_i
 \le \frac{u_j}{1-u_j/n}=u_{j+1}.
\]
If \(i_*\notin I_j\), the statistic is unchanged and the larger
threshold still applies. All the remaining rows agree across the
two inputs, so \(I_j\setminus\{i_*\}\) is valid on the second
input at threshold \(u_{j+1}\).
The returned set \(I'_{j+1}\) contains every valid set at this
threshold. Applying this fact in both directions gives
\begin{equation}\label{eq:ub-stable-interlacing}
 I_j\setminus\{i_*\}\subseteq I'_{j+1},
 \qquad I'_j\setminus\{i_*\}\subseteq I_{j+1}
 \qquad (0\le j<2k).
\end{equation}
The inclusion from an empty set holds trivially.

\paragraph{Score and weight bounds.}
If \(g\ge k-1\), then
\(g'\le k\le g+1\). Otherwise, \(j=g\le k-2\) and \(|I_j|\ge n-j\).
The first inclusion gives \(|I'_{j+1}|\ge n-(j+1)\).
Since \(j+1<k\), the definition of \(g'\) gives \(g'\le j+1\).
Exchanging the inputs proves \(|g-g'|\le1\).

The nesting gives \(\rho_i=1\) for \(i\in I_k\) on every
dataset. If \(g<k\), then \(|I_g|\ge n-g\), so
\(|I_k|\ge n-g>n-k\). In particular, \(I_k\) is nonempty
and \(T_{I_k}\succ0\). All weights are in \([0,1]\),
and the records in \(I_k\) have weight one, so
\(0\prec T_{I_k}\preceq B\preceq T_{[n]}\) and \(|I_k|\le b\le n\).

For an unchanged index \(i\ne i_*\), the first inclusion in
\eqref{eq:ub-stable-interlacing} shifts each indicator at levels
\(k+1,\ldots,2k-1\) to the next level on \(x'\).
The term at level \(2k\) contributes at most one, so
\[
 k\rho_i
 \le 1+\sum_{j=k+2}^{2k}\one_{\{i\in I'_j\}}
 \le 1+k\rho_i'.
\]
The reverse inclusion gives \(|\rho_i-\rho_i'|\le1/k\).

Now assume \(g,g'<k\). A positive difference
\(\rho_i-\rho_i'\) requires \(i\notin I'_k\), since the rows of \(I'_k\)
have weight one. The same observation holds with primes exchanged.
Each complement has fewer than \(k\) rows, so
\begin{equation}\label{eq:weight-change}
 \sum_{i\ne i_*}(\rho_i-\rho_i')_+\le1,
 \qquad
 \sum_{i\ne i_*}(\rho_i'-\rho_i)_+\le1,
\end{equation}
where \(a_+=\max\{a,0\}\).

\paragraph{Relative change in the weighted statistic.}
To turn the weight bound into a matrix bound, we bound \(\widetilde x_i^\top B^{-1}\widetilde x_i\) for records of positive weight. Every positive-weight row belongs to \(I_{2k}\).
Each row of this set has outer product at most
\((u_{2k}/n)T_{I_{2k}}\) in the positive semidefinite order.
Since \(|I_k|\ge n-g>n-k\), the difference
\(T_{I_{2k}}-T_{I_k}\) sums at most \(k\) such outer products.
The records in \(I_k\) have weight one, so
\[
 B\succeq T_{I_k}\succeq (1-ku_{2k}/n)T_{I_{2k}}.
\]
Since \(u_{2k}=\lambda/(1-2k\lambda/n)\) and \(n\ge16k\lambda\),
the coefficient is positive. Inverse monotonicity now yields
\[
 \rho_i>0
 \quad\Longrightarrow\quad
 \widetilde x_i^\top B^{-1}\widetilde x_i
 \le \frac{u_{2k}/n}{1-ku_{2k}/n}
 =\frac{\lambda}{n-3k\lambda}
 \le\frac{2\lambda}{n}.
\]
The same bound holds for positive-weight rows on the second input.

To separate decreases from increases in the weighted statistic,
write \(B'-B=E'-E\), where
\[
\begin{aligned}
 E &= \sum_{i\ne i_*}(\rho_i-\rho_i')_+
          \widetilde x_i\widetilde x_i^\top
       +\rho_{i_*}\widetilde x_{i_*}\widetilde x_{i_*}^\top,\\
 E'&= \sum_{i\ne i_*}(\rho_i'-\rho_i)_+
          \widetilde x_i\widetilde x_i^\top
       +\rho_{i_*}'\widetilde x_{i_*}'(\widetilde x_{i_*}')^\top .
\end{aligned}
\]
Here \(\widetilde x_{i_*}'\) is the record vector at \(i_*\)
on the second input. Both matrices are positive semidefinite.
The records contributing to \(E\) have positive weight on \(x\),
and those contributing to \(E'\) have positive weight on \(x'\).
Their quadratic forms use \(B^{-1}\) and \((B')^{-1}\), respectively.
By \eqref{eq:weight-change} and \(\rho_i,\rho_i'\le1\), the
coefficients of the rank-one terms in each matrix sum to at most two. The bound on the quadratic forms gives
\(\tr(B^{-1}E)\le4\lambda/n\) and
\(\tr(B'^{-1}E')\le4\lambda/n\).
The second trace bounds the largest eigenvalue of
\(B'^{-1/2}E'B'^{-1/2}\). Since \(n\ge16k\lambda\), it is at most
\(1/2\), so \(E'\preceq B'/2\).
To measure both changes relative to \(B\), note that
\(B=B'-E'+E\succeq B'/2\). Inverse monotonicity gives
\(B^{-1}\preceq2B'^{-1}\), so \(E'\succeq0\) implies
\(\tr(B^{-1}E')\le2\tr(B'^{-1}E')\). Both \(B^{-1/2}EB^{-1/2}\) and \(B^{-1/2}E'B^{-1/2}\)
are positive semidefinite, so their nuclear norms equal their traces.
The triangle inequality now gives
\[
 \|B^{-1/2}B'B^{-1/2}-I_{d+1}\|_*
 \le \tr(B^{-1}E)+\tr(B^{-1}E')
 \le 12\lambda/n \le 16\lambda/n .
\]

\paragraph{The survivor count at the base threshold.}
For any dataset, put \(a=n-|I_0|\). If \(a<k\), nesting gives
\(|I_a|\ge n-a\), so \(g\le a\). If \(a\ge k\), we have
\(g\le k\le a\). Thus pruning at \(u_0=\lambda\) returns at most
\(n-g\) survivors.

Under the affine change above, \(\widetilde x_i\) becomes
\(\widetilde F\widetilde x_i\) and \(T_I\) becomes
\(\widetilde F T_I\widetilde F^\top\). This preserves the rank of
\(T_I\). For nonsingular \(T_I\),
\[
 (\widetilde F\widetilde x_i)^\top
 (\widetilde F T_I\widetilde F^\top)^{-1}
 (\widetilde F\widetilde x_i)
 =\widetilde x_i^\top T_I^{-1}\widetilde x_i.
\]
Thus the pruning decisions, score, and weights are unchanged.
Summing the weighted outer products gives
\[
 B\mapsto\sum_i\rho_i(\widetilde F\widetilde x_i)
                   (\widetilde F\widetilde x_i)^\top
       =\widetilde F B\widetilde F^\top.
\] If \(T_{[n]}\) is singular,
all subset statistics are singular and the routine returns
\(g=k\) and \(B=0\).
\end{proof}

\paragraph{Deletion sets for the correction.}
When the test on a simulated reference dataset accepts, \(g<k\)
and the selector records \(A_*=[n]\setminus I_k\). There are at
most \(g\) such indices.
Because \(I_k\) is nonempty, they are precisely the records
removed at threshold \(u_k\ge\lambda\). Every record outside
\(A_*\) has weight one, which gives \(T_{A_*^c}\preceq B\).

When the reference test rejects, the selector needs \(a_0\)
deletions, with \(a_0\) chosen in \eqref{eq:parameter-choice}.
It takes the first \(a_0\) indices in the ordered deletion list
at \(u_0=\lambda\). The determinant bound
in Section~\ref{sec:exact-correction} uses the quadratic form for each record just before its deletion. The fixed choice of the smallest
eligible index makes the order unambiguous, even though the final
survivor set is independent of the order of eligible deletions.

To see why the list is long enough on rejection, recall that an
ordinary reference dataset is obtained by centering and invertibly
normalizing independent Gaussian records. Every \(d+1\) records
remain affinely independent almost surely by
Lemma~\ref{lem:prelim-gaussian-components}. The same holds under
the reweighted reference distribution, which has a density
relative to the ordinary one. Since \(n-k>d\), a run remains
nonsingular after any of its first \(k\) deletions. On rejection
\(g\ge a_0\), with \(1\le a_0<k\). Stopping before \(a_0\)
deletions would leave more than \(n-a_0\) survivors, contradicting
\(|I_0|\le n-g\). Thus the selector can take the required prefix
on every such reference run outside a probability-zero event.

\subsection{Normalization and parameter choices}
\label{app:upper-calculations}
\label{sec:det-sampler}\label{sec:upper-calibration}

The selector in Section~\ref{sec:exact-correction} samples a
reference dataset from a distribution reweighted by a power of
the deletion determinant ratio. Its normalizing constant \(Z_a\)
is needed to evaluate the selection probability. An upper bound
on \(Z_a\) helps show that the probability is at most one.
We derive both and then verify the conditions on the parameters.

\paragraph{Computing the normalizing constant \(Z_a\).}
Fix a guessed set \(A\subseteq[n]\) with \(a=|A|\le k\), and
set \(q=\lfloor n/3\rfloor\). In this calculation assume
\(n-a-d-2q\ge n/4\), as verified below. Let \(U\) have the
ordinary uniform centered reference distribution. Write \(T(U)\)
for the unweighted statistic of all rows and \(T_{A^c}(U)\) for
that of the retained rows. Since \(n-a>d\), the general-position
argument above gives \(T_{A^c}(U)\succ0\) almost surely.
Define \(D_A(U)=\det T(U)/\det T_{A^c}(U)\) and
\(Z_a=\E[D_A(U)^q]\), with the expectation taken over \(U\) for
this fixed \(A\). Exchangeability makes \(Z_a\) depend only on \(a\).
The reweighted distribution has density \(r_A(U)=D_A(U)^q/Z_a\)
relative to the uniform centered reference distribution.
For \(a=0\), \(Z_0=1\). For \(a>0\), use the block \(F_A\) from
Section~\ref{sec:exact-correction}. Under the ordinary reference
distribution it has the density of the first \(a\) rows of a
uniform \((n-1)\times d\) frame. By \eqref{eq:determinant-frame-block},
multiplying this density by \(D_A^q\) introduces the constant
\((n/(n-a))^q\) and changes the determinant exponent to that of
a block from a frame with \(n-1-2q\) rows, on the same support.
Write \(c_{N,a,d}\) for the density constant in
Lemma~\ref{ub-bg:haar-block-density}. The determinant factor with
the new exponent integrates to \(1/c_{n-1-2q,a,d}\). Thus
\[
 Z_a=\left(\frac n{n-a}\right)^q
       \frac{c_{n-1,a,d}}{c_{n-1-2q,a,d}}.
\]
For each $i\in[d]$, put $x_i=(n-i)/2$ and $y_i=(n-a-i)/2$.
The ratio of the constants contributes
$\Gamma(x_i)\Gamma(y_i-q)/(\Gamma(x_i-q)\Gamma(y_i))$
at index $i$. Applying $\Gamma(x+1)=x\Gamma(x)$ yields the finite
product
\begin{equation}\label{eq:full-det-exact}
 Z_a=\left(\frac n{n-a}\right)^q
       \prod_{i=1}^d\prod_{j=1}^q
       \frac{n-i-2j}{n-a-i-2j}.
\end{equation}
The same expression gives one for $a=0$. Each denominator in the
product is at least $n/4$, as is $n-a$. Consequently,
\[
 \begin{aligned}
 \log Z_a
 &=q\log\left(1+\frac{a}{n-a}\right)
   +\sum_{i=1}^d\sum_{j=1}^q
        \log\left(1+\frac{a}{n-a-i-2j}\right)\\
 &\le \frac{4q(d+1)a}{n}
 \le 2(d+1)a.
 \end{aligned}
\]
The selector uses \eqref{eq:full-det-exact} to evaluate \(Z_a\)
and the bound \(Z_a\le e^{2(d+1)a}\) to control the selection
probability.

Section~\ref{sec:exact-correction} shows how to complete the
reweighted block \(F_A\) to obtain the simulated dataset, using
\eqref{eq:weighted-frame-recipe} when \(a>0\). When \(a=0\), the
selector draws an ordinary uniform centered frame.

\paragraph{Choosing parameters.}
Let $d,m\ge1$, $0<\alpha,\delta<1$, and $0<\varepsilon\le1$.
Set $\eta=\alpha/(2m)$ and define
\[
 L=1+\log\frac{dm}{\alpha\delta\varepsilon},
 \qquad
 F=mL+\frac{(d+L)(\sqrt{mL}+L)}{\varepsilon}.
\]
Let $C_{\rm priv}$ be the universal constant in the private sampler's
privacy bound in Section~\ref{sec:joint-privacy}. Increase it if
needed so that $t\ge1$ below.
For a sufficiently large universal constant $C_0$, take
\begin{equation}\label{eq:parameter-choice}
 \begin{gathered}
 n=\lceil C_0^2F\rceil,\qquad
 \lambda=\lceil C_0(d+L)\rceil,\qquad
 a_0=\left\lceil\frac{16m\log(1/\eta)}{\lambda}\right\rceil,\\
 \ell=1+\frac2\varepsilon\log\frac{32}{\delta},\qquad
 k=a_0+\lceil 2\ell\rceil+1,\qquad
 t=\log\frac{C_{\rm priv}(d+1)}{\delta}.
 \end{gathered}
\end{equation}

The test rejects when the score plus the clipped noise is at
least \(a_0+\ell\). The noise lies in \([-\ell,\ell]\), so
rejection gives \(g\ge a_0\), which certifies \(a_0\) deletions
on a reference dataset. If the test accepts on either of two
neighboring inputs, the score on that input is less than
\(a_0+2\ell\). The scores differ by at most one, so both are
below \(k\), as required for the joint comparison.

We verify three sample-size requirements. The first bounds the
size needed for privacy of the private sampler, the second gives the
size condition for the stability bounds, and the third gives room for the
frames and the determinant exponent in the correction. Since
$t=O(L)$, $\ell=O(L/\varepsilon)$, and $\log(1/\eta)\le L$,
the ceilings in \eqref{eq:parameter-choice} give
\[
 \begin{aligned}
 k\lambda
 &\le 16m\log(1/\eta)+(2\ell+3)\lambda=O(C_0F),\\
 \frac{\lambda(\sqrt{mt}+t)}{\varepsilon}&=O(C_0F),\\
 m+d+t&=O(F).
 \end{aligned}
\]
The constants here are independent of $C_0$. Taking $C_0$ large
ensures
\[
 n\ge C_{\rm priv}\left(k\lambda+m+d+t+
          \frac{\lambda(\sqrt{mt}+t)}{\varepsilon}\right),
 \qquad
 n\ge16k\lambda,
 \qquad
 n\ge256(m+d+1).
\]

The third bound gives \(q-m\ge n/4\) and \(q\ge5n/16\) because
\(q=\lfloor n/3\rfloor\ge n/3-1\).
For the correction on rejection, the choice of \(a_0\) gives
\(a_0\lambda\ge16m\log(1/\eta)\). To check the requirement
\(\lambda\ge16(d+2+\log n)\), first bound \(F\le5e^{3L}\), since
$d,m,L,1/\varepsilon\le e^L$. Hence
$\log n\le 3L+2\log C_0+\log 10$ for $C_0\ge1$, and
$\lambda\ge16(d+2+\log n)$ for a sufficiently large $C_0$.
The same choice gives $\lambda\ge d+1$. Since
$L\ge1+\log(1/\delta)$, it also gives
$\lambda\ge8\log(4/\delta)$, so $e^{-\lambda/8}\le\delta/4$.
The selector includes each index in its guessed set with probability
\(p=e^{-\lambda/8}/n\), independently across indices. The guessed
set is empty with probability \(w=(1-p)^n\). Applying the selection
identity in Section~\ref{sec:exact-correction} to the whole proposal
space shows that the private branch is taken with probability \(w\).
Thus \(1-w\le np=e^{-\lambda/8}\le\delta/4\).

Finally, $k\le n/16$ and $\max\{m,d\}\le n/256$ give
$n-k>\max\{m,d\}$. For every $a\le k$, the reduced frame obeys
\[
 n-a-d-2q\ge n/3-k-d\ge n/4.
\]
In particular, $n-1-2q\ge a+d$, so its block density and the
normalizer calculation above are valid. The original frame has
$n-1\ge m+d+1$ rows, which makes the candidate density's
determinant exponent nonnegative.
This establishes the parameter conditions used in
Section~\ref{sec:complete-sampler}.

\section{Supporting estimates for the lower bound}
\label{app:lower-estimates}\label{app:lower}

\subsection{Concentration of the ideal batch direction}
\label{app:alignment-concentration}

In the Gaussian family of Section~\ref{sec:batch-alignment},
\(B\in\R^{k\times k}\) is the unknown parameter and \(k\) is the
dimension of each Gaussian block. That section uses total variation
to transfer an alignment event from an ideal batch to the released
batch for each fixed \(B\). Here we bound the probability of the
event for the ideal batch, averaging over \(B\) and the ideal
draws. With noise variance \(\sigma^2=km\), the inner product
with \(B\) still reaches order \(\sqrt{k}\).

\begin{lemma}[A correlated direction from an ideal batch]\label{lem:lb:direction}
Let \(k,m\ge1\), and let \(B\in\R^{k\times k}\) have independent
standard Gaussian entries. Conditional on \(B\), draw an ideal batch
of \(m\) independent pairs \((z,w)\), where \(z,g\) are independent
standard Gaussian vectors in \(\R^k\) and \(w=Bz+\sigma g\).
Write \(Z^\star,W^\star\in\R^{k\times m}\) for the matrices with
columns \(z,w\), respectively. Define \(A^\star=W^\star(Z^\star)^\top\)
and \(T^\star=A^\star/\|A^\star\|_F\) when \(A^\star\ne0\), with
\(T^\star=0\) otherwise. If \(\sigma^2=km\), the probability
over \(B\) and the ideal batch satisfies
\begin{equation}\label{eq:lb:ideal}
 \Prob\{\langle B,T^\star\rangle\ge\sqrt{k}/8\}\ge 1-26/k.
\end{equation}
\end{lemma}

\begin{proof}
Set \(S=Z^\star(Z^\star)^\top\) and \(t=\tr S\).
Let \(G\) have the batch noise vectors \(g\) as its columns.
Then
\[
 A^\star=BS+\sigma G(Z^\star)^\top,
\]
with \(B,G,Z^\star\) mutually independent. The trace \(t\) is a sum
of \(km\) independent squared standard Gaussians, so it has mean
\(km\) and variance \(2km\). By Chebyshev's inequality,
\(\Prob\{t<km/2\}\le 8/(km)\). Fix \(Z^\star\) with
\(t\ge km/2\), so \(\sigma^2\le2t\). We bound the numerator and the
denominator of the normalized inner product conditional on \(Z^\star\).

For the numerator \(R=\langle B,A^\star\rangle\), let \(b\in\R^k\)
and \(h\in\R^m\) be the transposes of corresponding rows of \(B\)
and \(G\). They are independent standard Gaussians, and the row
contributes \(b^\top Sb+\sigma b^\top Z^\star h\). The quadratic
term has mean \(t\) and variance \(2\tr(S^2)\), by diagonalizing
\(S\). The second term has conditional mean zero given \(b\), and
averaging its conditional variance gives \(\sigma^2 t\).
The two terms are uncorrelated. Summing over the \(k\) independent
row contributions yields
\[
 \E[R\mid Z^\star]=kt,\qquad
 \operatorname{Var}(R\mid Z^\star)
   =k\bigl(2\tr(S^2)+\sigma^2t\bigr)\le4kt^2.
\]
The bound uses \(\tr(S^2)\le t^2\) because \(S\succeq0\), and
\(\sigma^2\le2t\) on the event under consideration.
Chebyshev's inequality gives \(\Prob\{R<kt/2\mid Z^\star\}\le16/k\).

For the denominator, condition only on \(Z^\star\), so \(B\)
and \(G\) are still random. The rows of \(A^\star\) are then
independent centered Gaussians with covariance
\(D=S^2+\sigma^2S\).
In an eigenbasis of \(D\), the squared norm of each row is the sum
of independent \(\chi_1^2\) variables weighted by its eigenvalues.
The mean and variance of each \(\chi_1^2\) variable are \(1\) and \(2\),
respectively. Summing over all \(k\) rows gives
\[
 V:=\E[\|A^\star\|_F^2\mid Z^\star]=k\tr D\le3kt^2,\qquad
 \operatorname{Var}(\|A^\star\|_F^2\mid Z^\star)
   =2k\tr(D^2)\le 2V^2/k.
\]
Thus \(\|A^\star\|_F^2\le 2V\le6kt^2\) except with conditional
probability \(2/k\). On the two good events \(R\ge kt/2>0\), so
\[
 \langle B,T^\star\rangle
 \ge\frac{kt/2}{\sqrt{6k}\,t}
 =\frac{\sqrt{k}}{2\sqrt{6}}>\frac{\sqrt{k}}8.
\]
The total failure probability is at most \(8/(km)+16/k+2/k\le26/k\).
\end{proof}

\subsection{Gaussian tails for fixed signed selections}
\label{app:fixed-selection-tails}\label{sec:lb:sums}

Section~\ref{sec:lower-privacy} rounds the coefficients of a
Gaussian matrix sum to a random sign pattern \(V\).
For each deterministic pattern \(v\), we bound the norm of its
matrix sum and its expected contribution on \(A=\{V=v\}\).
This event may depend on all the Gaussian vectors.

\begin{lemma}[A fixed signed Gaussian sum]\label{lem:lb:fixed-selection}
Let \(k,N\ge1\) and let \((g_i,z_i)_{i=1}^N\) be mutually
independent pairs of independent standard Gaussian vectors in \(\R^k\).
Fix \(v\in\{-1,0,1\}^N\) with \(j\ge1\) nonzero entries and put
\(Y_v=\|\sum_i v_i g_i z_i^\top\|_F\). For universal constants \(C_0,C>0\),
\begin{equation}\label{eq:lb:fixed-tail}
 \Prob\{Y_v>C_0[k\sqrt{j+u}+j+u]\}\le3e^{-u},
 \qquad u\ge0.
\end{equation}
For any event \(A\) with \(p=\Prob(A)>0\), including one that
depends on the Gaussian vectors,
\begin{equation}\label{eq:lb:event-selection}
 \E[Y_v\one_A]\le
 Cp\bigl[k\sqrt{j+\log(1/p)}+j+\log(1/p)\bigr].
\end{equation}
When \(v=0\), \(Y_v=0\).
\end{lemma}

\begin{proof}
For the tail bound, put the vectors \(v_i g_i\) and \(z_i\)
for indices with \(v_i\ne0\) into the columns of \(E\) and \(F\),
respectively. Each \(v_i g_i\) is standard Gaussian, so \(E\) and
\(F\) are independent standard Gaussian matrices in \(\R^{k\times j}\),
and \(Y_v=\|EF^\top\|_F\).
For fixed \(F\), the map \(E\mapsto\|EF^\top\|_F\) is
\(\|F\|_\op\)-Lipschitz in Frobenius norm. Its conditional mean
is at most \(\sqrt{k}\,\|F\|_F\), since
\(\E[\|EF^\top\|_F^2\mid F]=k\|F\|_F^2\).
Gaussian concentration for Lipschitz functions
\cite[Theorem~5.2.2]{Vershynin2018} therefore gives, with conditional
probability at least \(1-e^{-u}\),
\[
 \|EF^\top\|_F\le\sqrt{k}\,\|F\|_F+C_1\sqrt{u+1}\,\|F\|_\op
\]
for a universal constant \(C_1\).

Both norms of \(F\) on the right are \(1\)-Lipschitz in Frobenius
norm. Their expectations satisfy \(\E\|F\|_F\le\sqrt{kj}\) and
\(\E\|F\|_\op\le\sqrt{k}+\sqrt{j}\)
\cite[Theorem~7.3.1]{Vershynin2018}. Applying the same concentration
bound to the two norms gives
\(\|F\|_F\le\sqrt{kj}+C_1\sqrt{u+1}\) and
\(\|F\|_\op\le\sqrt{k}+\sqrt{j}+C_1\sqrt{u+1}\), each with
failure probability at most \(e^{-u}\).
Integrating the conditional failure probability over \(F\) bounds
the unconditional failure probability in the first estimate by
\(e^{-u}\). A union bound over the three estimates gives
\[
 \|EF^\top\|_F\le
 \sqrt{k}(\sqrt{kj}+C_1\sqrt{u+1})
 +(\sqrt{k}+\sqrt{j}+C_1\sqrt{u+1})C_1\sqrt{u+1}
\]
with probability at least \(1-3e^{-u}\). Using \(k,j\ge1\) and
\(2\sqrt{j(u+1)}\le j+u+1\) gives \eqref{eq:lb:fixed-tail}.

To bound the contribution on \(A\), note that conditioning on it can
increase a tail probability by at most \(1/p\).
Set \(f(u)=C_0[k\sqrt{j+u}+j+u]\) and choose
\(u_0=\log(3/p)\), so \(3e^{-u_0}=p\) cancels this factor.
Then \eqref{eq:lb:fixed-tail} gives
\[
 \Prob\{Y_v>f(u_0+s)\mid A\}\le e^{-s}
 \qquad(s\ge0).
\]
Integrate this tail above \(f(u_0)\) using \(x=f(u_0+s)\).
Writing \(r=j+u_0\ge1\), we obtain
\begin{align*}
 \E[Y_v\mid A]
 &\le f(u_0)+\int_0^\infty e^{-s}f'(u_0+s)\,ds\\
 &\le C_0\left[k\sqrt r+r+\frac{k}{2\sqrt r}+1\right]
 \le 2C_0[k\sqrt r+r].
\end{align*}
Since \(j\ge1\), \(r\le(1+\log3)[j+\log(1/p)]\).
Multiplying by \(p\) proves \eqref{eq:lb:event-selection}.
\end{proof}

In the main proof, apply \eqref{eq:lb:event-selection} to
\(A=\{V=v\}\) with \(p_v=\Prob\{V=v\}>0\). The total weighted
cost \(\sum_{v\ne0:p_v>0}p_v\log(1/p_v)\) is bounded by the
entropy \(H(V)=\sum_{v:p_v>0}p_v\log(1/p_v)\).
\end{document}